\documentclass[11pt,a4paper]{article}

\usepackage{amsmath}
\usepackage{amssymb}
\usepackage{appendix}
\usepackage{multirow}
\usepackage{geometry}
\usepackage{float}
\usepackage{axodraw}
\usepackage{epsf,epsfig,psfrag,graphics,epstopdf}
\def\vys {{Vysostski$\breve{\rm{\i}}$}}

\usepackage{color}

\usepackage{cite}

\usepackage{graphicx,axodraw}

\usepackage{jheppub}

\title{Short-distance QCD corrections to {\boldmath $K^0\bar{K}^0$} mixing\\ at next-to-leading order in Left-Right models}

\author[a]{V\'eronique Bernard,}
\author[b]{S\'ebastien Descotes-Genon}
\author[a,b]{and Luiz Vale Silva}

\affiliation[a]{Groupe de Physique Th\'eorique, Institut de Physique Nucl\'eaire, UMR 8608,\\
CNRS, Univ. Paris-Sud, Universit\'e Paris-Saclay, 91406 Orsay Cedex, France}
\affiliation[b]{Laboratoire de Physique Th\'eorique, UMR 8627,\\
CNRS, Univ. Paris-Sud, Universit\'e Paris-Saclay, 91405 Orsay Cedex, France}

\emailAdd{bernard@ipno.in2p3.fr}
\emailAdd{sebastien.descotes-genon@th.u-psud.fr}
\emailAdd{luiz.vale@th.u-psud.fr}

\abstract{Left-Right (LR) models are extensions of the Standard Model where left-right symmetry
is restored at high energies, and which are strongly constrained by kaon mixing
described in the framework of the  $|\Delta S|=2$ effective Hamiltonian.
We consider the short-distance QCD corrections to this Hamiltonian
both in the Standard Model (SM)
and in LR models. The leading logarithms occurring in these short-distance corrections can be resummed within a rigourous
Effective Field Theory (EFT) approach integrating out heavy degrees of freedom progressively, or using an approximate simpler method of regions identifying the ranges of loop
momentum generating large logarithms in the relevant two-loop diagrams. We compare the
two approaches in the SM at next-to-leading order, finding a very good agreement when one scale dominates the problem, but only a fair agreement in the presence of a large logarithm at leading order. We compute the short-distance QCD
corrections for LR models at next-to-leading order using the method of regions, and we compare the results with the EFT approach for the $WW'$ box with two charm quarks (together with additional diagrams forming a gauge-invariant combination), where a large logarithm occurs already at leading order. We conclude by providing next-to-leading-order estimates for $cc$, $ct$ and $tt$ boxes in LR models.}


\arxivnumber{1512.00543}

\begin{document}

\maketitle

\flushbottom








A natural extension of the Standard Model (SM) is provided by  Left-Right (LR) symmetric models,
which explain the left-handed structure of the SM through the existence of a larger gauge group
$SU_C(3)\times SU_L(2)\times SU_R(2)\times U_Y(1)$, broken first at a scale
$\mu_R$ of the order of the TeV (inducing a difference between left and right sectors) followed by an electroweak symmetry breaking occurring at a scale $\mu_W$~\cite{Pati:1974yy,Mohapatra:1974hk,Mohapatra:1974gc,Senjanovic:1975rk,Senjanovic:1978ev}. This extension induces the presence of heavy spin-1 $W'$ and $Z'$ bosons predominantly coupling to right-handed fermions, introducing a new CKM-like matrix for right-handed quarks, as well as charged and neutral heavy Higgs bosons with an interesting pattern of flavour-changing currents~\cite{Chang:1982dp,Zhang:2007da}.
Such a framework has been revived in the recent years for its potential collider implications when parity restoration in the LHC energy reach is considered~\cite{Maiezza:2010ic,Guadagnoli:2010sd}.

Many different mechanisms can be invoked to trigger the breakdown of the left-right symmetry.
Historically, LR models (LRM) were first considered with doublets in order to break the left-right symmetry spontaneously. Later the focus was set on triplet models, due to their ability to generate both Dirac and Majorana masses for neutrinos and thus introducing a see-saw mechanism~\cite{Mohapatra:1980yp,Deshpande:1990ip}. LR models provide also interesting candidates for a $Z'$ boson as currently hinted at by $b\to s\ell\ell$ observables~\cite{ Descotes-Genon:2013wba,Descotes-Genon:2014uoa,Descotes-Genon:2015uva}. Stringent constraints come from electroweak precision observables~\cite{Hsieh:2010zr} and from direct searches at LHC~\cite{ATLAS:2012ak,Khachatryan:2014dka,Chen:2013fna,Dev:2015kca,Patra:2015bga}, pushing the limit for LR models to several TeV. Studies in the framework of flavour physics suggest also that the structure for the right-handed CKM-like matrix should be quite different from the left-handed one, far from the manifest or pseudo-manifest scenarios~\cite{Harari:1983gq,Beall:1981ze,Langacker:1989xa,Barenboim:1996nd,Barenboim:2001vu}.

In this setting, a particularly important indirect constraint comes from kaon-meson mixing, favouring a mass scale for the new scalar particles of a few TeV or beyond~\cite{Mohapatra:1977mj,Mohapatra:1983ae,Barenboim:1996wz,Blanke:2011ry,Bertolini:2014sua}. This comes from the very accurate measurement of kaon mixing together with the possibility of generating kaon mixing in the LR model by exchanging at tree level a heavy neutral Higgs boson with flavour-changing neutral couplings. As usual in flavour physics, such a process involves dynamics occurring at several different scales: the heavy degrees of freedom $W'$ of mass of order $\mathcal{O}(\mu_R)$, the degrees of freedom occurring at the electroweak symmetry breaking $\mu_W$, and the dynamics at low energies (around the charm quark mass or below). The first range is addressed 
directly in the LR model whereas  the last energy domain is tackled by lattice QCD computations, which now provide accurate kaon mixing matrix elements for the operators in the SM and beyond~\cite{Carrasco:2015pra}. The two  domains can be bridged thanks to the effective Hamiltonian approach, which also provides an elegant framework to take into account higher-order QCD corrections~\cite{Buchalla:1995vs}.

Indeed, short-distance QCD corrections prove to have an important impact on the computation of 
kaon mixing in the Standard Model, easily increasing or decreasing the contributions from the different diagrams to the amplitude by 50\%. This large impact stems from the multi-scale nature of the problem, leading to the presence of large logarithms (for instance $\alpha_s \cdot \log(m_c^2/M_W^2)$). This requires a resummation of the leading logarithms, which  can be obtained by applying an Effective Field Theory (EFT) approach to the problem. One considers a tower of effective Hamiltonians where heavy degrees of freedom are integrated out progressively and which can be matched onto each other. The renormalisation group equations provide the resummation of the 
large logarithms in a natural way, which requires dedicated computations of two-loop diagrams~\cite{Gilman:1982ap,Buras:1990fn,Buras:1991jm,Herrlich:1993yv,Herrlich:1994kh,Herrlich:1996vf,Buras:2000if}.

In the early days of these computations, an alternative method was proposed in Refs.~\cite{Vainshtein:1975xw,Vysotskii80}, attempting at catching the main effects of large logarithms by considering the relevant regions of momentum integration in the diagrams. This method of regions was applied to resum the leading logarithms both in the SM~\cite{Vysotskii80} and LR models~\cite{Ecker:1985vv,Bigi:1983bpa}, with a much more limited amount of computation, since most of the method relies on anomalous dimensions already known.

The aim of the present paper is to reconsider the evaluation of short-distance QCD corrections needed to evaluate neutral-meson mixing (and in particular kaon meson mixing) precisely in the case of LR models. In Sec.~\ref{sec:method}, we recall a few elements of the two methods in the SM case at Leading Order (LO), before illustrating how the method of regions of Refs.~\cite{Vysotskii80,Ecker:1985vv,Bigi:1983bpa} could be extended to Next-to-Leading Order (NLO) and comparing the results
with the EFT case. 
In Sec.~\ref{sec:LRRM}, we discuss the additional contributions arising in LR models and we
compute short-distance QCD corrections at NLO using  the method of regions. In Sec.~\ref{sec:EFT}, we use
the EFT approach to compute these corrections in the case of the $cc$ box with $W$ and $W'$ exchanges (together with additional diagrams to get a gauge-invariant contributions), where a large logarithm occurs already at leading order.  Our final
results for the short-distance corrections in LRM are gathered  
in Sec.~\ref{sec:FinalResult}. We provide our conclusions in Sec.~\ref{sec:concl}. Several appendices are devoted to more technical aspects of the computation.

\section{Short-distance QCD corrections in the Standard Model}\label{sec:method}

\subsection{Generalities on the EFT computation} \label{sec:SMEFT}

The analysis of kaon mixing is customarily performed in the framework of the effective Hamiltonian, separating short and long distances in the following way~\cite{Buchalla:1995vs}
\begin{eqnarray}\label{eq:SMEffHamiltonian}
&& H=\frac{G_F^2}{4 \pi^2}M_W^2
\Bigg[ \lambda_c^{LL} \lambda_c^{LL} \eta_{cc} S^{LL}(x_c)
    + \lambda_t^{LL} \lambda_t^{LL} \eta_{tt} S^{LL}(x_t) +2 \lambda_t^{LL} \lambda_c^{LL}\eta_{ct} S^{LL}(x_c,x_t)\Bigg]b(\mu_{h}) Q_{V} \nonumber\\
&& \qquad +h.c. ,
\end{eqnarray}
where the local $|\Delta S|=2$ operator involved is
\begin{equation}\label{eq:Qv}
Q_{V}=(\bar{s}^\alpha \gamma_\mu P_Ld^\alpha)(\bar{s}^\beta \gamma^\mu P_L d^\beta)
 =  \frac{1}{4}  (\bar{s}d)_{V-A}(\bar{s}d)_{V-A}\, .
\end{equation}
This result involves the short-distance QCD corrections $\eta_{cc}, \eta_{tt}, \eta_{ct}$
(note that in the literature these corrections are also called $\eta_1, \eta_2, \eta_3$, respectively).
$S^{LL}$ are related to the usual Inami-Lim functions depending on the quark masses through $x_i=m_i^2/M_W^2$ (see Eq.~\eqref{eq:inami}) and $\lambda_i^{LL}=V^{CKM}_{is} (V^{CKM}_{id})^{*}$ combines
two CKM matrix elements. 
The derivation of this result relies on the GIM mechanism to eliminate
the $\lambda_u^{LL}$ terms.

The matrix element $\langle\bar{K}^0 | H|K^0\rangle$ can be computed knowing   $\langle\bar{K}^0|Q_{V}|K\rangle$ from lattice QCD simulations at a low hadronic scale
$\mu_{h}$ of a few GeV~\cite{Carrasco:2015pra} and  $b(\mu_{h})$ is a function which combines with 
$\langle\bar{K}^0 | Q_{V} |K^0\rangle$ to form a renormalisation-group invariant quantity. This
function contains the scale dependence of the Wilson coefficient due to its running
down to the hadronic scale. Note that in the literature this function is sometimes 
absorbed into the definition of the QCD correction factor: 
\begin{equation}
\bar \eta = \eta \, b(\mu_{h}) ,
\end{equation}
which is thus scale and renormalisation-scheme dependent. In the
discussion of LR  models we will deal with the scale-dependent $\bar\eta$ factors, as it proves easier to deal with the latter in the case of several $|\Delta S|=2$ local operators mixing among each other. In the absence of the resummation of short-distance QCD corrections we would have $\eta_{ct}=\eta_{cc}=\eta_{tt}=1$. This 
clearly also holds for the scale-dependent terms $\bar \eta$.

The determination of the short-distance QCD contributions requires a detailed analysis of the effective Hamiltonian in the SM, performed in Ref.~\cite{Herrlich:1996vf}.
After integrating out the top quark and the $W$ boson we are left with
an effective five-flavour Hamiltonian of the form
\begin{eqnarray}
H &=&
-\frac{G_F}{\sqrt{2}} \lambda^{LL}_i\lambda^{LL}_j \sum_{k} C_{k} Q_k
-\frac{G_F^2}{2} \lambda^{LL}_i\lambda^{LL}_j \sum_{l} \widetilde{C}_{l} \widetilde{Q}_{l}.
\label{lgeneric}
\end{eqnarray}
The
$Q_k$, $\widetilde{Q}_l$ are local $|\Delta S|=1$\ and $|\Delta S|=2$\ operators and
the $C_k$, $\widetilde{C}_l$ are the corresponding Wilson coefficients.  The $|\Delta S|=1$\ operators
$Q_k$ are necessary since they contribute to the $|\Delta S|=2$ transition amplitude through four-point functions
with two operator insertions. The $|\Delta S|=2$\ operators $\widetilde{Q}_k$ can be obtained by shrinking the top-top box to a point. Yet the $\widetilde{Q}_k$'s are also needed for the light-quark contributions, since diagrams with two operator insertions are in general divergent and require counterterms proportional to $|\Delta S|=2$ operators.

The detailed structure of the effective Hamiltonian has been worked out in Ref.~\cite{Herrlich:1996vf}. We summarize the different steps of the 
calculation here following closely this reference:
\begin{enumerate}
\renewcommand{\labelenumi}{\roman{enumi})}
\item Find the minimal operator basis  in Eq.~\eqref{lgeneric} to describe the physics of $|\Delta S|=2$ transition and closing under renormalization. 
\item Consider the full SM Green function $\widetilde{G}$ describing 
the transition of interest (at the leading order of $m_c/M_{t,W}$, one can neglect the external momenta) and
match to the one obtained in the effective theory to obtain the Wilson coefficients 
 $C_k$ and $\widetilde{C}_l$ at the high scale $\mu=\mu_{tW}=\mathcal{O} (M_W,m_t)$.
\item Determine the RG evolution of the Wilson coefficients from the high scale
$\mu=\mu_{tW}$ down to the low scale $\mu=\mu_c=\mathcal{O} (m_c)$.  This must be obtained by considering
the  general RG equation for Green functions with double insertions and its solution.
The RG equation involves an anomalous dimension tensor in
addition to the familiar anomalous dimension matrices, requiring the  calculation of two-loop
diagrams.
\item If needed, perform the matching onto theories with fewer flavours when crossing a threshold, in particular the charm quark mass.
\end{enumerate}

The computation requires the choice of a regularisation scheme for the ultraviolet divergences arising in the theory
(typically, the NDR-$\overline{MS}$ scheme) and for the infrared divergences (usually by keeping small masses for the external quarks). Also, the simplification of operators in $D$ dimensions requires the introduction of evanescent operators, which can contribute to the physical quantities once inserted in loops.

Since in  the case of the LRM we will follow the same lines and
in order for the paper to be self-contained  
we recall  the main elements of the SM analysis of the $|\Delta S|=2$ 
Lagrangian performed in Ref.~\cite{Herrlich:1996vf} in App.~\ref{app:SMcomput},
borrowing heavily from that reference. We will just
summarise  a few important features for the determination of the short-distance corrections $\eta$ at the order of leading and next-to-leading logarithms in the next section.

\subsection{EFT computation: specific issues}

In the case of the $tt$ box~\cite{Buras:1990fn}, the Wilson coefficient can be obtained easily by integrating out both the $W$ boson and the $t$ quark at a high scale $\mu_{tW} =
\mathcal{O}(m_t,M_W)$ (the initial conditions of the Wilson coefficients are determined by integrating out the top quark and the $W$ boson simultaneously, thus neglecting the evolution between the
scales $\mu_t$ and $\mu_W$, see
Ref.~\cite{Herrlich:1996vf} for further discussion). The corresponding effective Hamiltonian consists of a single operator $Q_V$ multiplied by a Wilson coefficient obtained by matching at $\mu_{tW}$. The coefficient is then run down to $\mu_{h}$. The analytic expression for $\eta_{tt}$ can be found in App.~\ref{app:SMVy}

The $cc$ box~\cite{Herrlich:1993yv} has the additional complication that the charm quark cannot be integrated out at the same time as the $W$ boson. One first integrates out the $W$ boson, leading to a $|\Delta S|=1$ effective Hamiltonian of the form
\begin{equation} \label{eq:hc}
H^c=\frac{4 G_F}{\sqrt{2}}\sum_{U,V=u,c} V^{CKM}_{Us} (V^{CKM}_{Vd})^{*}
  (C_+ O^{UV}_+  + C_- O^{UV}_-)\,,
\end{equation}
involving the  $|\Delta S|=1$ operators which do not mix into each other under QCD when penguin operators are not present
\begin{equation}
O^{UV}_\pm=\frac{O_1^{UV}\pm O_2^{UV}}{2}\,,
\end{equation}
with
\begin{equation}
O_1^{UV}=  \frac{1}{4}  (\bar{s}^\alpha U^\alpha)_{V-A} (\bar{V}^\beta d^\beta)_{V-A},
\qquad O_2^{UV}=  \frac{1}{4}  (\bar{s}^\alpha U^\beta)_{V-A} (\bar{V}^\beta d^\alpha)_{V-A}\,,
\end{equation}
where $\alpha,\beta$ are colour indices. $|\Delta S|=2$ transitions occur through bilocal operators of
the form $\int d^4y \, T[H_c(x) H_c(y)]$
yielding a sum of four bilocal operators $O_{ij}$ (with $i,j=\pm$):
\begin{eqnarray}\label{eq:hcc}
H^{cc}&=&2 G_F^2 \lambda_c^2 \sum_{i,j=\pm} C_iC_j O_{ij} ,\\
O_{ij}(x)&=&-2 i \int d^4y \, T[O_i^{cc}(x) O_j^{cc}(y)+O_i^{uu}(x) O_j^{uu}(y) \nonumber\\
&& \qquad -O_i^{cu}(x) O_j^{uc}(y)-O_i^{uc}(x) O_j^{cu}(y)] . \, \label{eq:bilocSMcc}
\end{eqnarray}
The Wilson coefficients of the operators $O_{ij}$ (equal to the product $C_iC_j$) must be evolved  from $\mu_W=\mathcal{O}(M_W)$ down to $\mu_c$, before matching onto a theory without charm  containing the single operator $Q_V$,
see Eq.~\eqref{eq:Qv} (at NLO, the matching must be performed at $\mathcal{O}(\alpha_s)$). The resulting coefficient must be evolved down to $\mu_{h}$. 
Note that in some renormalisation schemes one could have to add a set of penguin operators in Eq.~\eqref{eq:hc}
(for more detail see Ref.~\cite{Buras:1991jm}).

Finally, the top-charm contribution $\eta_{ct}$ requires a more involved analysis of the renormalisation group structure of the theory~\cite{Herrlich:1996vf}. The first step consists in integrating out the $t$ and $W$ quarks, adding to the $|\Delta S|=1$ Hamiltonian Eq.~\eqref{eq:hc} a set of penguin operators.
The resulting expression is
\begin{eqnarray}
H^{ct}&=&2 G_F^2 \lambda_c\lambda_t 
      \left[ \sum_{i=\pm, \, j=1, \ldots, 6} C_iC_j O_{ij} + C_7 Q_7 \right]\,,\\
O_{ij}(x)&=&-2 i \int d^4y \, T[2O_i^{uu}(x) O_j^{uu}(y)-O_i^{cu}(x) O_j^{uc}(y)-O_i^{uc}(x) O_j^{cu}(y)] , \label{eq:bilocSMct}
\end{eqnarray}
for $ j=1,2 $, with a similar result for bilocal operators involving penguins $j=3, \ldots, 6$, and an additional $|\Delta S|=2$ operator
\begin{equation}
Q_7=\frac{m_c^2}{g^2 \mu^{2 \epsilon}}  \frac{1}{4}  (\bar{s}d)_{V-A} (\bar{s}d)_{V-A} \,,
\end{equation}
which is required as the bilocal operators $O_{ij}$ exhibit an ultraviolet divergence which has
to be regularised by a local counterterm (this problem does not occur for the $cc$ box as the divergences cancel due to the GIM mechanism). This results into the logarithmic contribution $-x_c\log x_c$ to the corresponding Inami-Lim function contained in $S^{LL}(x_c,x_t)$, not present in the $cc$ case. This means that there is a mixing between the bilocal operators $O_{ij}$ and the local operator
$(\bar{s}d)_{V-A} (\bar{s}d)_{V-A}$ at leading order, even before taking QCD corrections into account. 
This undesirable feature can be avoided by introducing the $1/g^2$ normalisation factor for
$Q_7$, so that this mixing is treated on the same footing as QCD radiative corrections and
a common RGE framework can be applied to discuss the mixing of all the operators~\cite{Gilman:1982ap,Buchalla:1995vs}. 
This theory can be evolved down to the charm quark mass, where it is matched onto a theory without charm, containing the single operator $Q_V$ once again, to be evolved down to $\mu_{h}$. Neglecting any effects of the five-flavour theory  and switching off the penguin operators  whose  contribution has been found to be of the order of $1\%$ allows one to write a relatively simple expression for $\eta_{ct}$~\cite{Herrlich:1996vf}.

In the SM case, the short-distance QCD correction is known at next-to-leading order (NLO) for the dominant top-quark contribution, $\eta_{tt}=0.5765\pm 0.0065$~\cite{Buras:1990fn,Herrlich:1996vf}. Since $\epsilon_K$ is the relevant
observable for kaon mixing and arises by considering the imaginary part of the $|\Delta S|=2$ matrix element, 
the small imaginary part of $\lambda_t^{LL}$ means that the top-top contribution can be of similar size to the charm-top and charm-charm contributions.
This led to an evaluation of these contributions at NNLO, leading to a significant 
positive shift compared to NLO for  $\eta_{cc}=1.87\pm 0.76$~\cite{Brod:2011ty} and a $7 \%$ increase for 
$\eta_{ct}=0.496\pm 0.047$~\cite{Brod:2010mj}  ($\eta_{tt}$ remaining almost 
unchanged). This illustrates the importance of higher orders in the evaluation of the short-distance 
QCD corrections.

\subsection{Method of regions at leading order}\label{sec:SMLO}

Historically, the first determination of 
$K^0 \bar K^0$ mixing  in the SM did not take into account the short-distance QCD 
corrections~\cite{VK83,Gaillard:1974hs}.
 A method to determine these corrections by
resumming  the leading logarithms was then developed  
in the case of the charm quark~\cite{Vainshtein:1975xw}, the inclusion of the top quark being studied in Ref.~\cite{Vysotskii80}.
It was further used to calculate the mixing in Left-Right symmetric models 
\cite{Ecker:1985vv,Bigi:1983bpa}. In the following this method 
will be called ``method of regions'' (MR) for reasons that will become clear soon.

Contrarily to more recent works which
use the EFT approach presented in Sec.~\ref{sec:SMEFT},  this method aims at  catching the main features in
an approximate way.
Let us summarise briefly the underlying idea, basically amounting to resum
the leading logarithms with the help of renormalisation group equations. We
consider first the calculation of the ${\cal O}(\alpha_s)$  corrections to the
one-loop $c$ quarks contribution to the Green function with the insertion of 
four weak currents ($cc$ box). This was done in Refs.~\cite{Buras:1990fn,Herrlich:1993yv}, taking into account the GIM mechanism and leading to
\begin{equation}
\langle H^{cc}(\mu)\rangle=\langle H^{cc}(\mu)\rangle^{(0)} + \frac{\alpha_s(\mu)}
{4 \pi} \langle H^{cc}(\mu)\rangle^{(1)} + {\cal O}(\alpha_s^2)\,,
\end{equation}
where $\langle H \rangle^{(n)}$ denotes the value of the matrix element between $K^0$ and $\bar{K}^0$ external states at $O(\alpha_s^n)$. We have
\begin{eqnarray}\label{eq:Hcc5f}
\hspace{-0.5cm}&\langle H^{cc}(\mu)\rangle^{(0)}&=\frac{G_F^2}{4 \pi^2} \lambda_c^2 m_c^2( \mu)\langle Q_{V} ( \mu) \rangle^{(0)} \,,
 \nonumber\\
\hspace{-0.5cm}&\langle H^{cc}(\mu)\rangle^{(1)}&= \frac{ 3 G_F^2}{2 \pi^2} \lambda_c^2 m_c^2( \mu)\langle Q_{V} ( \mu) \rangle^{(0)} \biggl[ - C_F \log\left(\frac{m_c^2}{\mu^2}\right) \biggr. \\
&&\qquad\qquad + \frac{N-1}{2 N} \left(2 \log\left(\frac{m_c^2}{M_W^2}\right) - \log\left(\frac{m_c^2}{\mu^2}\right)\right)\biggr] 
+ \cdots
\nonumber
\end{eqnarray}
where $N$ denotes the number of colours and the ellipsis contains constant terms  proportional to 
$ \langle Q_{V} (\mu) \rangle^{(0)}$ and contributions from unphysical
operators that are not relevant here. Indeed, in the leading-logarithm approximation
one only keeps track of the logarithms  in Eq.~\eqref{eq:Hcc5f} and resums them
to all orders in perturbation theory.

Instead of performing the 
whole calculation,  it was  rather proposed in Refs.~\cite{Vainshtein:1975xw,Vysotskii80} to analyse all the possible ways of dressing the box diagrams with gluons.
The one-loop momentum $k$ of the original graph  is kept fixed, and one has to identify the region for the gluon  momentum $q$ leading to a logarithmic behaviour. These logarithms are then resummed at fixed $k$ and finally the integration over $k$ is  performed. Let us illustrate this procedure in the case of the $ \alpha_s \cdot \log (m_c^2/M_W^2) $ contribution  in Eq.~\eqref{eq:Hcc5f}.

\vys\  showed that the integration over $q^2$ in the range $[k^2,M_W^2]$ in the left diagram in Fig.~\ref{fig:diagVys}  leads to a term  $\log(k^2/M_W^2)$, 
responsible for the second logarithm  (for $k^2 =\mathcal{O}(m_c^2$))  in Eq.~\eqref{eq:Hcc5f}. Cutting this graph  along the two internal quark lines yields the set of 
multiplicatively renormalised operators contributing to each half of the diagram, giving rise to the bilocal operators $O_{ij}$ introduced in Eq.~\eqref{eq:bilocSMcc}. Using RGE over the relevant range of momentum for $q^2$ provides the resummation of logarithms as required

\begin{center}
\begin{figure}[t!]
\begin{center}
\begin{picture}(200,50)(0,0)
\ArrowLine(80,25)(40,60)
\Text(60,55)[]{$d$}
\ArrowLine(40,-55)(80,-20)
\Text(60,-50)[]{$s$}
\ArrowLine(125,25)(80,25)
\Photon(80,25)(80,-20){4}{8}
\GlueArc(50,-20)(60,0,75){4}{10}
\ArrowLine(160,60)(125,25)
\Text(145,55)[]{$s$}
\ArrowLine(125,-20)(160,-55)
\ArrowLine(80,-20)(125,-20)
\Photon(125,25)(125,-20){4}{8}
\Text(145,-50)[]{$d$}
\Text(90,40)[]{$\searrow$}
\Text(95,45)[]{$q$}
\Text(70,0)[]{$\uparrow$}
\Text(60,0)[]{$k$}
\Text(105,-30)[]{$c$}
\Text(110,35)[]{$c$}
\end{picture}
\begin{picture}(200,50)(0,0)
\ArrowLine(80,25)(40,60)
\ArrowLine(40,-55)(80,-20)
\ArrowLine(125,25)(80,25)
\Photon(80,25)(80,-20){4}{8}
\GlueArc(80,-20)(80,-21,114){4}{20}
\ArrowLine(160,60)(125,25)
\ArrowLine(125,-20)(160,-55)
\ArrowLine(80,-20)(125,-20)
\Photon(125,25)(125,-20){4}{8}
\Text(155,30)[]{$\searrow$}
\Text(160,35)[]{$q$}
\Text(70,0)[]{$\uparrow$}
\Text(60,0)[]{$k$}
\Text(105,-30)[]{$c$}
\Text(110,35)[]{$c$}
\end{picture}
\end{center}
\vspace{2.cm}
\caption{\small\it Typical SM $cc$ box diagram leading to the contributions $\log({m_c^2}/{M_W^2})$  (four possibilities for gluon exchanges in total, left) and $\log({m_c^2}/{\mu_{h}^2})$ (two possibilities in total, right) in the computation of short-distance QCD corrections to kaon mixing.}\label{fig:diagVys}
\end{figure}
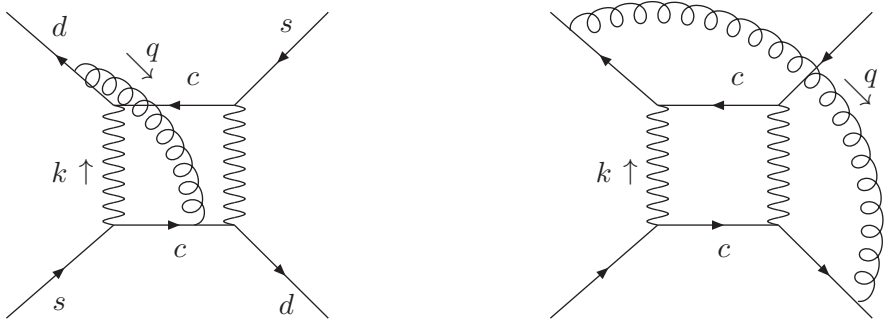
\end{center}
\vspace{-0.2cm}
\begin{equation}
\frac{1}{2} \biggl (\frac{\alpha_s(k^2)}{\alpha_s(M_W^2)} \biggr)^{8/\beta_0}
- \biggl (\frac{\alpha_s(k^2)}{\alpha_s(M_W^2)} \biggr)^{2/\beta_0}
+\frac{3}{2} \biggl (\frac{\alpha_s(k^2)}{\alpha_s(M_W^2)} \biggr)^{-4/\beta_0}
= \sum_{i,j=\pm} t_{ij} \biggl (\frac{\alpha_s(k^2)}{\alpha_s(M_W^2)} 
\biggr)^{d_{ij}}\,,
\label{eq:resumv}
\end{equation}
where the exponents  $ d_{ij}=\gamma_{ij}^{(0)}/(2 \beta_0)$ come from the anomalous dimensions $\gamma_{ij}$  of the bilocal operators
$O_{ij}$ involved (corresponding to the sum of the anomalous dimensions for the individual $|\Delta S|=1$ operators), $\beta_0=(11 N -2 f)/3$ is the first term in the expansion of the usual renormalisation group function that governs the evolution of the QCD
coupling constant (with $f$ the number of active flavours),
and 
\begin{equation}\label{eq:tauSM}
t_{ij}=\frac{1}{4}(1+i+j+N \  ij)\,, \qquad i,j=\pm\,,
\end{equation}
is a factor arising from the matching of the bilocal operators $O_{ij}$ onto the $|\Delta S|=2$ local operator, leading to the same integral but with different coefficients due to the different projectors involved.

After having introduced the resummation of large logarithms coming from the operator evolution, we still have to perform
the remaining integration over the momentum $k$, typically
\begin{equation}
\int d^4 k\ f(k^2)  \biggl (\frac{\alpha_s(k^2)}{\alpha_s(M_W^2)} 
\biggr)^{\gamma}\,,
\end{equation}
($\gamma=0$ corresponds to the original loop integral without radiative corrections), which is treated 

in two different ways depending on the behaviour of the one-loop integral.
If it has a power law behaviour dominated by a single mass scale $m$  i.e. ($ a \neq 0 $)
\begin{equation}
\int d^4 k \, f(k^2) \sim (m^2)^a\,,
\end{equation}
we can replace the integral as follows
\begin{equation}\label{eq:integonescale}
\int d^4 k\ f(k^2)  \biggl (\frac{\alpha_s(k^2)}{\alpha_s(\mu^2)} 
\biggr)^{\gamma} \sim (m^2)^a \biggl (\frac{\alpha_s(m^2)}{\alpha_s(\mu^2)} 
\biggr)^{\gamma}\,.
\end{equation}
This is our case in Eq.~\eqref{eq:Hcc5f} since $\langle H^c(\mu_W)\rangle^{(0)}\propto m_c^2$,
and we obtain a sum of contributions to the
Wilson coefficient of the form
\begin{equation}\label{eq:SMresummed}
m_c^2 \biggl (\frac{\alpha_s(m_c^2)}{\alpha_s(M_W^2)} \biggr)^{d_{ij}}\,.
\end{equation}
If we expand it at leading order in $\alpha_s \log({m_c^2}/{M_W^2}) $ 
using the evolution of $\alpha_s$ between two scales
\begin{equation}
\alpha_s(m_1)=\frac{\alpha_s(m_2)}{1 -\beta_0 \frac{\alpha_s(m_2)}{2 \pi} 
\log\left(\frac{m_2}{m_1}\right)} \,,
\end{equation}
we obtain
\begin{equation}
 \frac{\alpha_s}{4 \pi} \log\left(\frac{m_c^2}{M_W^2}\right) \sum_{i,j=\pm} 
\frac{\gamma_{ij}^{(0)}}{2} t_{ij} \, , \quad \quad \sum_{i,j=\pm}  \frac{\gamma_{ij}^{(0)}}{2} t_{ij}=12\frac{N-1}{2N} \,,
\end{equation}
showing that the resummed expression Eq.~\eqref{eq:SMresummed} indeed reproduces the large logarithm  in Eq.~\eqref{eq:Hcc5f}.

The resummations leading to the two other logarithms in Eq.~\eqref{eq:Hcc5f} is performed in
a similar way. The last logarithm comes from a diagram where the gluon is attached to two external quarks of same flavour, see the right diagram in Fig.~\ref{fig:diagVys}. The relevant range of integration of $q^2$ is $[\mu^2_{h} ,k^2]$, where $\mu_{h}$ is the low hadronic scale. The relevant anomalous dimension is then the one attached to the $|\Delta S|=2$ local operator. Once again, the remaining integration over $k^2$ can be simplified by noticing that only the scale $k^2=\mathcal{O}(m_c^2)$ is relevant (for more detail, see Refs.~\cite{Vysotskii80,Ecker:1985vv}). The first logarithm in Eq.~\eqref{eq:Hcc5f}
comes from the evolution of the charm quark mass from the $m_c$ scale down to $\mu_{h}$. Finally, we take also into account the diagrams with a gluon with both ends attached to the same internal quark line, leading to a renormalisation of the corresponding quark masses $m_q$ (to be evaluated at the scale $\mu=m_q$).
In the SM, taking into account the GIM mechanism, all the box diagrams with internal quark lines of the same flavour exhibit 
such a power law behaviour for which  the procedure Eq.~(\ref{eq:integonescale}) holds.

In the case of the top-charm box, matters are a bit more complicated. 
Indeed the corresponding original integral has not a simple power law behaviour, but instead a logarithmic behaviour as stated before, i.e.
$\int_{m_1^2}^{m_2^2} d k^2 f(k^2)=\log(m_2^2/m_1^2)$. In 
this case one defines the LO averaging weight $R(\gamma,m_1,m_2)$ such that
\begin{equation} \label{eq:Rlog}
(\log(m_2^2/m_1^2))^{-1}\int_{m_1^2}^{m_2^2} \frac{d k^2}{k^2}  \biggl (\frac{\alpha_s(k^2)}{\alpha_s(\mu^2)} 
\biggr)^{\gamma} =R(\gamma,m_1,m_2)  \biggl (\frac{\alpha_s(m_1^2)}{\alpha_s(\mu^2)} 
\biggr)^{\gamma}\,.
\end{equation}
The method of regions amounts thus to computing the Wilson coefficients at
the lower scale $m_1^2$ and to multiply them by the appropriate factors $R$. 

One should in principle also consider contributions coming from the graphs where one or both $W$ bosons are replaced by Goldstone bosons. Actually, the sum of those diagrams ($WW$, $WG$, $GG$) is independent of the gauge chosen for the electroweak bosons, and the discussion can be performed in the unitarity gauge where only the $WW$ diagram should be considered.

An additional comment is in order concerning the anomalous dimensions and the number of active flavours.
In the EFT approach one performs a matching onto an effective Hamiltonian 
valid between two scales determined by the  
number of flavours involved, integrating out a quark flavour each time the scale gets lower than the corresponding quark threshold. One then runs the Wilson coefficient from one scale to 
the other. In \vys's original  procedure, it is assumed that the
$t$ and $b$ quarks do no appear in large logarithms so that $f$ could be chosen as $3$ or $4$, arguing that  the difference between the numerical values of $\beta_0$ (involved in the running of the operators) for  $f=5$ and $f=4$
would anyway be very small~\cite{Vysotskii80}. Thus only two scales have to be considered, $\mu_c$ and the 
low scale $\mu_{h}$ at which the matrix element of the relevant operator is computed.
In a similar vein, in the case of the presence of the logarithm in 
$\langle H^c(\mu_{h})\rangle^{(0)}$ \vys\  did not distinguish the anomalous 
dimension of the $|\Delta S|=2$ local operator between the
scale $\mu_c$ and $\mu_W$ and below $\mu_c$. A later reference~\cite{Datta:1995he} showed how to include the effect of these thresholds.

In Ref.~\cite{Ecker:1985vv}, the same method was reexpressed in a slightly different language. Expressed in the SM case, it amounts to considering the bilocal operators Eqs.~\eqref{eq:bilocSMcc} and \eqref{eq:bilocSMct}, running them from the high scale $\mu_W^2$ to a scale $k^2$, and multiplying the evolution factors given by the RGE with the evolution factor coming from the local $|\Delta S|=2$ operator from 
the scale $k^2$ down to $\mu_{h}^2$. This provides the two contributions to large logarithms from the diagrams displayed in Fig.~\ref{fig:diagVys}. The integration with respect to $k^2$ is then performed by the procedure outlined in Eqs.~\eqref{eq:integonescale} and \eqref{eq:Rlog}.

The LO values of the short-distance QCD corrections in the 
SM for the kaon system using this method are given in Tab.~\ref{tab:etaSM}
 and compared
with the values obtained from a systematic EFT approach~\cite{Herrlich:1996vf}.  We included the flavour thresholds neglected by \vys. We do not provide $\eta_{cc}$ as it  turns out that its expression  is identical  in both
 approaches up to NLO, see Eq.~(XII.31) in Ref.~\cite{Buchalla:1995vs} for example, for 
the expression in the EFT approach.
The numerical results are obtained using the same inputs as in Ref.~\cite{Herrlich:1996vf},
namely \linebreak $m_t(m_t) =167$~GeV,
$m_c(m_c)=\mu_c=1.3$~GeV, $M_W=80$~GeV, $\Lambda^{(4)}=0.310$~GeV. The matchings onto the effective theories are performed at
$\mu_b =4.8$ GeV, whereas the high scale $\mu_W$ is chosen differently depending on the
box considered: $\mu_W=130$ GeV when a $t$ quark is involved in order 
to take care of the fact that in the EFT approach the top quark and the 
$W$ boson are integrated out at the same time (hence $\mu_W$ is an average of the two masses), whereas  $\mu_W=M_W$ when only $c$ and $u$ quarks are involved and only the $W$ boson has to be integrated out in the diagram. As can be seen in Tab.~\ref{tab:etaSM}, the method of regions works very well  at leading order.

\begin{table}[t!]
\begin{center}
{\renewcommand{\arraystretch}{1.4}
\begin{tabular}{|c| c c  |}
\hline
& $\eta_{tt}$& $\eta_{ct}$\\
\hline
MR  &$ 0.591 - 0.010 = 0.581$& $0.345-0.011=0.334$ \\
\hline
EFT &$0.612-0.038=0.574$&$0.368+0.099=0.467$\\
\hline
\end{tabular}}
\end{center}
\caption{\small\it Comparison of the SM short-distance QCD corrections using 
the method of regions (MR)  and a systematic EFT approach. The first number corresponds to 
the LO (resummation of $(\alpha_s \log(m_c/\mu))^n$) and the second to the NLO (resummation of $\alpha_s (\alpha_s \log(m_c/\mu))^n $). Note that in the
case of $\eta_{ct}$ the LO in the four-quark theory corresponds to a resummation of 
$(\alpha_s \log(m_c/M_W))^n  \log(m_c/M_W)$ and the NLO
to $(\alpha_s \log( m_c/M_W))^n$.
Flavour thresholds are taken into account. Both approaches lead to an identical result
in the case of $\eta_{cc}$, not shown here.}
\label{tab:etaSM}
\end{table}

\subsection{Method of regions at next-to-leading order}\label{sec:MRNLO}

We will now extend the method of regions to determine the short-range corrections $\eta$ at NLO taking advantage that the anomalous dimensions of 
all (most of) the operators involved have been determined for the SM (LRM~\footnote{Some additional anomalous
dimensions  needed for the LRM will be discussed in the EFT approach,
Sec.~\ref{sec:EFT}.}) \cite{Buras:2000if}.  Following closely what is done in the
EFT approach one  uses 
the renormalisation group equations for the Wilson coefficients to determine 
them at $\mathcal{O}(\alpha_s)$ (requiring to know both matching and anomalous dimensions at this order). Second one should calculate the $\mathcal{O}(\alpha_s)$
corrections to the operators involved. Indeed considering both kinds of corrections is mandatory in order to get a scheme-independent result.

We can check that extending the method of regions at NLO is appropriate by applying it to the SM case first. We use the result of Ref.~\cite{Herrlich:1993yv} for 
the calculation of the $\mathcal{O}(\alpha_s)$ 
corrections of the $|\Delta S|=2$ local operator $Q_V$ appearing
in the effective four- and three-quark theories for the computation of  $\eta_{cc}$.
The expressions of  $\eta_{tt}$ and $\eta_{ct}$ at NLO are given in App.~\ref{app:SMVy} and are obtained by including the same diagrams and integration ranges as in the LO case, but considering the additional $\mathcal{O}(\alpha_s)$ corrections for the matching and evolution and modifying the averaging procedure to take them into account. The numerical results are gathered in Tab.~\ref{tab:etaSM}.

In the case of $\eta_{cc}$ (which is identical in the EFT MR approaches), let us just stress the importance of the $\mathcal{O}(\alpha_s)$
corrections  $\beta_{ij}$, ($i,j=\pm$) coming from the matching of the product of operators $O_\pm$ onto the $|\Delta S|=2$ local operators. We obtain, using the
same input as before except by setting $ \mu_W = M_W $:
\begin{equation}
\eta_{cc}=0.89 + (0.62 - 0.19) \qquad {\rm (EFT)}\,,
\end{equation}
where the first number corresponds to the LO result (in Ref.~\cite{Buchalla:1995vs},
the LO result corresponds to a calculation with the LO value of $\alpha_s$
leading to $\eta_{cc}=0.74$), the second 
and third numbers are the NLO contributions, the former coming from $\beta_{ij}$ and the latter corresponding to the remaining contributions. The matching at $\mu_W$ is also important: neglecting the scheme-invariant quantity  $\alpha_s(\mu_W)(B_i+B_j
-J_{ij})$ (where $B_\pm$ comes from the matching of the SM to $O_\pm$ operators at $\mu_W$
and $J$ comes from the anomalous dimension matrix of these operators)  would lead to a  $7\%$ increase coming almost 
entirely from the $B_i$ terms.
 
In Tab.~\ref{tab:etaSM} the NLO contributions obtained with the method of regions are compared to the EFT approach. The agreement is quite good for the short-distance corrections with two same quarks in the loop, which do not involve any large logarithm 
in the calculation without QCD corrections. A small discrepancy is obtained in the case of $\eta_{tt}$
which can be traced back to the fact that the top quark is not integrated
out at the same time as the $W$ boson contrarily to the EFT case. The MR 
method is much less accurate at NLO for $\eta_{ct}$, where large logarithms are present and the top quark is not treated on the same footing as the $W$ though both are heavy degrees of freedom: our way of extending \vys 's method yields a result with a $30\%$ discrepancy.

\section{QCD corrections for Left-Right models}\label{sec:LRRM}

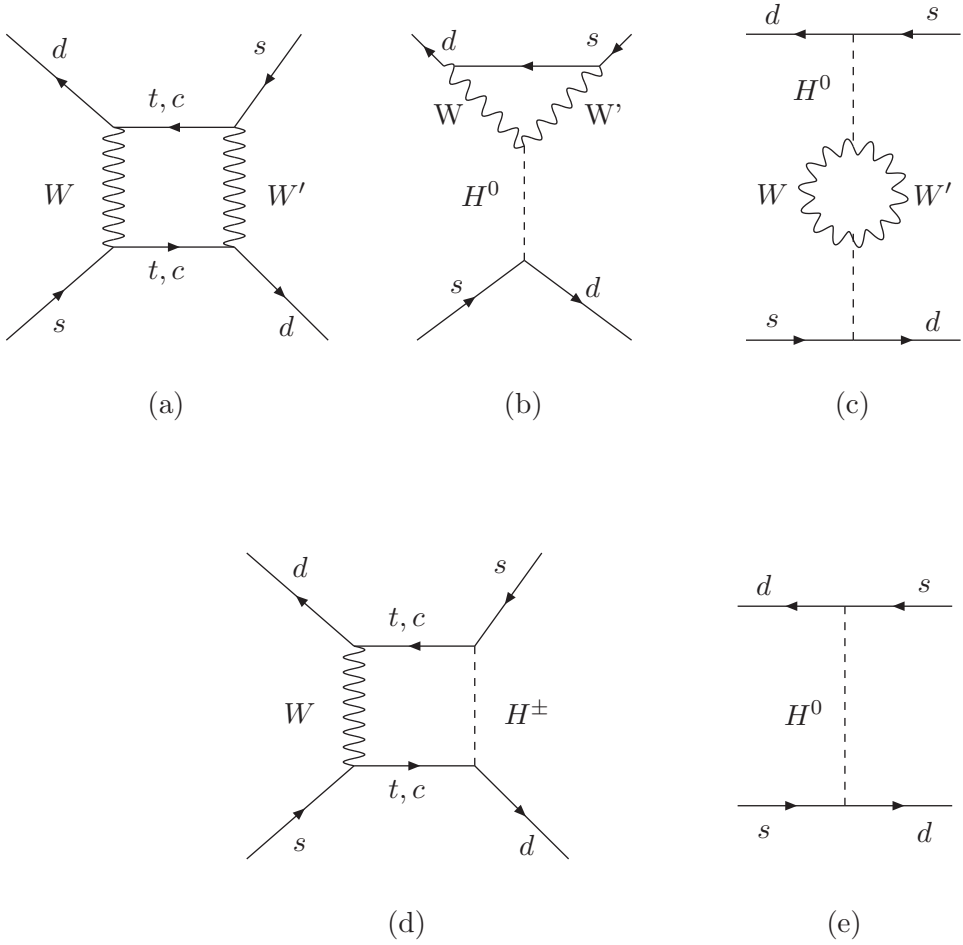
\begin{figure}[t!]
\begin{center}
\begin{picture}(180,50)(90,0)
\ArrowLine(80,25)(40,60)
\Text(60,55)[]{$d$}
\ArrowLine(40,-55)(80,-20)
\Text(60,-50)[]{$s$}
\ArrowLine(125,25)(80,25)
\Photon(80,25)(80,-20){4}{8}
\ArrowLine(150,60)(125,25)
\Text(135,55)[]{$s$}
\ArrowLine(125,-20)(160,-55)
\ArrowLine(80,-20)(125,-20)
\Photon(125,25)(125,-20){4}{8}
\Text(145,-50)[]{$d$}
\Text(60,0)[]{$W$}
\Text(100,-30)[]{$t,c$}
\Text(100,35)[]{$t,c$}
\Text(100,-80)[]{(a)}
\Text(145,0)[]{$W'$}
\end{picture}
\begin{picture}(150,50)(120,0)
\ArrowLine(50,48)(38,60)
\Text(52,58)[]{$d$}
\ArrowLine(40,-55)(80,-25)
\ArrowLine(120,60)(108,48)
\Text(64,0)[]{$H^0$}
\ArrowLine(108,48)(54,48)
\Photon(80,18)(50,48){3}{5}
\Photon(108,48)(80,18){3}{5}
\Text(52,30)[]{W}
\Text(110,30)[]{W'}
\DashLine(80,18)(80,-25){3}
\ArrowLine(80,-25)(120,-55)
\Text(106,58)[]{$s$}
\Text(55,-35)[]{$s$}
\Text(106,-35)[]{$d$}
\Text(80,-80)[]{(b)}
\end{picture}
\begin{picture}(140,50)(-90,-51)
\ArrowLine(80,60)(40,60)
\ArrowLine(120,60)(80,60)
\DashLine(80,60)(80,20){3}
\PhotonArc(80,0.225)(17.5,90,270){3}{7}
\PhotonArc(80,0.225)(17.5,270,450){3}{7}
\DashLine(80,-15)(80,-55){3}
\ArrowLine(40,-55)(80,-55)
\ArrowLine(80,-55)(120,-55)
\Text(80,-80)[]{(c)}
\Text(65,40)[]{$H^0$}
\Text(50,0)[]{$W$}
\Text(110,0)[]{$W'$}
\Text(50,-48)[]{$s$}
\Text(110,-48)[]{$d$}
\Text(50,68)[]{$d$}
\Text(110,68)[]{$s$}
\end{picture}

\vspace{4cm}

\begin{picture}(180,50)(30,-20)
\ArrowLine(80,25)(40,60)
\Text(60,55)[]{$d$}
\ArrowLine(40,-55)(80,-20)
\Text(60,-50)[]{$s$}
\ArrowLine(125,25)(80,25)
\Photon(80,25)(80,-20){4}{8}
\ArrowLine(150,60)(125,25)
\Text(135,55)[]{$s$}
\ArrowLine(125,-20)(160,-55)
\ArrowLine(80,-20)(125,-20)
\DashLine(125,25)(125,-20){3}
\Text(145,-50)[]{$d$}
\Text(60,0)[]{$W$}
\Text(100,-30)[]{$t,c$}
\Text(100,35)[]{$t,c$}
\Text(100,-80)[]{(d)}
\Text(145,0)[]{$H^{\pm}$}
\end{picture}
\begin{picture}(90,50)(30,-20)
\ArrowLine(80,40)(40,40)
\ArrowLine(120,40)(80,40)
\DashLine(80,40)(80,-35){3}
\ArrowLine(40,-35)(80,-35)
\ArrowLine(80,-35)(120,-35)
\Text(80,-80)[]{(e)}
\Text(65,0)[]{$H^0$}
\Text(50,-45)[]{$s$}
\Text(110,-45)[]{$d$}
\Text(50,48)[]{$d$}
\Text(110,48)[]{$s$}
\end{picture}
\end{center}
\vspace{2.5cm}
\caption{\small\it Diagrams for kaon mixing in Left-Right models: the sum of the first row (a)+(b)+(c)  is gauge invariant, whereas the second row corresponds to additional diagrams of interest. We do not show the diagrams where one or several gauge bosons are replaced by the corresponding Goldstone bosons. Diagrams with $u$-quarks in the loop are suppressed by powers of $m_u$ and are thus not considered. }\label{fig:diagall}
\end{figure}

\subsection{Contributions to kaon mixing in Left-Right models}

The LRM generates corrections for kaon mixing compared to the SM case. We will exploit the hierarchy between the left-right and electroweak symmetry breaking scales, reflected by the hierarchy of masses between $W$ and $W'$ bosons (as well as heavy Higgs bosons), and we keep only the first correction in $\beta=(M_W/M_{W'})^2$ (and assuming  $\omega=(M_{W'}/M_H)^2=\mathcal{O}(1)$). 

The problem differs from the SM on several points due to the different structure of $W'$ couplings. First, the GIM mechanism cannot be invoked since the two different CKM-like matrices are involved (one for left-handed quarks, the other one for right-handed quarks). Second, the effective theory at the low scale involves two different $|\Delta S|=2$ operators which are not multiplicatively renormalised.
Third, the $WW'$ box together with the contributions from Goldstone bosons is 
not gauge invariant (in contrast with the SM case), which means that 
additional diagrams involving heavy neutral Higgs exchanges together with 
a $W$ and a $W'$ must be considered~\cite{Chang:1984hr,Basecq:1985cr,GagyiPalffy:1997hh}, shown in the first row of Fig.~\ref{fig:diagall}. Additional diagrams
are given in the second row of the same figure. Note that we do not 
consider diagrams suppressed by powers of $\beta$.

We will give the results for the method of regions in the t'Hooft-Feynman gauge for the gauge bosons (the complete result being in principle gauge invariant, even though individual contributions are not~\cite{Basecq:1985cr,Kenmoku:1987fm}). The contributions from the gauge bosons and their associated Goldstone bosons at the scale $\mu_W$, diagram \ref{fig:diagall}(a), are given by Refs.~\cite{Ecker:1985vv,Zhang:2007da,Hou:1985ur,Chang:1984hr,Basecq:1985cr}
\begin{eqnarray}\label{eq:boxlr}
A^{({\rm box})}&=&\frac{G_F^2 M^2_W}{4\pi^2} 2 \beta h^2 \langle Q_2^{LR}\rangle\\
&& \times\sum_{UV=c,t}\lambda_U^{LR}\lambda_V^{RL} \sqrt{x_Ux_V} [(4+x_Ux_V \beta)I_1(x_U,x_V,\beta)-(1+\beta)I_2(x_U,x_V,\beta)]\,,
\nonumber
\end{eqnarray}
where the $|\Delta S|=2$ scalar operator $Q_2^{LR}=(\bar{s}^\alpha P_L d^\alpha)(\bar{s}^\beta P_R d^\beta)$ appears. The quark masses enter as $x_i=(m_i/M_W)^2$, and are evaluated at the scale $m_i$ for heavy quarks (we set $m_u = m_d =m_s= 0$). $\lambda_i^{PQ}=(V^{P}_{id})^{*} V^Q_{is}$ collects the product of CKM-like matrices, the couplings from $SU(2)_L$ and $SU(2)_R$ gauge groups
appear through $h=g_R/g_L$, and $I_1$ and $I_2$ are modified Inami-Lim functions which can be expanded at leading order in $\beta$:
\begin{eqnarray}\label{eq:InamiLimfunctionsGeneral}
I_1 & = & \frac{x_U\log x_U}{(1-x_U)(x_U-x_V)} + (U\leftrightarrow V) + \mathcal{O}(\beta) , \nonumber\\
I_2 & = & \frac{x_U^2\log x_U}{(1-x_U)(x_U-x_V)} + (U\leftrightarrow V) - \log\beta + \mathcal{O}(\beta)\,.
\end{eqnarray}
In the t'Hooft-Feynman gauge, one can identify the various contributions to Eq.~\eqref{eq:boxlr} coming from $WW'$ (term proportional to $ I_1(x_U,x_V,\beta) $), $GG'$ (term $ \propto x_U x_V \beta \cdot I_1(x_U,x_V,\beta) $, of higher order in $ \beta $), $GW'$ (term $ \propto I_2(x_U,x_V,\beta) $) and $WG'$ (term $ \propto \beta \cdot I_2(x_U,x_V,\beta) $, of higher order).

We rewrite the transition amplitude Eq.~\eqref{eq:boxlr} in a different form and keep the leading term in $x_c$:
\begin{eqnarray}\label{eq:boxlrexp}
A^{({\rm box})} &=& \frac{G_F^2 M_W^2}{4 \pi^2} 2 \beta h^2 \langle Q_2^{LR}\rangle\\
&& \times \biggl( \lambda_c^{LR}\lambda_c^{RL}S^{({\rm box})}(x_c) + \lambda_t^{LR}\lambda_t^{RL}S^{({\rm box})}(x_t)+  (\lambda_c^{LR}\lambda_t^{RL}+\lambda_t^{LR}\lambda_c^{RL})S^{({\rm box})}(x_c,x_t)\biggr)\,,\nonumber
\end{eqnarray}
at one-loop order in the absence of QCD corrections and at leading order in $\beta$, with
\begin{eqnarray}\label{eq:Sbox}
S^{({\rm box})} (x_c, x_t) \!\!&=& \!\! \sqrt{x_c x_t}\left[\frac{x_t - 4}{x_t - 1} \log (x_t) + \log (\beta) \right] +  \mathcal{O} (\beta, x^{3/2}_c)\,, \label{eq:AmpccexpCT} \\
S^{({\rm box})} (x_t) \!\!&=&\!\! x_t\left( \frac{x_t^2 - 2 x_t + 4}{(x_t - 1)^2} \log (x_t) + \frac{x_t - 4}{x_t - 1} + \log (\beta) \right) + \mathcal{O} (\beta)\,,\label{eq:AmpccexpTT} \\
S^{({\rm box})} (x_c) \!\!&=&\!\! x_c  \left( 4 \log (x_c) + 4 + \log (\beta) \right) + \mathcal{O} (\beta, x^2_c)\,.
\label{eq:Ampccexp}
\end{eqnarray}
We notice that a large $\log(x_c)$ arises for the $cc$ box, whereas $ct$ 
and $tt$ boxes are dominated by the single scale $m_t$. The extra $\log(\beta)$ present  in these equations comes from the $I_2$ function which is due to boxes with one Goldstone boson $ G $ exchanged in the t'Hooft-Feynman gauge.

The contributions from the vertex correction \ref{fig:diagall}(b) and self-energy diagrams \ref{fig:diagall}(c) read
\begin{eqnarray}
A^{\rm (vert)}&=&
-32\beta\omega h^2\frac{G_F^2 M_W^2}{4\pi^2}  \langle Q_2^{LR}\rangle S_V(\beta,\omega) \sum_{U,V=c,t} \lambda_U^{LR}\lambda_V^{RL}  \sqrt{x_U x_V} \,,
\nonumber \\
A^{\rm (self)}&=&
-2\beta\omega h^2\frac{G_F^2 M_W^2}{4\pi^2} \langle Q_2^{LR}\rangle S_S(\beta,\omega)  \sum_{U,V=c,t} \lambda_U^{LR}\lambda_V^{RL}  \sqrt{x_U x_V} \,,
\end{eqnarray}
with the two functions~\cite{ Basecq:1985cr,GagyiPalffy:1997hh,Bertolini:2014sua}
\begin{eqnarray}
S_S(\beta,\omega)&=& \left[\frac{\omega^2+1}{\omega}  [I_a(0)-I_a(M_H^2)] +\left(\frac{\omega-1}{\omega}\right)^2 \frac{M_W^2}{\beta} I_b(M_H^2)\right] +\mathcal{O}(\beta)\,,\\
S_V(\beta,\omega)&=&  [I_a(0)-I_a(M_H^2)] +\mathcal{O}(\beta^{1/2})\,.
\end{eqnarray}
We only kept the leading power of $\beta$ in the above expressions, so that for
$m_i,M_W\ll M_{W'}$ and an arbitrary $M_{W'}/M_H$
\begin{eqnarray}
I_a(0)-I_a(M_H^2)&\simeq & -1+(1-\omega)\log\left|\frac{1-\omega}{\omega}\right|+\mathcal{O}(\beta)\,,\\
I_b(M_H^2)&\simeq &\frac{\beta}{M_{W}^2}\left[\omega+\omega^2\log\left|\frac{1-\omega}{\omega}\right|\right]+\mathcal{O}(\beta^2)\,.
\end{eqnarray}
As can be seen no logarithms in $\beta$ are generated by these diagrams in the t'Hooft-Feynman gauge. 

Another  contribution must be considered, the one represented in Fig.~\ref{fig:diagall}(e). In these models, heavy neutral Higgs bosons can
exhibit flavour-changing neutral couplings generating $|\Delta S|=2$ transitions at tree level.  The corresponding transition
has the form
\begin{equation}\label{eq:neutralhiggstree}
A^{(H^0)} =
 -\frac{4 G_F}{\sqrt{2}} u\beta\omega   \langle Q_2^{LR}\rangle \sum_{i,j=c,t} \lambda_i^{LR}\lambda_j^{RL} \sqrt{x_i(\mu_H)x_j(\mu_H)}\,,
\end{equation}
with $u=(1+r^2)^2/(1-r^2)^2$ and $r=|\kappa_1/\kappa_2|$ the ratio of Higgs vacuum expectation values triggering electroweak symmetry breaking. 

Finally, we have contributions coming from the box with a W boson and a heavy charged Higgs (of a mass similar to the neutral Higgs boson considered above), Fig.~\ref{fig:diagall}(d):
\begin{eqnarray}
A^{(H^{\pm} \, {\rm box})}&=&
 \frac{G_F^2 M_W^2}{4\pi^2}\langle Q_2^{LR}\rangle \sum_{U,V=c,t} \lambda_U^{LR}\lambda_V^{RL} S^H_{LR}(x_U,x_V, \beta\omega)\,,
\end{eqnarray}
with
\begin{equation}\label{eq:chargedHiggsLoopFunction}
S^H_{LR}(x_U,x_V, \beta\omega)= 2\omega \beta u \sqrt{x_U x_V} [x_Ux_V I_1(x_U,x_V, \beta\omega)-I_2(x_U,x_V, \beta\omega)]\,,
\end{equation}
the first term coming from boxes with a Goldstone boson (relevant only for $tt$ boxes) and the second term from boxes with a $W$ boson in the t'Hooft-Feynman gauge. 

We remark that in the above expressions, there are no contributions from $u$-quarks as they always come multiplied by $m_u=0$.
We should notice that in principle, another set of diagrams is necessary to obtain gauge invariance, namely the diagrams Fig.~\ref{fig:diagall}(b) and (c) where $W'$ is replaced by a heavy charged Higgs. However, as noticed in Ref.~\cite{GagyiPalffy:1997hh}, these contributions are suppressed by powers of $M_W/M_{H^\pm}$ compared to the diagrams considered here.

In the above expressions, we assumed that the breakdown of the left-right symmetry is triggered only by non-vanishing vacuum expectation values of scalar fields charged under $SU(2)_R$ (the structure remains similar, but the prefactor is modified in the case of non-vanishing v.e.v. for scalar fields charged under $SU(2)_L$ and further effects due to the mixing among the various scalars must be taken into account~\cite{BDV:2016}).

\begin{table}[!t]
\begin{center}
{\small {\renewcommand{\arraystretch}{1.4}
\begin{tabular}{|c| c c c |}
\hline
& $\bar \eta_{tt}$& $\bar \eta_{ct}$& $\bar \eta_{cc}$ \\
\hline
$(W'1)$ & $4.65+0.99=5.64$ & $2.42+0.27=2.69$ & $1.46+0.16-0.28=1.34$ \\
\hline
$(W'2)$& $ 4.66 + 0.98 = 5.64 $ & $ 2.42 + 0.27 = 2.69 $ & $ 1.26 + 0.01 = 1.27 $ \\
\hline
$(H^0)$, (vert), (self) & $4.66+0.98=5.64$ & $2.42 + 0.27=2.69$ & $1.26+0.02=1.28$  \\
\hline
$(H1)$& $ 4.66 + 1.00 = 5.66 $ & - & - \\
\hline
$(H2)$& $ 4.66 + 0.98 = 5.64 $ & $ 2.42 + 0.27 = 2.69 $ & $ 1.26 + 0.02 = 1.28 $ \\
\hline
\end{tabular}}}
\end{center}
\caption{\small\it Short-distance QCD corrections at NLO for the LR contributions to kaon mixing with the method of regions. Flavour thresholds are taken into account. The $\bar \eta$ are calculated at the hadronisation scale $\mu_{{h}}=1$~GeV with the parameters given in the text. The first (second) number corresponds to the LO (NLO, respectively) result. $\alpha_s$ is always evaluated up to NLO. In the case of $\bar\eta_{cc}$ the next-to-leading order is split into the NLO corrections to $\log(x_c)$ (second number) and the NLO contribution to the non-logarithmic piece (third number). We do not indicate the value for $(H1)$ when it corresponds to a higher order term in $ x_c $ in the effective Hamiltonian.}
\label{tab:etaLR}
\end{table}

\begin{table}[t]
\begin{center}
{\small {\renewcommand{\arraystretch}{1.4}
\begin{tabular}{|c|ccc|}
\hline
& $\bar \eta_{tt}$ & $\bar \eta_{ct}$ & $\bar \eta_{cc}$ \\
\hline
$(W'1)$ & $ 4.68 + 0.96 = 5.64 $ & $ 2.43 + 0.26 = 2.69 $ & $ 1.55 + 0.16 - 0.31 = 1.40 $ \\
\hline
$(W'2)$ & $ 4.86 + 7.32 - 5.26 = 6.92 $ & $ 2.52 + 1.91 - 1.51 = 2.92 $ & $ 1.31 - 0.02 = 1.29 $ \\
\hline
$(H1)$ & $ 4.66 + 0.99 = 5.65 $ & - & - \\
\hline
$(H2), \, \omega = 0.1$ & $ 4.86 + 4.11 - 2.65 = 6.33 $ & $ 2.53 + 1.17 - 0.86 = 2.83 $ & $ 1.31 - 0.02 = 1.29 $ \\
\hline
$(H2), \, \omega = 0.8$ & $ 4.84 + 6.70 - 4.76 = 6.79 $ & $ 2.52 + 1.77 - 1.40 = 2.89 $ & $ 1.31 - 0.03 = 1.28 $ \\
\hline
\end{tabular}}}
\end{center}
\caption{\small\it Same results as in Tab.~\ref{tab:etaLR} using the $ \log(\beta) $ approach.
Note that in this case $ (H2) $ is sensitive to the value of $ \omega $.}
\label{tab:etaLRlogbeta}
\end{table}

\subsection{Method of regions}\label{sec:MRLR}

Short-distance QCD corrections, denoted $\bar\eta_{UV}$, will correct the previous expressions. We are now in a position to compute these corrections at NLO since the anomalous dimensions needed for the calculation have been determined in Ref.~\cite{Buras:2000if} and are summarised
in App.~\ref{app:anomdimgen} for completeness. 

Ref.~\cite{Ecker:1985vv} considered the LO case, following the same steps as
 in Sec.~\ref{sec:SMLO}, with the following modifications: when considering a $ W W' $ box,  the bilocal operators involve
one left-handed and one right-handed $|\Delta S|=1$ operators ($O_l^{VLL}$  and $O_r^{VRR}$), which are matched onto the LRM at different scales ($\mu_W$ versus $\mu_R$), and the matching 
has to be performed onto two $|\Delta S|=2$ local operators rather than a single one.  
Note that in Ref.~\cite{Ecker:1985vv} the two additional diagrams involving 
heavy neutral Higgs exchanges together with
$W$ and $W'$ bosons,  diagrams  \ref{fig:diagall}(b) and \ref{fig:diagall}(c), have been neglected arguing that in the t'Hooft-Feynman gauge
their contributions are small for large enough neutral Higgs masses.

We adapt Ref.~\cite{Ecker:1985vv} to include the NLO contributions, even though the treatment of the energy range between $\mu_W$ and $\mu_{W',H}$ is not appropriate when these scales are very different (which is the situation in practice) since all the heavy particles ($ W, W', H $) are integrated out simultaneously.
Note that $\alpha_s(\mu_W) \sim 0.1$ so that the contributions $\alpha_s(\mu_W)  \log(\beta)/\pi$ are of the order of $20$ to $30 \%$ for typical values of $M_{W'}$ between 1 TeV and 10 TeV. We thus expect an uncertainty of this order on our results, which we will take into account in our final error budget.  
This is in fact sufficient at the present time considering the level
of accuracy needed for phenomenological applications. 

The expression for $\bar\eta_{UV}$ at NLO within the method of regions without flavour
thresholds (it is rather trivial to take these thresholds into account, but the expressions are somewhat lengthy and will not be
given here though we took them into account in our numerical calculation) are easily derived.
One gets 
\begin{eqnarray}\label{eq:boxlrNLO}
A^{({\rm box})}&=&\frac{G_F^2 M^2_W}{4\pi^2} 2 \beta h^2 
\sum_{a=1,2} 
\langle Q_a^{LR}\rangle\\
&&\quad \times\sum_{UV=c,t}\lambda_U^{LR}\lambda_V^{RL} \sqrt{x_Ux_V} [4\bar\eta^{(W'1)}_{a,UV} I_1(x_U,x_V,\beta)-\bar\eta^{(W'2)}_{a,UV} I_2(x_U,x_V,\beta)]\,,
\nonumber
\end{eqnarray}
with the two $|\Delta S|=2$ local operators
\begin{eqnarray}
Q_1^{LR}&=&(\bar{s}^\alpha \gamma_\mu P_L d^\alpha)(\bar{s}^\beta \gamma^\mu P_R d^\beta)\,,
\qquad
Q_2^{LR}=(\bar{s}^\alpha P_L d^\alpha)(\bar{s}^\beta P_R d^\beta)\,.
\end{eqnarray}
In order to express the short-distance QCD correction $\bar\eta^{(W' 1)}_{a,UV}$ ($U$ and $V$ denote the quarks in the loop with $m_U \leq m_V$), we start by defining
\begin{eqnarray}\nonumber
&&\xi^{(W'1)}_{a,UV} [R]=\sum_{r,l=\pm \, ,i=1,2}
 \left(\frac{\alpha_s(m_{V})}{\alpha_s(\mu_{h})}\right)^{-d_l-d_r+d_i+ d_m} 
  \left(\frac{\alpha_s(m_U)}{\alpha_s(\mu_{h})}\right)^{-d_m}  \quad \left(\frac{\alpha_s(\mu_W)}{\alpha_s(\mu_{h})}\right)^{d_l}
    \left(\frac{\alpha_s(\mu_R)}{\alpha_s(\mu_{h})}\right)^{d_r} \nonumber\\
&&\times 
\left[\left(1+\frac{\alpha_s(\mu_{h})}{4\pi} \hat{K}\right) \hat{W}\right]_{ai}\nonumber\\
&& \quad\times 
 R^{NLO}\Bigg(-d_l-d_r+d_i+2d_m,\nonumber\\
&&  \qquad
  \left[\hat{W}^{-1}\left(1-\frac{\alpha_s(\mu_W)}{4\pi}[J_l-B_l] -\frac{\alpha_s(\mu_R)}{4\pi}[J_r-B_r]
   +\frac{\alpha_s(m_U)+\alpha_s(m_V)}{4\pi}J_m\right)
 \left(\begin{array}{c} \tau_{1}^{rl}\\\tau_{2}^{rl}\end{array}\right)
\right]_i, \nonumber\\
&&   \qquad\quad
   \left[\hat{W}^{-1}\left(-\hat{K}+J_l+J_r-2J_m\right) \left(\begin{array}{c} \tau_{1}^{rl}\\\tau_{2}^{rl}\end{array}\right)
\right]_i , m_{V},\mu_W \Bigg)\,,\label{eq:etaLRW1}
\end{eqnarray} 
with $d_{l,r}$ determined from the anomalous dimensions of the $|\Delta S|=1$ current-current operators,
$d_i$ from the corresponding $|\Delta S|=2$ local operator, $d_m$ from the evolution of the masses, $J_{l,r,i,m}$ the corresponding terms from the anomalous dimension matrix at NLO and
$\hat{W}$ a diagonalisation matrix (see App.~\ref{app:anomdimgen} for a 
definition of all these quantities). Finally the values of the Wilson coefficients coming from the matching between the bilocal operators $O_{rl}$ and the local $|\Delta S|=2$ operators are
\begin{eqnarray}
\tau_{1}^{rl}=\tau_{rl}/4 \, ,\qquad \qquad \tau_{2}^{rl}=1/4 \, , \qquad\qquad \tau_{rl}=-(r+l+Nrl)/2\,.\label{eq:taurl}
\end{eqnarray}
For $\bar\eta_{a,ct}^{(W'1)}$ and $\bar\eta_{a,tt}^{(W'1)}$, there are no large logarithms in the contribution from $I_1$ in equation \eqref{eq:AmpccexpCT}-\eqref{eq:AmpccexpTT}, the integral is dominated by $k^2=\mathcal{O}(m_V^2)$  and  we have
\begin{equation}
\bar\eta_{a,ct}^{(W'1)}=\xi_{a,ct}^{(W'1)}[R^{NLO}\to R^{NLO}_1] , \qquad
\bar\eta_{a,tt}^{(W'1)}=\xi_{a,tt}^{(W'1)}[R^{NLO}\to R^{NLO}_1]\qquad
\end{equation}
where
$R^{NLO}$ should be replaced by $R^{NLO}_1$ defined in Eq.~\eqref{eq:rnlo1}.

$\bar\eta_{cc}^{(W'1)}$ should in principle be obtained by taking $\xi_{a,ct}^{(W'1)}$ and replacing
 $R^{NLO}$ by $R^{NLO}_{\log}$ given in equation \eqref{eq:rnlo}.
However, Eq.~\eqref{eq:etaLRW1}
resums the $\log \left( \frac{m_c}{M_W} \right) \left(\alpha_s \log\left( \frac{m_c}{M_W} \right) \right)^n$ terms (counted as LO), plus
some of the terms as  $\left(\alpha_s \log\left( \frac{m_c}{M_W} \right) \right)^n$ (counted as NLO).
Since $I_1= \log x_c +1 +\mathcal{O}(x_c)$ provides contributions both at LO ($\log x_c$, with an average $R^{NLO}_{\log}$)
and NLO ($1$, with an average $R^{NLO}_1$), we should separate the two contributions. This procedure\footnote{A similar separation can be performed in the SM case for $\eta_{ct}$, as explained in App.~\ref{app:SMVy}.} yields the modified expression
\begin{eqnarray}
\bar \eta^{(W'1)}_{a,cc}
&=&\frac{1}{1+\log x_c} \biggl(\xi^{(W'1)}_{a,cc}\log(x_c)  +\sum_{r,l=\pm,i=1,2}
 \left(\frac{\alpha_s(m_c)}{\alpha_s(\mu_{h})}\right)^{-d_l-d_r+d_i}
\biggr.
\\\nonumber
&& \biggl.\qquad\qquad\qquad\qquad\qquad\qquad\times
    \left(\frac{\alpha_s(\mu_W)}{\alpha_s(\mu_{h})}\right)^{d_l}
    \left(\frac{\alpha_s(\mu_R)}{\alpha_s(\mu_{h})}\right)^{d_r}
\hat W_{ai}\left[\hat{W}^{-1}\left(\begin{array}{c} \tau_{1}^{rl}\\\tau_{2}^{rl}\end{array}\right)
\right]_i \biggr)\,.
\end{eqnarray}

Similar expressions are obtained for the other short-distance QCD corrections given above, which are gathered in App.~\ref{lrregion}. They collect the short-distance QCD corrections $ \bar{\eta}_{UV} $ for the other diagrams:
\begin{eqnarray}
A^{(H^0)} &=&
 -\frac{4 G_F}{\sqrt{2}} u\beta\omega \sum_{a=1,2} \langle Q_a^{LR}\rangle \sum_{UV=c,t} \bar\eta^{(H)}_{a,UV} \lambda_U^{LR}\lambda_V^{RL} \sqrt{x_Ux_V}\,,\\
A^{\rm (vert)}&=&
-32\beta\omega h^2\frac{G_F^2 M_W^2}{4\pi^2}  \sum_{a=1,2} \langle Q_a^{LR}\rangle \sum_{UV=c,t} \bar\eta^{(H)}_{a,UV} \lambda_U^{LR}\lambda_V^{RL}  \sqrt{x_U x_V} 
 S_V(\beta,\omega)\,, \nonumber\\
A^{\rm (self)}&=&
-2\beta\omega h^2\frac{G_F^2 M_W^2}{4\pi^2}  \sum_{a=1,2} \langle Q_a^{LR}\rangle \sum_{UV=c,t} \bar\eta^{(H)}_{a,UV} \lambda_U^{LR}\lambda_V^{RL}  \sqrt{x_U x_V} S_S(\beta,\omega)\,,\nonumber\\
A^{(H^{\pm} \, {\rm box})}&=&\frac{G_F^2 M^2_W}{4\pi^2} 
\sum_{a=1,2} 
\langle Q_a^{LR}\rangle \times \sum_{U,V=c,t} \lambda_U^{LR}\lambda_V^{RL} \nonumber\\
&& \times 2\beta \omega u \sqrt{x_U x_V}
  [ \bar\eta^{(H1)}_{a,UV} x_Ux_V I_1(x_U,x_V,\beta\omega)- \bar\eta^{(H2)}_{a,UV} I_2(x_U,x_V, \beta\omega)]\,,\nonumber
  \end{eqnarray}
\noindent where we followed Ref.~\cite{Ecker:1985vv} to attribute the same scaling to the three contributions related to neutral Higgs exchanges (the momenta relevant for the method of regions are smaller than the high scales $ M_{W, W', H} $).

The results for $\bar\eta_{2, UV} \equiv \bar\eta_{UV} $ are shown in Tab.~\ref{tab:etaLR} with the following inputs: \linebreak
$m_t(m_t)= \mu_t =170$~GeV,
$m_c(m_c)=\mu_c=1.3$~GeV, $M_W=\mu_W=80.385$~GeV, \linebreak $\mu_b =4.8$~GeV, $M_{W'}=1$~TeV,  $\omega=0.1$
and $\Lambda^{(4)}=0.325$~GeV. They include the flavour thresholds. 
The LO results are in fairly good agreement 
with the calculation of Ref.~\cite{Bertolini:2014sua}. The short-distance corrections $\bar\eta_{1,UV}$ are at least an order of magnitude smaller than 
$\bar\eta_{2,UV}$ and will not be considered further, in agreement with Refs.~\cite{Blanke:2011ry,Bertolini:2014sua}.

Two contributions $(W'2)$ and $(H2)$ contain $\log\beta$, which can be considered either large or small depending on the hierarchy of the gauge bosons (for the above input values, we have the intermediate case $\log\beta\simeq -5$). In Tab.~\ref{tab:etaLR} and in App.~\ref{lrregion}, we provide the expressions without resumming this logarithm (``small $\log\beta$ approach''). One may however be worried that for significant hierarchies between the left and right gauge sectors, a resummation would be needed also for $\log\beta$ (even though this term would come with a suppressing factor $\alpha_s(\mu_W)$). Treating it in a similar way to $\log x_c$, we obtain the results for the ``large  $\log\beta$ approach'' gathered in Tab.~\ref{tab:etaLRlogbeta} and in App.~\ref{lrregion}. The results, obtained for the same input values as in Tab.~\ref{tab:etaLR},  indicate a typical 10\%-20\% variation compared to the previous case for $tt$ (and a smaller variation for $ct$ and $cc$).
We also show the mild dependence of the result on $M_H$.

Up to now we have given the short-range contributions diagrams by diagrams without assessing the uncertainties. We will come back 
to these short-range contributions and their uncertainty in Sec.~\ref{sec:result}.

\begin{figure}
\begin{center}
\includegraphics[width=8.3cm,angle=0]{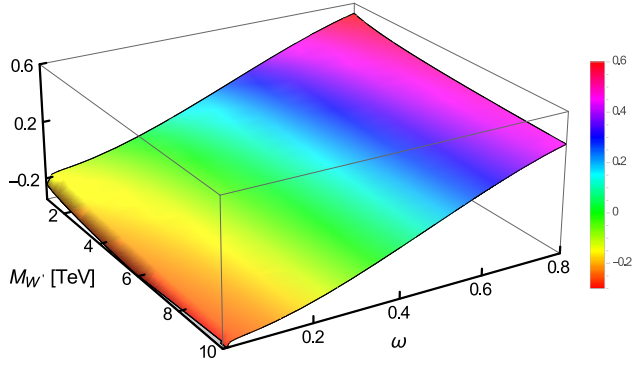}
\end{center}
\caption{$(S^{LR}-\log(x_c))/\log(x_c)$ as a function of typical values of $M_{W'}$ and $\omega$.}
\label{fig:3Df}
\end{figure}

\section{NLO computation of $\bar\eta_{cc}^{LR}$ in the EFT approach}
\label{sec:EFT}

In Section \ref{sec:method}, we have shown that  the method of regions gives results
in good agreement with those obtained using EFT in the SM ($cc$ and $tt$ boxes), when we start from diagrams exhibiting no large logarithms at leading order. The agreement is less satisfying in the case of the $ct$ box where a large logarithm occurs and where the heavy degrees of freedom ($t$~and~$W$) are not treated in the same way. Moving to the LRM, one may thus worry that the $WW'$ box with two charm quarks (exhibiting large $\log(x_c)$ contributions) might not be computed accurately within the MR due to the presence of a large logarithm. We will thus determine the corrections also in the EFT framework. In this setting, it is more natural to discuss the short-distance QCD corrections to the gauge-invariant sum of the diagrams~\ref{fig:diagall}(a), (b) and (c) involving two $c$ quarks:
\begin{equation}
A^{LR}_{cc}=\frac{G_F^2 M^2_W}{4\pi^2} 2 \beta h^2 \langle Q_2^{LR}\rangle \lambda_c^{LR} \lambda_c^{RL} 4x_cS^{LR}(x_c,\beta,\omega)\,,
\end{equation}
with 
\begin{eqnarray}
S^{LR}(x_c,\beta,\omega)&=&1 +\log(x_c) +\frac{1}{4} \log(\beta) +\frac{1}{4} F(\omega)\,,
\label{eq:ccSLR}\\
F(\omega)&=&18 \, \omega- (1+16 \, \omega-17 \omega^2)\log\bigl|(1 - \omega)/\omega\bigr|\, .
\label{eq:fomega}
\end{eqnarray}
We will calculate $\bar \eta_{cc}^{(LR)}$ within the 
EFT approach,  following the cases of the $cc$~\cite{Herrlich:1993yv} and $ct$ boxes  
\cite{Herrlich:1996vf} in the SM.  Since we have also computed the short-distance QCD corrections for the LRM using the method of regions in Section \ref{sec:LRRM} we will
be able to  compare both results.

The EFT computation will allow us to determine the mixing
between the $|\Delta S|=1$ and $|\Delta S|=2$ operators in the four-quark
theory, as well as  the ${\cal O}(\alpha_s)$ 
contributions to the $|\Delta S|=2$  operator in the effective four- and three-quark theories. In fact the latter contributions appear at NNLO thus beyond the order at which we work.
The comparison between the two methods and the consideration of higher orders (part of NNLO contributions, variation of the scales) will provide an estimate of the remaining uncertainties that we will discuss at the end of our evaluation.
This piece of information will be used also when discussing the uncertainties of the short-distance QCD corrections in the $ct$ and $tt$ case in the LRM.

\subsection{Operator basis in the effective four-quark theory}

\subsubsection{Physical operators}

Before entering the calculation within the EFT framework, it is worth studying Eq.~\eqref{eq:ccSLR} more closely.
In Fig.~\ref{fig:3Df}, the quantity $(S^{LR}(x_c,\beta,\omega)-\log(x_c))/\log(x_c)$ is shown as a function of
$M_{W'}$ and $\omega$ for phenomenologically relevant values of these two quantities. In most of this region,
the $\log(x_c)$ term is significantly dominant over the rest of $S^{LR}(x_c,\beta,\omega)$. On the other hand, as discussed in Sec.~\ref{sec:MRLR}
the $\alpha_s \log(\beta)/\pi$ contributions can reach 20 to $30 \%$. We will thus ignore the resummation of these terms 
occurring between $\mu_H$ and $\mu_W$, so that we can match directly the LRM onto an EFT at $\mu_W$ (to be varied somewhat between $M_H$ and $M_W$) in order to focus on the resummation of $\log(x_c)$ terms from $\mu_W$ to $\mu_c$. In this case, the counting is similar to the one for $\eta_{ct}$ in the SM:
 one resums the $\log (x_c) \, (\alpha_s \log(x_c))^n$ terms at LO, and the $(\alpha_s \log(x_c))^n$ ones at NLO.
Consequently in our EFT approach the $\alpha_s \log(\beta)$ terms will only appear at NNLO (in contrast with
 the MR case where a partial resummation of these terms has been performed).

We thus integrate out both $W$ and $W'$ simultaneously
and consider the complete set of diagrams necessary for gauge invariance shown
in Fig.~\ref{fig:diagall}(a), (b), (c). This leads to the following effective Hamiltonian~\cite{Witten:1976kx}
\begin{equation}
H^{cc}  = 8 G_F^2   \beta h^2 \lambda_c^{LR} \lambda_c^{RL} \left[\sum_{i,j=\pm} C_{ij} (\mu) O_{ij} (\mu) + C_{1}^r (\mu)  Q_1 (\mu)  + C_{2}^r (\mu) Q_2 (\mu)  + \cdots\right]
\label{eq:Hamil}
\end{equation}
where we have considered only the lowest-dimension operators necessary to perform a consistent matching and RGE. This situation is similar to the case of the $ct$ box in the SM, as recalled in App.~\ref{app:SMcomput}.
The operators $O_{ij}$ correspond to one insertion of $ \gamma_\mu~P_L~\otimes~\gamma^\mu~P_L $ and one of $ \gamma_\nu P_R \otimes \gamma^\nu P_R $ (these operators suffice to describe the sum of the diagrams Fig.~\ref{fig:diagall} (a), (b), (c), since contributions from other operator structures, in particular scalar ones, correspond to higher-order operators~\cite{Basecq:1985cr}).
The last two terms on the right-hand side are required to absorb one-loop divergences, with the local dimension-eight $|\Delta S|=2$ operators $Q_a$  
defined as
\begin{equation}
Q_1 = \frac {m_c^2}{g^2 \mu^{2 \epsilon}} (\bar s \gamma_\mu P_L d) (\bar s \gamma^\mu P_R d) \, , 
\quad \quad  Q_2 = \frac {m_c^2}{g^2 \mu^{2 \epsilon}}(\bar s P_L d) (\bar s  P_R d) \, .
\label{eq:opQi}
\end{equation} 
According to the usual convention~\cite{Buchalla:1995vs,Herrlich:1996vf}, two inverse powers of the strong  coupling
constant have been introduced compared to $Q_{1,2}^{LR}$ in order to avoid mixing of the operators already at $ \mathcal{O} (\alpha_s^0)$.
The Wilson coefficients $C_{ij}$ are given by
$C_{ij}(\mu)=C_i(\mu)C_j(\mu)$ i.e., the product of $|\Delta S|=1$ Wilson coefficients \cite{Witten:1976kx,Lee:1979vs,Herrlich:1993yv}, whereas
$C_{a}^r(\mu_W)$ can be determined from a matching  at
the scale $\mu_W$.  
 The ellipsis in Eq.~\eqref{eq:Hamil} denotes the contribution of 
penguin operators which we will neglect
in the following. The $|\Delta S|=1$ ones are proportional to 
 $\lambda_i^{XY}$  and one can distinguish two different types: the
ones which come with $X=Y$ and those with
$X \neq Y$. In the former case the GIM cancellation operates in the same way as in the Standard Model \cite{Buras:1991jm}, and the only penguin operators which survive are
proportional to $\lambda_t^{XX}$ thus contributing only to $\eta_{ct}^{LR}$. In
the latter case GIM cannot be used anymore and one could in principle have
contributions from penguin operators for any of the $\eta$'s. However the
QCD penguin contributions do not contribute at the order we are working while
the Higgs ones
will be suppressed by powers of $\beta$ which, as already stated, we consistently drop. This same latter reason suppresses  the $|\Delta S|=2$ Higgs penguin
contributions.

Following Ref.~\cite{Herrlich:1993yv}, we will work in the $\overline{MS}$ scheme,
with an anticommuting $\gamma_5$ (NDR scheme) in $ D = 4 - 2 \epsilon $ dimensions, and
we use an arbitrary QCD $R_\xi$ gauge. We keep non-vanishing strange and down quark masses to regularise infrared singularities (this regularisation leads to the 
appearance of unphysical operators which however do not affect the outcome of the computation~\cite{Herrlich:1993yv}). By analogy with the SM case, we can indicate explicitly the renormalisation matrices $Z$ needed here
\begin{eqnarray}
H^{cc} &=&8 G_F^2 \beta h^2 \lambda^{LR}_{c} \lambda^{RL}_{c} \\
&&
\times \Bigg[
\sum_{i,j=\pm}\! C_i C_j\!
\underbrace{\left(
	\sum_{i',j'=\pm} 
	Z^{-1}_{ii'} Z^{-1}_{jj'}
	O^{{\rm bare}}_{i'j'}
	+
	\sum_{k=1,2} {Z}_{ij,k}^{-1} Q_k^{{\rm bare}}
	\right)}
	_{\displaystyle \equiv O_{ij}}
+  \sum_{k,l=1,2} C_k^r Z_{kl}^{-1} Q_l^{{\rm bare}} \Biggr]\,.\nonumber
\end{eqnarray}
The matrices $Z^{-1}$ are known from $|\Delta S|=1$ and $|\Delta S|=2$ operator mixings, whereas the
mixing tensor ${Z}_{ij,k}^{-1}$ corresponding to the mixing between the two kinds of operators must be determined.

As discussed in particular  in Refs.~\cite{Buras:1989xd, Dugan:1990df,Herrlich:1994kh} and briefly mentioned in App.~\ref{app:SMcomput}, we need to consider also a type of unphysical operators which appear in dimensional regularisation and are necessary to renormalise the theory: these are the so-called evanescent operators, which
appear in the ellipsis  in Eq.~\eqref{eq:Hamil} and will be discussed now.
    
\begin{center}
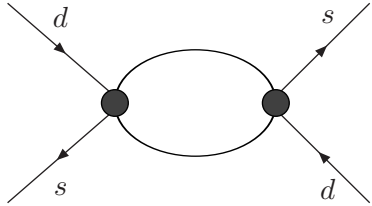
\begin{figure}[!t]
\begin{picture}(300,50)(-120,0)
\ArrowLine(40,60)(80,25)
\Text(60,55)[]{$d$}
\ArrowLine(80,20)(40,-15)
\Text(60,-10)[]{$s$}
\GOval(110,22.5)(20,30)(360){1}
\GCirc(80,22.5){5}{0.25}
\ArrowLine(140,25)(175,60)
\Text(160,55)[]{$s$}
\ArrowLine(175,-15)(140,20)
\Text(160,-10)[]{$d$}
\GCirc(140,22.5){5}{0.25}
\end{picture}
\vspace{1.0cm}
\caption{\small\it Diagram $D_0$ in the effective four-flavour theory. Black circles denote the insertions of $|\Delta S|=1$ current-current operators.}\label{fig:diagsD0}
\end{figure}
\end{center}

\subsubsection{Evanescent operators}\label{sec:evanescent}

Evanescent operators appear in the discussion of the RGE evolution of the effective Hamiltonian. These operators occur in the definition of the Dirac algebra in $D$ dimensions:
they vanish for $D = 4$ dimensions, but they appear as counterterms to physical operators multiplied by $1/\epsilon$. In principle, at each order of perturbation theory, new sets of evanescent operators are required, arising in the computation of radiative corrections to the physical and evanescent operators already present in the theory. In the context of the RGE for the effective Hamiltonian, the evanescent operators play a role in two different issues: first, the matrix elements of evanescent operators can affect the matching equation allowing one to determine the Wilson coefficients in the effective theory~\cite{Buras:1989xd},
and second, the presence of evanescent operators in counterterms for physical 
operators (and the other way around) means that both set of operators may mix under renormalisation~\cite{Dugan:1990df}. In Refs.~\cite{Buras:1989xd,Dugan:1990df} it was shown that a finite renormalisation of the evanescent operators 
could make their matrix elements vanish and that evanescent operators could 
not mix into physical ones at the level of the anomalous dimension matrix 
$\gamma$, so that evanescent operators do not contribute to the Wilson 
coefficients through matching or evolution. On the other hand, the 
renormalisation matrix $Z$ of evanescent operators do contribute to the 
computation of the anomalous dimension matrix $\gamma$ for physical operators, 
and thus must be taken into account to renormalise the effective theory and to determine its running.

In our case, we will need the following evanescent operators $E_i[O]$ when we consider QCD corrections for the bilocal operators
\begin{eqnarray}
\gamma_\nu \gamma_\mu P_R \otimes  \gamma^\nu \gamma^\mu P_L &=& (4 + a_5 \epsilon ) P_R \otimes P_L + E_5[O]\,,
\nonumber \\ 
 \gamma_\rho \gamma_\nu \gamma_\mu P_R \otimes  \gamma^\rho \gamma^\nu \gamma^\mu P_L &=& (4 + a_3 \epsilon ) \gamma_\mu P_R \otimes \gamma^\mu P_L + E_3[O]\,,
\nonumber \\ 
\gamma_\alpha \gamma_\rho \gamma_\nu \gamma_\mu P_R \otimes \gamma^\alpha 
\gamma^\rho \gamma^\nu \gamma^\mu P_L &=& ((4 + a_5 \epsilon)^2+ b \epsilon) P_R \otimes P_L + E_7[O]\,,
\nonumber \\
\bigl( \bar s^\alpha P_L d^\beta\bigr)\bigl(\bar s^\beta P_R d^\alpha\bigr) +1/2 Q^{LR}_1 &=& E_1[O]\,,
\nonumber \\
\bigl(\bar s^ \alpha \gamma_\mu \gamma_\nu P_L d^\beta\bigr)\bigl(\bar s^\beta \gamma^\mu \gamma^\nu P_R d^\alpha\bigr) + (4+a_5 \epsilon)/2 Q^{LR}_1 &=& E_6[O]\,.
\label{eq:eva1}
\end{eqnarray}  
In the equations for $E_{1,6}$ $\alpha$ and $\beta$ are colour indices. Note that 
the quark fields have been written explicitly only for these two evanescent
operators which involve both colour singlet and
anti-singlet operators.  In all other cases the operators are
colour singlets and each choice of colour structure and external quark fields define a particular evanescent operator. Most of these definitions can be found in Ref.~\cite{Buras:2000if}. 
As discussed in Ref.~\cite{Herrlich:1994kh}, the definition of these evanescent operators is not unique (as illustrated by the presence of arbitrary constants $a_i$) and one has to ensure
that one uses the same definitions in all steps of the calculation, so
that the physical observables are independent of this choice. 
The definition of $E_7[O]$ has been chosen in relation with that of $E_5[O]$, introducing a coefficient $b$ in addition to the coefficient $a_5$ introduced for the latter. This is a consistent choice for the two evanescent operators since $E_7[O]$ may be seen as the evanescent operator coming from an evanescent operator (for instance, when inserting $E_5[O]$ in loop diagrams). It was shown in Ref.~\cite{Herrlich:1994kh} that such a consistent scheme led the anomalous dimensions to be independent of $b$. 

A few more evanescent operators will be relevant in the four-quark theory when we dress the $|\Delta S|=2$ operators $Q_{1,2}$ with gluons. 
These are written in a similar way as the previous ones up to a 
factor $m_c^2/g^2$ multiplying the Dirac structure (see the end of Sec.~\ref{sec:SMEFT}). For instance one has for $\hat{E}_5[Q]$ and $\hat{E}_1[Q]$:
\begin{eqnarray}
\frac {m_c^2}{g^2} \bigl(\gamma_\nu \gamma_\mu P_R \otimes  \gamma^\nu \gamma^\mu P_L \bigr) &=& \frac {m_c^2}{g^2}(4 + \bar{a}_5 \epsilon  ) P_R \otimes P_L + \hat{E}_5[Q]\,,
\nonumber \\ 
\frac {m_c^2}{g^2} \bigl( \bar s^\alpha P_L d^\beta\bigr)\bigl(\bar s^\beta P_R d^\alpha\bigr) + 1/2 Q_1 &=& \hat{E}_1[Q]\,,
\label{eq:eva2}
\end{eqnarray} 
and similarly for the other combinations considered in Eq.~\eqref{eq:eva1}.
The parameter associated with the $\epsilon$ term is denoted with a bar
since its value does not need to be the same as the one used in Eq.~\eqref{eq:eva1} and the same is true for the other evanescent operators (in the following we use $\bar{a}_i = a_i$  and $\bar{b}=b$ for simplicity). Finally when evaluating loop diagrams with the insertion of QCD counterterms we will need the following evanescent operator:
\begin{equation}
\gamma_\rho \gamma_\nu \gamma_\mu P_L \otimes  \gamma^\rho \gamma^\nu \gamma^\mu P_L = (16 + a_2 \epsilon ) \gamma_\mu P_L \otimes \gamma^\mu P_L + E_2[O]\,.
\end{equation}

In order to check our results we have thus performed the calculation for
arbitrary values of $a_i$ and $b$ (clearly no Fierz transformations
have been used since they are only valid for a special choice of values). However, unless specified and for simplicity, we will quote our results for 
\begin{equation}\label{eq:stvalai}
a_5= 4,\qquad a_3=4,\qquad b=96, \qquad a_2=-4\,.
\end{equation} 
Indeed, these values have been used in the determination
of the anomalous dimensions~\cite{Buras:2000if} which were relevant for the renormalisation group calculations
of the Wilson coefficients recalled in App.~\ref{app:anomdimgen}, and choosing different $a_i$ would require us to recompute these anomalous dimensions with the corresponding set of evanescent operators. Moreover, Fierz transformation can be applied in $D$ dimensions
with the choice $a_5=a_3=4$.

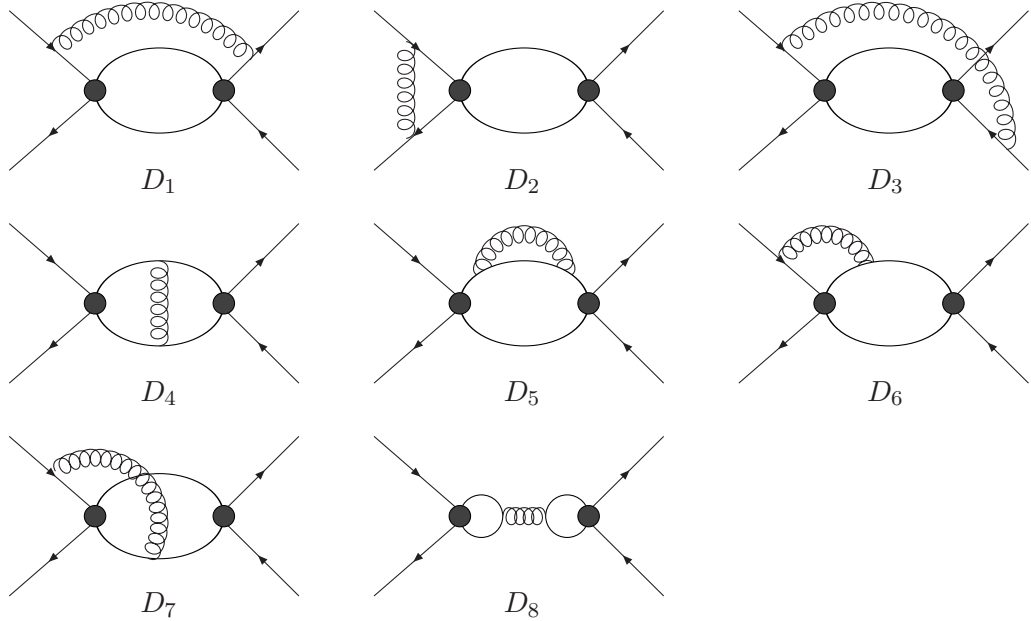
\begin{figure}[t]
\begin{center}
\begin{picture}(550,250)(0,-200)
\SetScale{0.8}\setlength{\unitlength}{0.8pt}
\ArrowLine(40,60)(80,25)
\ArrowLine(80,20)(40,-15)
\GlueArc(105,0)(60,38,137){4}{15}
\GOval(110,22.5)(20,30)(360){1}
\GCirc(80,22.5){5}{0.25}
\ArrowLine(140,25)(175,60)
\ArrowLine(175,-15)(140,20)
\Text(110,-20)[]{$D_1$}
\GCirc(140,22.5){5}{0.25}

\ArrowLine(210,60)(250,25)
\ArrowLine(250,20)(210,-15)
\Gluon(225,45.5)(225,0){4}{6}
\GOval(280,22.5)(20,30)(360){1}
\GCirc(250,22.5){5}{0.25}
\ArrowLine(310,25)(345,60)
\ArrowLine(345,-15)(310,20)
\Text(280,-20)[]{$D_2$}
\GCirc(310,22.5){5}{0.25}

\ArrowLine(380,60)(420,25)
\ArrowLine(420,20)(380,-15)
\GlueArc(445,0)(60,-5,137){4}{20}
\GOval(450,22.5)(20,30)(360){1}
\GCirc(420,22.5){5}{0.25}
\ArrowLine(480,25)(515,60)
\ArrowLine(515,-15)(480,20)
\Text(450,-20)[]{$D_3$}
\GCirc(480,22.5){5}{0.25}

\ArrowLine(40,-40)(80,-75)
\ArrowLine(80,-80)(40,-115)
\GOval(110,-77.5)(20,30)(360){1}
\GCirc(80,-77.5){5}{0.25}
\Gluon(110,-57.5)(110,-97.5){4}{6}
\ArrowLine(140,-75)(175,-40)
\ArrowLine(175,-115)(140,-80)
\Text(110,-120)[]{$D_4$}
\GCirc(140,-77.5){5}{0.25}

\ArrowLine(210,-40)(250,-75)
\ArrowLine(250,-80)(210,-115)
\GlueArc(280,-65)(20,0,180){4}{10}
\GOval(280,-77.5)(20,30)(360){1}
\GCirc(250,-77.5){5}{0.25}
\ArrowLine(310,-75)(345,-40)
\ArrowLine(345,-115)(310,-80)
\Text(280,-120)[]{$D_5$}
\GCirc(310,-77.5){5}{0.25}

\ArrowLine(380,-40)(420,-75)
\ArrowLine(420,-80)(380,-115)
\GlueArc(420,-65)(20,0,160){4}{10}
\GOval(450,-77.5)(20,30)(360){1}
\GCirc(420,-77.5){5}{0.25}
\ArrowLine(480,-75)(515,-40)
\ArrowLine(515,-115)(480,-80)
\Text(450,-120)[]{$D_6$}
\GCirc(480,-77.5){5}{0.25}

\ArrowLine(40,-140)(80,-175)
\ArrowLine(80,-180)(40,-215)
\GOval(110,-177.5)(20,30)(360){1}
\GCirc(80,-177.5){5}{0.25}
\GlueArc(80,-180)(30,-35,128){4}{16}
\ArrowLine(140,-175)(175,-140)
\ArrowLine(175,-215)(140,-180)
\Text(110,-220)[]{$D_7$}
\GCirc(140,-177.5){5}{0.25}

\ArrowLine(210,-140)(250,-175)
\ArrowLine(250,-180)(210,-215)
\Gluon(270,-177.5)(290,-177.5){4}{4}
\ArrowLine(310,-175)(345,-140)
\ArrowLine(345,-215)(310,-180)
\Text(280,-220)[]{$D_8$}
\GCirc(260,-177.5){10}{1.}
\GCirc(300,-177.5){10}{1.}
\GCirc(250,-177.5){5}{0.25}
\GCirc(310,-177.5){5}{0.25}
\end{picture}
\end{center}
\vspace{-0.5cm}
\caption{\small\it Diagrams $D_i$ contributing at ${\cal O}(\alpha_s)$ to the operators
$O_{ij}$ in the effective four flavour theory. The curly lines denote gluons 
and the black circles the insertions of $|\Delta S|=1$ current-current operators.}\label{fig:diagsDi}
\end{figure}

\begin{figure}[t]
\begin{picture}(300,85)(0,0)
\ArrowLine(40,60)(80,25)
\Text(54,58)[]{$d$}
\ArrowLine(80,20)(40,-15)
\Text(54,-13)[]{$s$}
\Gluon(54,48)(108,48){4}{6}
\ArrowLine(82,25)(120,60)
\Text(106,58)[]{$s$}
\ArrowLine(120,-15)(82,20)
\Text(106,-13)[]{$d$}
\Text(80,-25)[]{$L_1$}
\GCirc(80,22.5){5}{0.25}

\ArrowLine(180,60)(220,25)
\Text(194,58)[]{$d$}
\ArrowLine(220,20)(180,-15)
\Text(194,-13)[]{$s$}
\Gluon(190,50)(190,-5){4}{7}
\ArrowLine(222,25)(260,60)
\Text(246,58)[]{$s$}
\ArrowLine(260,-15)(222,20)
\Text(246,-13)[]{$d$}
\Text(220,-25)[]{$L_2$}
\GCirc(220,22.5){5}{0.25}

\ArrowLine(320,60)(360,25)
\Text(334,58)[]{$d$}
\ArrowLine(360,20)(320,-15)
\Text(334,-13)[]{$s$}
\GlueArc(335,-3.7)(52,-0.9,94){4}{11}
\ArrowLine(362,25)(400,60)
\Text(386,58)[]{$s$}
\ArrowLine(400,-15)(362,20)
\Text(386,-13)[]{$d$}
\Text(360,-25)[]{$L_3$}
\GCirc(360,22.5){5}{0.25}
\end{picture}

\vspace{0.8cm}
\caption{\small\it Diagrams $L_i$
contributing at ${\cal O}(\alpha_s)$ in the effective four flavour theory. The curly lines denote gluons 
and the black circles the insertions of $|\Delta S|=2$ local operators.
}\label{fig:diagsLi}
\end{figure}
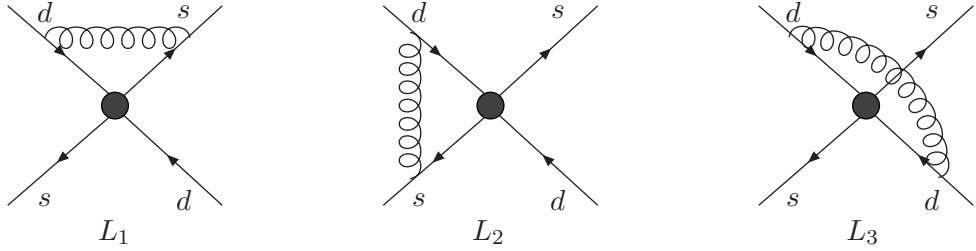

The NLO QCD corrections will correspond to two different kinds of diagrams: 
first, the one-loop diagram involving two $|\Delta S|=1$ operators and leading 
to the operators $O_{ij}$ can be dressed with a gluon (Fig.~\ref{fig:diagsDi}),
 then the $|\Delta S|=2$ local operators (counterterms or evanescent operators)
 can also be dressed  (Fig.~\ref{fig:diagsLi}). We will consider both types 
of contributions in the following.

\subsection{Matching at the high scale}

We will start by determining the value of the Wilson coefficients at the high scale.
The coefficients $C_{ij}$ for the bilocal operators are the product of $C_i$ Wilson coefficients, known from the matching of $O_\pm$ operators onto the underlying theory, and they are given in App.~\ref{app:anomdimgenS1}. On the other hand, we have to determine the value of the Wilson coefficients for $C^r_{1,2}$ for the $|\Delta S|=2$ local operators.

Let us consider the LO diagram in Fig.~\ref{fig:diagsD0}, 
giving in $D$ dimensions:
\begin{eqnarray}
D_0&=&i\frac{ m_c^2}{16 \pi^2} \biggl(\frac{1}{\epsilon} -\log \left(\frac{m_c^2}{\mu^2} \right) -1 - \frac{a_5}{4} \biggr) \bigl(P_R \otimes P_L +\tau_{rl} \gamma_\mu P_R \otimes 
\gamma^\mu P_L  \bigr)
\nonumber\\ 
&&-i\frac{ m_c^2}{64 \pi^2} \frac{1}{\epsilon} \biggl (E_5 + 2 \tau_{rl} (-
E_6 + 8  E_1)
\biggr) 
\label{eq:D0} 
\end{eqnarray}  
$\tau_{rl}$ is defined in Eq.~\eqref{eq:taurl} as
\begin{eqnarray}
\tau_{1}^{rl}=\tau_{rl}/4 \, ,\qquad \qquad \tau_{2}^{rl}=1/4 \, , \qquad\qquad \tau_{rl}=-(r+l+Nrl)/2\,,
\end{eqnarray}
where $r,l$ are equal to $\pm 1$ depending on the operator $O_{rl}$  considered.
The two antisinglets evanescent operator $E_1$ and $E_6$ are needed to
translate  the antisinglet operators into 
$\gamma_\mu P_R \otimes \gamma^\mu, P_L $ while $E_5$ appears in the 
calculation of $D_0$ as can be seen from the presence of the term $a_5$ in 
Eq.~\eqref{eq:D0}.
As already noted 
it is important to keep track of these operators: they contribute at two loops even in four dimensions, since
their one-loop matrix element yield contributions proportional to the physical operators $Q^{LR}_i$ (see below).

The LO contribution to the part of the amplitude proportional to the Wilson coefficient $C_{ij}$
in the effective four-quark theory Eq.~\eqref{eq:Hamil} thus reads:
\begin{equation}
A^{(WW')} (\mu)= 8 G_F^2 \beta h^2 \lambda^{LR}_{c} \lambda^{RL}_{c} \sum_{i,j=\pm} C_{ij}(\mu) \langle O_{ij}(\mu)\rangle^{(0)}\,, \label{eq:Amp4f}
\end{equation}
with 
\begin{equation}\label{eq:Oijprod}
\langle O_{ij} (\mu) \rangle^{(0)}=\frac{m_c^2 (\mu)}{ 4 \pi^2} \biggl(2 + \log \left(\frac{m_c^2}{\mu^2}\right) \biggr) \sum_{k=1,2} \tau^{ij}_{k} \langle Q^{LR}_k (\mu) \rangle^{(0)}\,,
\end{equation}
where from now on we use the
value $a_5=4$. The $1/\epsilon$ contribution in Eq.~\eqref{eq:D0} determines  the renormalisation tensor
\begin{equation}
Z^{-1,(1)}_{ij,k}=\frac{\alpha_s}{4\pi} \frac{1}{\epsilon} 4\tau^{ij}_k
\label{eq:z-11}
\end{equation}
see App.~\ref{app:SMcomput} for the notation of renormalisation quantities. 

We can match Eq.~\eqref{eq:Amp4f} to Eq.~\eqref{eq:Ampccexp} at the high scale $\mu_W$ (the precise value to be chosen for the high scale $\mu_W$ will be discussed in Sec.~\ref{sec:result}), which leads to
the following values of the Wilson coefficients $C_i^r$ for the local $|\Delta S|=2$ operators:
\begin{eqnarray}\label{eq:cirmuw}
C_1^r(\mu_W)&=& \mathcal{O} (\alpha^2_s)\,,
\\ \nonumber
C_2^r(\mu_W)&=& - \frac{\alpha_s(\mu_W)}{4 \pi}\times 4 \biggl[1 + \log\left(\frac{M_W^2}{\mu_W^2}\right)-\frac{1}{4}( \log\beta + 
F(\omega))\biggr] + \mathcal{O} (\alpha^2_s)\,,
\end{eqnarray}
with
$F(\omega)$ given in Eq.~\eqref{eq:fomega}  (using
the fact that $C_{ij}(\mu_W)=1$ at LO  and $\sum_{ij} \tau_{ij}=0$). This calculation is in fact sufficient to obtain $\bar \eta_{cc}$ at NLO.  At NNLO 
which we will also briefly consider, the corrections
to these equations will be very small since  of $O(\alpha_s(\mu_{W}))^2$ 
and we will not consider them further.

\subsection{RG evolution from the high scale down to $\mu=m_c$}\label{sec:anomdim}

The next step consists in determining the Wilson coefficients at a scale below $\mu_W$. This can be achieved once
we know the anomalous dimensions of all the operators involved.
Most of them have been determined in Ref.~\cite{Buras:2000if}. However, in the case
of $\bar\eta_{cc}$, we need to determine the anomalous dimension tensor $\gamma_{rl,i}$
which enters the renormalisation group equations for the $C_i$ coefficients and governs the
mixing from double insertions into the $C^r_i$ coefficients.
Eq.~\eqref{eq:Hamil} yields (see Ref.~\cite{Herrlich:1996vf} and App.~\ref{app:SMcomput} for more detail)
\begin{equation}
\mu \frac{d}{d \mu} C^r_i(\mu) =   \sum_j C^r_j(\mu) \gamma_{ji} + \sum_{r,l=\pm} C_r(\mu) C_l(\mu) \gamma_{rl,i} \,,
\label{eq:RGCi}
\end{equation}
where 
\begin{equation}
 \gamma_{rl,i}= \frac{\alpha_s}{4 \pi} \gamma^{(0)}_{rl,i} +\left(\frac{\alpha_s}{4 \pi} \right)^2 \gamma^{(1)}_{rl,i} +\cdots
\end{equation}

Using the relations from App.~\ref{app:SMcomput} and the result Eq.\eqref{eq:z-11}, we get
for the LO term  $\gamma^{(0)}_{rl,i}$ 
\begin{equation}
\gamma^{(0)}_{rl,i}= 2 [\tilde{Z}^{-1,(1)}_{1}]_{rl,i}=8 \tau_{rl,i}\,,
\end{equation}
while the $\gamma^{(1)}_{rl,i}$ are obtained from the divergences stemming from the diagrams in Fig.~\ref{fig:diagsDi} and Fig.~\ref{fig:diagsLi}. Some
intermediate results for different classes of diagrams are given in App.~\ref{app:indgraph} while the final result is
\begin{equation}
\gamma^{(1)}_{rl,i}=-4 h^{rl,i}(1/2)
\label{eq:anonlo}
\end{equation}
with 
\begin{eqnarray}
-h^{rl,1}(\lambda)&=&\frac{\lambda}{32 N}  \left((\bar{b}-96) \left(N^2-2\right) \beta_{rl}+\left(8(\bar{b}-48)
  -6 (\bar{b}-96) N^2\right) \tau_{rl}+ 6 N (\bar{b}-80)\right)\,
\nonumber \\
&&-(\bar{b}-280)
   \left(N^2-2\right)\frac{ \beta_{rl}}{64 N}+\left(3 \bar{b} N^2-4\bar{b}-152 N^2+48\right)
\frac{ \tau_{rl}}{32 N}+\frac{1}{32} (376-3 \bar{b}) ,
\nonumber \\
-h^{rl,2}(\lambda)&=& \frac{ \lambda}{8 N }  \left(3 \left(\bar{b}-16 \left(N^2+4\right)\right)
   +\left(48 -\frac{\bar{b}}{2}\right)N \beta_{rl}+\left(\bar{b} +96\right) N
   \tau_{rl}\right)
\nonumber \\
&& +\frac{1}{16 N}(-3 \bar{b}+72 N^2+304)+ 
( \bar{b}-280) \frac{\beta_{rl}}{32}-\left(\frac{\bar{b}}{8}+13\right) \frac{ \tau_{rl}}{2}\,,
\end{eqnarray}
where $ \beta_{rl} = r + l $ and the contribution from the evanescent operators is multiplied by a factor $\lambda$ which is set to $\lambda = 1/2$ in Eq.~\eqref{eq:anonlo}.  Indeed 
as discussed in Ref.~\cite{Buras:1989xd}, exploited in Ref.~\cite{Herrlich:1994kh}, and recalled in App.~\ref{app:SMcomput},
the contribution of evanescent operators to the NLO physical anomalous dimension corresponds to $1/\epsilon$ terms originating from $1/\epsilon^2$ poles in the tensor integrals multiplying a factor proportional to $\epsilon$ coming from the evanescent Dirac algebra. In each two-loop diagram the former are related to the corresponding one-loop counterterm diagrams by a factor of 1/2, because the non-local $1/\epsilon$-poles cancel in their sum in the expression for $ \gamma^{(1)}_{rl,i} $. Therefore, the correct contribution of the evanescent operators is obtained by inserting the evanescent counterterms with a factor of 1/2 into the one-loop diagrams.

It is easy to check that these anomalous dimensions are independent
of $\bar  b$ as demonstrated in Ref.~\cite{Herrlich:1994kh}. This provides 
an important check of our calculation.
In the case  $N=3$ one obtains:
\begin{eqnarray}
\gamma^{(1)}_{++,1}&=&-251/6 \, , \quad \quad \gamma^{(1)}_{+-,1}= \gamma^{(1)}_{-+,1}=169/2 \, , \quad  \quad \gamma^{(1)}_{--,1}=-355/6\,,
\nonumber \\
\gamma^{(1)}_{++,2}&=&-41/3 \, , \quad \quad \gamma^{(1)}_{+-,1}= \gamma^{(1)}_{-+,2}=73/3 \, , \quad \quad \gamma^{(1)}_{--,2}=223/3\,.
\end{eqnarray}
In order to solve Eq.~\eqref{eq:RGCi} we can rewrite the problem as a $6 \times 6$ 
homogeneous renormalisation group equation 
\begin{equation}
\mu \frac{d}{d \mu} \vec D = \tilde \gamma^T \cdot \vec D\,,\qquad\qquad\qquad
\vec D = \left( {
\begin{array}{c}
C_r C_l  \\
C_1 \\
C_2 
\end{array}} 
\label{eq:Dmat}
\right) ,
\end{equation}
with 
\begin{equation}\tilde \gamma^T = \left( {
\begin{array}{ccc}
  (\gamma_r+ \gamma_l) \cdot \mathbf{1}_{4 \times 4} & 0  \\
 \gamma_{rl}  & \gamma^T \\   
\end{array}} \right)\,,
\qquad\qquad \gamma_{rl} = \left( {
\begin{array}{cccc}
\gamma_{++,1}& \gamma_{+-,1}&\gamma_{-+,1}& \gamma_{--,1}
\\
\gamma_{++,2}& \gamma_{+-,2}&\gamma_{-+,2}& \gamma_{--,2}
\end{array}} 
\label{eq:Dmat1}
\right),
\end{equation}
and
\begin{equation}\gamma^{(i)} = \left( {
\begin{array}{cc}
\hat\gamma^{(i)}_{LR,11}-2 (\gamma_m^{(i)}-\beta_i)&\hat\gamma^{(i)}_{LR,12}
\\
\hat\gamma^{(i)}_{LR,21} & \hat\gamma^{(i)}_{LR,22}-2 (\gamma_m^{(i)}-\beta_i )
\end{array}} 
\right),
\end{equation}
with $\hat\gamma^{(i)}_{LR}$ the anomalous dimension at LO ($i=0$) or NLO ($i=1$) of the two  $|\Delta S|=2$ operators $Q_j^{LR}$ given in Eq.~\eqref{eq:anoQLR} and  $\beta_i$ the $\beta$ functions which govern
the evolution of the QCD coupling constant. $\beta_0$ is given below 
Eq.~\eqref{eq:resumv} and $\beta_1=102 -38/3 f$. The solution for $\vec D$ can
be straightforwardly obtained and will be given below at the scale $\mu_c$ in
Eq.~\eqref{eq:vecD}.

\subsection{Matching between the four- and the three-quark effective theories}

\subsubsection{Expression in the four-quark theory}

After running the Wilson coefficients from the high scale $\mu_W$ to the scale $m_c$, we have to match this theory onto a three-flavour effective theory with no charm. In order to perform this matching and determine the value of the Wilson coefficients in the three-flavour theory, we must compute $\langle H^{cc}\rangle$ in both theories. We will thus consider the computation in the four-flavour theory, which requires the finite part of the previous diagrams, given in App.~\ref{app:indgraph}.

Adding up the two-loop calculation of the diagrams $D_i$
and the contribution from the (evanescent and physical) counterterms, 
one  obtains finally
for the matrix element in the  effective four-quark theory
\begin{equation}\label{eq:heff4}
\langle H^{cc}\rangle =\frac{ 2 G_F^2}{ \pi^2} \beta h^2 m_c^2  \lambda^{LR}_c\lambda^{RL}_c\sum_i \biggl[ \sum_{rl} C_r C_l\left[ \left(2 +\log 
\frac{m_c^2}{\mu^2}\right) \tau_i^{rl} + \frac{\alpha_s}{4 \pi} c_i^{rl} \right] +C^r_i\biggr] \langle Q^{LR}_i\rangle^{(0)}
+ \cdots
\end{equation}
with   
\begin{eqnarray}
4 c_1^{rl}&=&-\frac{3}{2 }\log ^2\left(\frac{m_c^2}{\mu ^2}\right) \left(\frac{ \left( N^2-2\right) \beta_{rl}}{2 N }+ N \tau_{rl}+1 \right)
\nonumber\\
&&-4 \log
   \left(\frac{m_c^2}{\mu ^2}\right) \left(-\frac{11 \left(N^2-2\right) \beta_{rl}}{16 N}+\left(\frac{N}{2}+\frac{1}{N}-\frac{3}{8}\right) \tau_{rl}-\frac{3}{16 N}-1\right)
\nonumber \\
&&+ \frac{3}{ N} \tau_{rl} \, R \, \left(2+\log\left(\frac{m_c^2}{\mu ^2}\right)\right)
\nonumber \\
&&- \frac{3\left(N^2-2\right) \beta_{rl}}{16 N}-\frac{1}{8} \left(-71
   N+\frac{114}{N}-24\right) \tau_{rl}+\frac{3}{2 N}-\frac{41}{8}\,,
\label{eq:c1rl}
\end{eqnarray}
\begin{eqnarray}
4 c_2^{rl}&=&-\frac{3}{ N} R \left( \left(N^2-1\right)- 2 N \tau_{rl}\right)  \left(2+\log\left(\frac{m_c^2}{\mu ^2}\right)\right)- 3 \log ^2\left(\frac{m_c^2}{\mu ^2}\right) \left(\frac{1}{N}-\frac{ \beta_{rl}}{2}+ \tau_{rl}\right)
\nonumber\\
&&-4\log \left(\frac{m_c^2}{\mu
   ^2}\right) \left(3\left(1-\frac{1}{4 N}\right) \tau_{rl} + \frac{1}{8} \left(-2
   N-\frac{14}{N}-3\right)+\frac{11 \beta_{rl}}{8}\right)
\nonumber \\
&&- 4\left(\frac{43}{16}-\frac{3}{2 N}\right) \tau_{rl}-\frac{ 1}{4}(-19 N +\frac{60}{N} -12)+\frac{ 3 \beta_{rl}}{8}\,,
\label{eq:c2rl}
\end{eqnarray}
and the values for $C^r_i$ are given in Eq.~\eqref{eq:cirmuw}.
These gauge-independent terms have a remaining dependence on the regularisation
through the $R$ infrared-regularising terms defined as
\begin{equation}
R= \frac{1} {m_s^2- m_d^2} \bigl( m_s^2  \log(m_s^2/\mu^2) - m_d^2 \log(m_d^2/\mu^2)\bigr)\,.
\end{equation}
The gauge-dependent terms are
\begin{eqnarray}
4  c_{(1,\xi)}^{rl}&=&-\biggl[\biggl(\log \left(\frac{m_d^2 m_s^2}{\mu ^4}\right) \left(\frac{1}{2} 
   +\frac{\tau_{rl}}{N}\right)+\left(1-\frac{1}{2 N}\right)
+  R \left(-1+ 2 \tau_{rl} \left( N-\frac{2}{N}\right)\right) \biggr. \biggr.
\nonumber\\
&&\biggl. \biggl. \qquad + \tau_{rl} \left(-2 N+\frac{4}{N}-1\right) \biggr)  \biggl(1+\frac{1}{2}\log\left(\frac{m_c^2}{\mu ^2}\right)\biggr) \biggr]\,,
\nonumber\\
4  c_{(2,\xi)}^{rl}&=&-\biggl[\biggl( \log \left(\frac{m_d^2 m_s^2}{\mu ^4}\right) \left(2  \tau_{rl}+
\frac{1}{N} \right)  + 2 R \left(
   N-\frac{2}{N} -2  \tau_{rl}\right)\biggr. \biggr.
\nonumber \\
&&\biggl. \biggl. \qquad+\left(2(2 -\frac{1}{N}) \tau_{rl}-2
   N-1+ \frac{4}{N}\right) \biggr)  \biggl(1+\frac{1}{2}\log\left(\frac{m_c^2}{\mu ^2}\right)\biggr) \biggr]\,.
\label{eq:cirlgauge}
\end{eqnarray}
It is interesting to notice that all the regularisation and gauge-dependent terms
in equations \eqref{eq:c1rl}-\eqref{eq:cirlgauge} are multiplied
by the same quantity $2+\log(m_c^2/\mu^2)$ which is up to a constant
the LO amplitude in the four-quark theory,  Eq.~\eqref{eq:Oijprod}.
We will come back to this point while discussing the matching but it 
already indicates that these terms will cancel against similar terms from the 
effective three-quark theory in the final result, which is an important
test of our calculation.

\subsubsection{Matching onto the effective three-quark theory}\label{sec:effectivethree}

Below the scale $\mu_c \sim m_c$ the effective Hamiltonian is much simpler 
\begin{equation}\label{eq:heff3}
H^{cc}= \frac{2 G_F^2}{\pi^2} \beta h^2 m_c^2 (\mu) \lambda^{LR}_c\lambda^{RL}_c\sum_{i=1,2} \tilde C_i (\mu) \tilde Q_i^{LR}(\mu)\,,
\end{equation}
where the $|\Delta S|=2$ local operators $\tilde{Q}_i^{LR}$ are defined as 
\begin{equation}
\tilde Q_1^{LR} = (\bar s \gamma_\mu P_R d) (\bar s \gamma^\mu P_R d)\,, \qquad
\tilde Q_2^{LR} =  (\bar s P_L d) (\bar s  P_R d)\, .
\end{equation}
They differ from the corresponding ones in the effective four-quark theory
only through a normalisation.

 The matrix element of these operators can be written in the following way:
\begin{equation}
\langle \tilde{Q}_i^{LR} (\mu) \rangle^{(1)} = \langle \tilde{Q}_i^{LR} (\mu) \rangle^{(0)} +\frac {\alpha_s(\mu)}{4 \pi} 
\left( \sum_{j}  a(\mu)_{ji} \langle \tilde{Q}_j^{LR} (\mu) \rangle^{(0)} + \cdots \right) \,,
\end{equation}
where the ellipsis represents possible contributions from other operators.
The determination of $\langle \tilde{Q}_i^{LR} (\mu) \rangle^{(1)}$ is sketched
in App.~\ref{app:indgraph}. 
Adding up the contributions detailed there and taking   into account the colour factors (and  the other members
of each class obtained by left-right and up-down reflections), we obtain:
 \begin{equation}a(\mu)= \left( {
\begin{array}{cc}
 -\frac{3 N^2 -3 N -4}{2 N} + \frac{3}{N} R -\xi a_g& \frac{3(2 N +1)}{4 N}  +\xi \frac{b_g}{4} \\
 \frac{ N + 3}{ N} + 6 R +\xi b_g  & \frac{ 2 N^2+ 3 N +4}{2 N} -\frac{3( N^2 -1)}{ N} R  -\xi  a_g
\end{array}} \label{eq:Lieff3}\right) ,
\end{equation}
with the gauge-dependent parts given by
\begin{eqnarray}
a_g&=&\frac{ N^2 -2}{ N} R - \frac{2 N^2 +N-4 }{2 N }+\frac{1}{2 N} \log \left(\frac{m_s^2m_d^2}{\mu ^4}\right)\,,
\nonumber \\
b_g&=& 2 R -\frac{ 2N-1}{ N }- \log \left(\frac{m_s^2m_d^2}{\mu ^4}\right)\,.
\end{eqnarray}

At NLO the matching of the effective four-quark theory, Eq.~\eqref{eq:heff4}, to 
the three-quark theory, Eq.~\eqref{eq:heff3}, at the scale $\mu_c$ leads to
\begin{equation}
\tilde C_i(\mu_c) = \sum_{rl} C_r(\mu_c) C_l(\mu_c)  \left(2+\log\left(\frac{m_c^2}{\mu_c^2}\right)\right) \tau^{rl}_i + C_i^r(\mu_c) 
\frac{\pi}{ \alpha_s(\mu_c)} , \label{eq:matchLO}
\end{equation}
which we will use in the following. The running of the Wilson coefficients below the scale $\mu_c$ is provided in App.~\ref{app:anomdimgenS2}.

\subsubsection{Estimate of NNLO corrections}

In addition, our results also provide  an 
estimate of the size of NNLO corrections. Indeed,
at NNLO several new contributions appear, one of them coming from the ${\cal O}(\alpha_s)$ corrections
to the operators discussed previously. In particular, the
previous equation is modified as follows:
\begin{equation}
\tilde C_i^{\rm NNLO}(\mu_c)= \sum_{rl} C_r (\mu_c) C_l (\mu_c) \biggl[
 \left(2+\log\left(\frac{m_c^2}{\mu_c^2}\right)\right)
  \tau_{rl}^i +\frac{\alpha_s (\mu_c)}{4 \pi}   C_i^{{\rm op}} \biggr] +
\frac{\pi}{ \alpha_s(\mu_c)} C_i^r(\mu_c) + \cdots
\end{equation}
with
\begin{equation}
C_i^{{\rm op}}=c_i^{rl}-\frac{1}{8}\left( 2 + \log\left(\frac{m_c^2}{\mu_c^2}\right)\right) a_i^{rl}
\, ,\quad 
a_i^{rl}= \sum_{k=1,2} \tau^{rl}_k a_{ki}(\mu) \, ,
\end{equation}
and the dots stand for all other NNLO contributions. 
Using the expressions from Eq.~\eqref{eq:Lieff3} the $a_i^{rl}$ read
\begin{eqnarray}
a_1^{rl}&=&
\frac{6}{N} R \tau_{rl}-  \left(3 N-\frac{4}{N}-3\right) \tau_{rl}+
\frac{3}{2 N}+3 +\xi \biggl[-\left(\frac{1}{N} \tau_{rl}+\frac{1}{2}\right) \log \left(\frac{m_d^2 m_s^2}{\mu^4}\right)
\nonumber\\
&+&  R 
\left( 2 \tau_{rl}(\frac{2}{N}- N)+1\right) + \tau_{rl}(2 N -\frac{4}{N}+1)+\frac{1}{2 N}-1\biggr]\,,
\nonumber\\
a_2^{rl}&=&
-6 R \left(\frac{ \left(N^2-1\right)}{N}-2 \tau_{rl}\right)+\frac{2 (N+3)}{N} 
\tau_{rl}+2 N+\frac{4}{N}+3
\nonumber\\
&+&\xi \biggl[-  \left(\frac{1}{N}+2 \tau_{rl}\right) \log \left(\frac{m_d^2 m_s^2}{\mu^4}\right)+ 2 R 
\left(- N+\frac{2}{N}+2 \tau_{rl}\right)
\nonumber\\
&+&2 \left(\frac{1}{N}-2\right) 
\tau_{rl}+2 N-\frac{4}{N}+1 \biggr]\,.
\end{eqnarray}
It is easy to check that  the gauge-dependent terms as well as the terms
involving small quark masses $m_s$ and $m_d$ are
 canceled at the matching scale $\mu_c$  for any choice of the coefficients $a_i$ in the definition
of the evanescent operators. This provides additional powerful checks of the calculation and shows 
that our results are indeed independent of the choice of the QCD gauge and the infrared regularisation.

For completeness we give the final results in terms of 
$a_2=-4 +\epsilon_2$, $a_3=4+\epsilon_3$, $a_5=4+\epsilon_5$, $ \bar{b}=96 + \epsilon_b $,
where $\epsilon_i=0$ corresponds to the most widely used definitions of the
evanescent operators
\begin{eqnarray}
8 C_1^{\rm op}&=&\log \left(\frac{m_c^2}{\mu ^2}\right) \left[\epsilon_2 \left(\frac{\left(N^2-2\right) \beta_{rl}}{4 N}+\left(\frac{1}{N}-N\right) \tau_{rl}+1\right)
-\frac{\epsilon_3 \tau_{rl}}{N}-\frac{\epsilon_5}{2}\right.
\nonumber \\
&&\left.+\frac{11
   \left(N^2-2\right) \beta_{rl}}{2 N}-\frac{\left(N^2+12\right) \tau_{rl}}{N}+5\right]
\nonumber\\
&&+\log ^2\left(\frac{m_c^2}{\mu ^2}\right) \left(\left(\frac{3}{N}-\frac{3 N}{2}\right) \beta_{rl}-3 N \tau_{rl}-3\right) 
\nonumber\\
&&+\epsilon_5^2
   \left(-\frac{\left(N^2-2\right) \beta_{rl}}{32 N}+\left(\frac{3 N}{16}-\frac{1}{4
   N}\right) \tau_{rl}-\frac{3}{16}\right)
\nonumber\\
&&+\epsilon_b \left(\frac{3
   \left(N^2-2\right) \beta_{rl}}{64 N}-\frac{\left(N^2-2\right) \tau_{rl}}{32
   N}+\frac{1}{8}\right)
\nonumber\\
&&+\epsilon_5 \left(\epsilon_2 \left(\frac{\left(N^2-2\right) \beta_{rl}}{16 N}-\frac{\left(N^2-1\right) \tau_{rl}}{4
   N}+\frac{1}{4}\right) \right.
\nonumber\\
&&\left.+\left(\frac{2}{N}-N\right) \beta_{rl}+\left(\frac{21
   N}{4}-\frac{11}{2 N}\right) \tau_{rl}-\frac{45}{8}\right)
\nonumber\\
&&
+\epsilon_2
   \left(\left(\frac{N}{2}-\frac{1}{N}\right) \beta_{rl}+\left(\frac{2}{N}-2 N\right)
   \tau_{rl}+2\right)-\frac{\epsilon_3 \tau_{rl}}{N}
\nonumber\\
&&-\frac{3
   \left(N^2-2\right) \beta_{rl}}{8 N}+\left(\frac{95 N}{4}-\frac{73}{2 N}\right) \tau_{rl}-\frac{65}{4}\,,
\end{eqnarray}
\begin{eqnarray}
8 C_2^{\rm op}&=&\log ^2\left(\frac{m_c^2}{\mu ^2}\right) \left(-\frac{6}{N}+3 \beta_{rl}-6 \tau_{rl}\right)
\nonumber\\
&&+\log \left(\frac{m_c^2}{\mu ^2}\right) \left(\epsilon_2
   \left(\frac{2}{N}-\frac{\beta_{rl}}{2}\right)-\frac{\epsilon_5}{N}
-2 \epsilon_3 \tau_{rl}+\frac{10}{N}-11
   \beta_{rl}-26 \tau_{rl}\right)-2 \epsilon_3 \tau_{rl}
\nonumber\\
&&+\epsilon_5
   \left(-\frac{3 \left(N^2+14\right)}{4 N}+2 \beta_{rl}-\frac{\tau_{rl}}{2}\right)+\epsilon_5^2 \left(-\frac{3}{8 N}+\frac{\beta_{rl}}{16}-\frac{\tau_{rl}}{8}\right)
\nonumber\\
&&+\epsilon_b \left(\frac{1}{4N}-\frac{3 \beta_{rl}}{32}+\frac{\tau_{rl}}{16}\right)
\nonumber\\
&&+\epsilon_2
   \left(\epsilon_5 \left(\frac{1}{2 N}-\frac{\beta_{rl}}{8}\right)+\frac{4}{N}-\beta_{rl}\right)+\frac{11 N}{2}-\frac{38}{N}+\frac{3 \beta_{rl}}{4}-\frac{51
   \tau_{rl}}{2}\,.
\end{eqnarray}
The physical observables should not 
depend on the values chosen for $\epsilon_i$. In the following, we will set 
$\epsilon_{i}=0 $ since this is consistent with the values used
for the anomalous dimensions.

\begin{figure}[t!]
\begin{minipage}[b]{0.5\linewidth}
\includegraphics[width=7.5cm,angle=0]{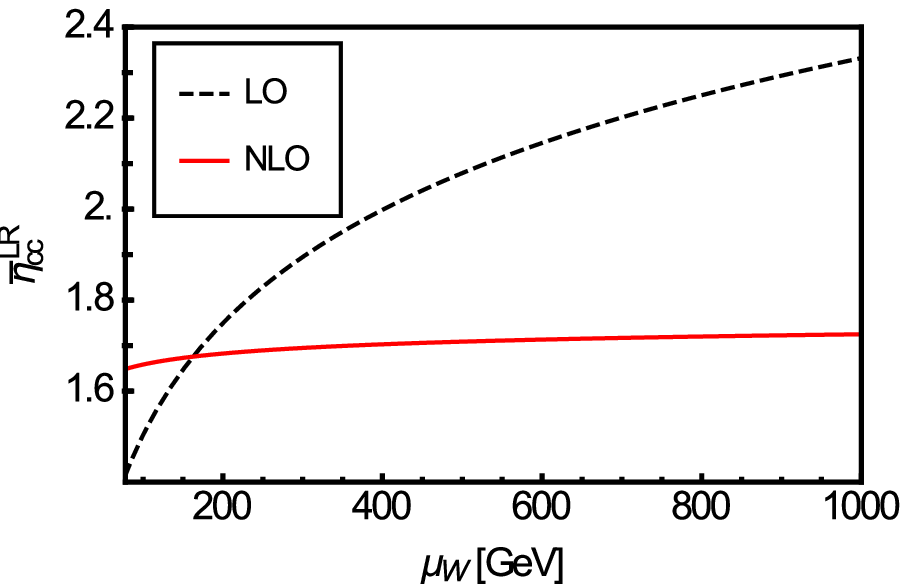}
\end{minipage}
\begin{minipage}[b]{0.5\linewidth}
\includegraphics[width=7.5cm,angle=0]{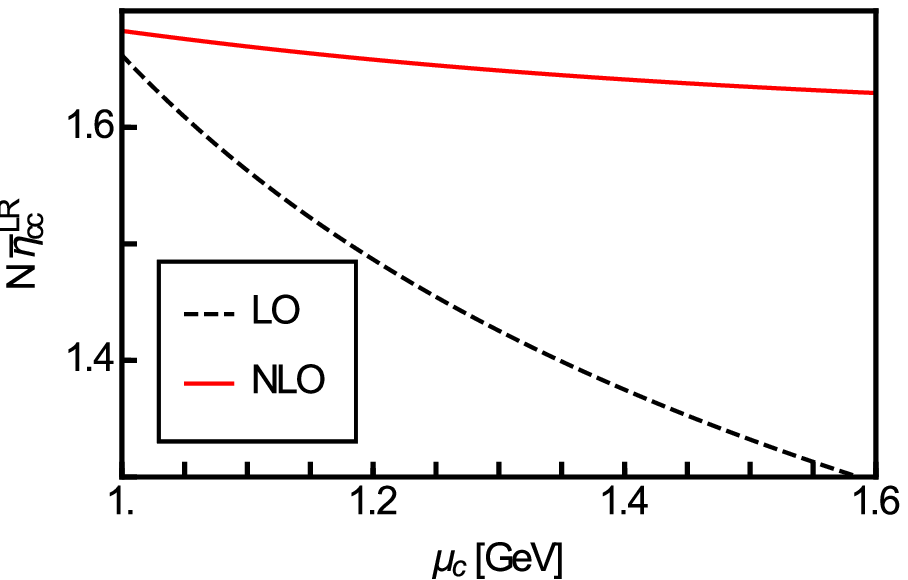}
\end{minipage}
\caption{\small\it
 Dependence of $\bar \eta_{cc}$ on the high (left panel) and on the low (right panel) scale  in the EFT approach for $M_{W_R}=1$ TeV and respectively for $\mu_c=m_c$ and
$\mu_W=M_W$. The other parameters are given in the text. The 
relevant 
quantity when $\mu_c \neq m_c$ is $ N \bar \eta_{cc}$ with $N$ defined in 
Eq.~\eqref{eq:noretamuc}.}
\label{figure:etaccmuw}
\end{figure}

\subsection{Short-distance corrections in EFT} \label{sec:result}

Combining Eq.~\eqref{eq:matchLO} with the renormalisation equation for $\vec D$ down to the low scale $\mu$ below $m_c$,
we obtain the final result for $\bar\eta_{a,cc}^{(LR)}$ at NLO in the EFT approach,
corresponding to the gauge-invariant combination of diagrams shown in the first row of Fig.~\ref{fig:diagall}:
 \begin{eqnarray}
&& \bar\eta_{a,cc}^{(LR)}=\frac{1}{S^{LR}(x_c (\mu_c),\beta,\omega)} \\
&& \sum_{j=1,2}\biggl( \bigl(1 +\frac{\alpha_s(\mu)}{4 \pi} K^{[3]}\bigr)
 \exp \biggl[ d^{[3]} \cdot \log\frac{\alpha_s(\mu_c)}{\alpha_s(\mu)}
\biggr] \bigl(1 -\frac{\alpha_s(\mu_c)}{4 \pi} K^{[3]}\bigr)\biggr)_{aj}
F_j(\mu_c)\,, \nonumber
\label{eq:etaeft}
\end{eqnarray}
with $S^{LR}(x_c,\beta,\omega)$ given in Eq.~\eqref{eq:ccSLR} and
\begin{eqnarray}
F_a(\mu_c)&=&\biggl( \frac{\pi}{\alpha_s(\mu_c)}  C_a^r(\mu_c)+  \sum_{r,l=\pm} 
\left( r_{rl,a}(\mu_c) +\frac{\alpha_s(\mu_c)}{4 \pi} C_a^{\rm op}(\mu_c) \right)C_r(\mu_c) C_l(\mu_c)\biggr) ,
\nonumber\\
&& \quad r_{rl,a}(\mu_c)= \bigl(2+\log(m_c^2/\mu_c^2)\bigr) \tau_a^{rl}  \, ,\qquad\qquad a=1,2\,,
 \label{eqmatch}
\end{eqnarray}
where  the values of $C_a^r(\mu_c)$, $C_r(\mu_c)$ and $C_l(\mu_c)$ are given by
the evolution of $\vec D$ down to $\mu_c$
\begin{eqnarray}
\vec D(\mu_c)&=&\left(1 +\frac{\alpha_s(\mu_c)}{4 \pi} \tilde J^{[4]}\right) \cdot \exp \biggl[ \tilde d^{[4]} \cdot \log\frac{\alpha_s(\mu_b)}{\alpha_s(\mu_c)}
\biggr]
\cdot \left(1 +\frac{\alpha_s(\mu_b)}{4 \pi} (\delta \tilde r^T(\mu_b) +\tilde J^{[5]}- \tilde J^{[4]})\right)
\nonumber\\
&& \cdot \exp \biggl[ \tilde d^{[5]} \cdot \log\frac{\alpha_s(\mu_W)}{\alpha_s(\mu_b)}\biggr]
\cdot \left(1 -\frac{\alpha_s(\mu_W)}{4 \pi} \tilde J^{[5]}\right)\cdot \vec D(\mu_W)\,.
\label{eq:vecD}
\end{eqnarray}
In order to get an estimate of the error due to neglected higher-order contributions, we have added in Eq.~\eqref{eqmatch} the contribution $C_a^{\rm op}$ which 
first appears at the next order.
The $C^r_i (\mu_W)$ are defined in Eq.~\eqref{eq:cirmuw} while $C_{\pm}(\mu_W)$
is defined in Eq.~\eqref{eq:cpmuw}.
The contribution $\delta \tilde r^T(\mu_b)$  cancels in the absence of penguin operators, which is the case here. 

Finally the matrices $\tilde d= \tilde d^{[f]}$, $\tilde J = \tilde J^{[f]}$ and $d=d^{[3]}$, $K=K^{[3]}$ encode respectively the 
$6 \times 6$  anomalous dimension matrix $\tilde \gamma$ defined in Sec.~\ref{sec:anomdim} and the $2 \times 2$ one
$\hat\gamma_{LR}$ defined in App.~\ref{app:anomdimgenS2}, with the additional definition
\begin{equation}
\tilde d=\frac{({\tilde\gamma}^{(0)})^T}{2 \beta_0} \, , \quad \quad \tilde{J} +[\tilde{d} ,\tilde{J}]=-\frac{(\tilde\gamma^{(1)})^T}{2 \beta_0}+\frac{\beta_1}{ \beta_0} \tilde{d}\, .
\end{equation}
Simplified expressions for $D_i(\mu_c)$ where effects from the 
five-flavour theory have been neglected and which are extremely good approximations
to the complete results read
\begin{eqnarray}
F_1&=&  
\frac{3}{104} \frac{\pi}{\alpha_s} \left(2 A^{--}-39 A^{+-}-26 A^{++}
+63 A_1\right)
\nonumber \\
&&\biggl. -\frac{1}{8} \left(
\log \left(\frac{m_c{}^2 (\mu_c)}{\mu _c{}^2}\right)+2\right) \left(A^{--}-6 A^{+-}+5
   A^{++}\right) 
\biggr. 
\nonumber \\
&&
\biggl. +\frac{1}{4}\biggl( -\frac{1761281}{390000} A^{--}+\frac{587029}
{220000} A^{+-}+\frac{16120889}{1110000}A^{++}
-\frac{4789827}{260000}A_1+\frac{1737}{296} A_2
 \biggr. \biggr.
\nonumber\\
&& \biggl. \biggl. 
+A \left(A^{--} \left(-\frac{12}{13} \log \left(\frac{\mu
   _W}{M_W}\right)-\frac{10181}{16250}\right)+A^{+-} 
\left(\frac{9}{2} \log \left(\frac{\mu
   _W}{M_W}\right)+\frac{39993}{10000}\right)  \right.
\nonumber\\
&&  \biggl. \left. 
+A^{++} \left(-6 \log \left(\frac{\mu
   _W}{M_W}\right)-\frac{7031}{2500}\right)+A_1 \left(\frac{63}{26} \log \left(\frac{\mu
   _W}{M_W}\right)-\frac{974889}{1430000}\right)\right)
\biggr)\,,\nonumber\\
\end{eqnarray}
\begin{eqnarray}
F_2&=&
 \frac{3}{1924} \frac{\pi}{\alpha_s} \left(2590 A^{--}-481 A^{+-}-182 A^{++}+777 A_1-2704 A_2\right) \biggr.
\nonumber \\
&&\biggl. + \frac{1}{4} \left(\log \left(\frac{m_c{}^2 (\mu_c)}{\mu _c{}^2}\right)+2\right) \left(A^{--}+2
   A^{+-}+A^{++}\right)
\nonumber \\
&&
 +\frac{1}{4}\biggl(-\frac{101273 A^{--}}{9750}+\frac{3969529
   A^{+-}}{330000}+\frac{6590729 A^{++}}{555000}-\frac{5219109 A_1}{130000}+\frac{21963
   A_2}{3700}
 \biggr.
\nonumber\\
&&  \biggl. 
+A \left(-\frac{7}{1625} A^{--} \left(15000 \log \left(\frac{\mu
   _W}{M_W}\right)+10181\right)+A^{+-} \left(3 \log \left(\frac{\mu
   _W}{M_W}\right)+\frac{13331}{5000}\right)  \right.\biggr. 
\nonumber\\
&&  \biggl. \left. 
-\frac{7}{46250} \left(15000 \log \left(\frac{\mu
   _W}{M_W}\right)+7031\right) A^{++}
\right .\biggr.
\nonumber \\
&&\biggl. \biggl. \left.
+A_2 \left(2\log \left(\frac{M_W{}}{M_{W'}{}}\right) +F(\omega)+\frac{2600}{37} \log \left(\frac{\mu
   _W}{M_W}\right)+\frac{1318747}{22200}\right)\biggr. \biggr.\right.
\nonumber\\
&& \biggl. \left. +A_1 \left(\frac{21}{13} \log \left(\frac{\mu
   _W}{M_W}\right)-\frac{324963}{715000}\right)\right)
\biggr) \,,
\end{eqnarray}
with
\begin{eqnarray}
&& A=\frac{\alpha_s(\mu_W)}{\alpha_s(\mu_c)}\,, \qquad \qquad 
A_1=\biggl(\frac{\alpha_s(\mu_W)}{\alpha_s(\mu_c)}\biggr)^{\frac{2}{25}}\,, \quad \quad
A_2=\biggl(\frac{\alpha_s(\mu_W)}{\alpha_s(\mu_c)}\biggr)^{-1}\,,
\\
&& A^{++}=\biggl(\frac{\alpha_s(\mu_W)}{\alpha_s(\mu_c)}\biggr)^{\frac{12}{25}}\,, \quad \quad
A^{+-}=\biggl(\frac{\alpha_s(\mu_W)}{\alpha_s(\mu_c)}\biggr)^{-\frac{6}{25}}\,, \quad \quad
A^{--}=\biggl(\frac{\alpha_s(\mu_W)}{\alpha_s(\mu_c)}\biggr)^{-\frac{24}{25}}\,. \nonumber
\end{eqnarray}

The value of  $\bar\eta_{cc}^{(LR)}\equiv\bar\eta_{2,cc}^{(LR)}$ at the scale $\mu=1$~GeV is 
\begin{equation}
\left.\bar\eta_{cc}^{(LR)}\right|_{EFT}=\frac{1}{1-0.0294 \, F(\omega)}[1.562  +(0.604 -0.037 F(\omega))-0.473]\,,
\label{eq:etaEFT}
\end{equation}
where $F(\omega)$ is defined in Eq.~\eqref{eq:fomega} and we have
taken $M_{W'}=1$ TeV (for \linebreak $M_{W'}=\mathcal{O} (1-10)$~TeV, the dependence on this parameter is very weak). 
The first and second terms in the brackets are the LO and NLO contributions stemming from the first term in Eq.~\eqref{eqmatch}, whereas the
last term comes from the $r_{rl,a}$ term in the same equation (the term $C_a^{\rm op}$ in Eq.~\eqref{eqmatch} being higher order).

\begin{figure}[t!]
\begin{minipage}[b]{0.5\linewidth}
\includegraphics[width=7.5cm,angle=0]{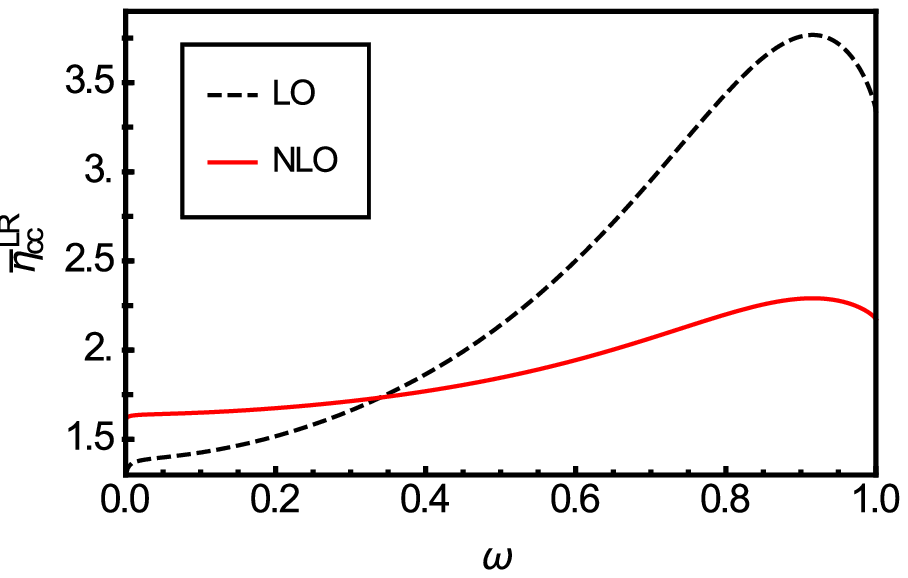}
\end{minipage}
\begin{minipage}[b]{0.5\linewidth}
\includegraphics[width=7.5cm,angle=0]{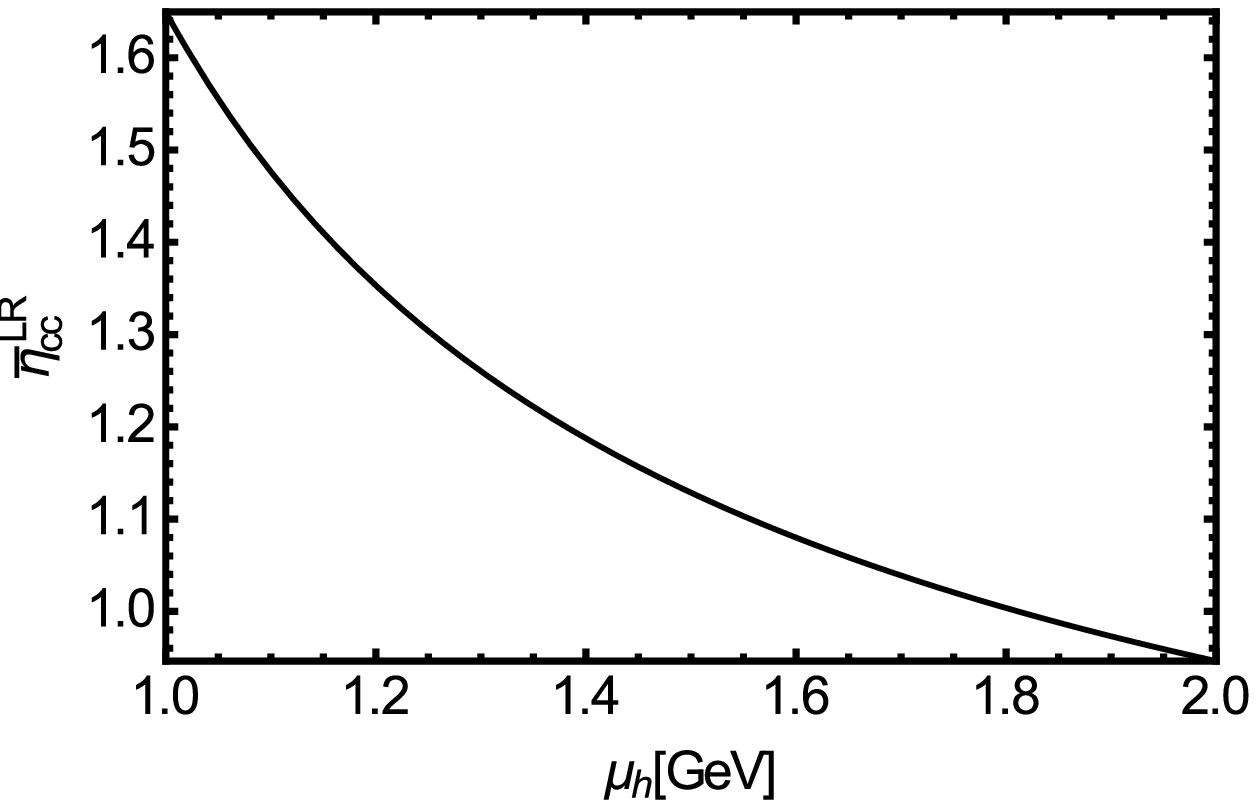}
\end{minipage}
\caption{\small\it
 Dependence of $\bar \eta_{cc}$ on  $\omega=M_{W'}^2/M_H^2$ (left panel) and on the hadronic scale  $\mu_h$  (right panel) in the EFT approach.
\label{figure:etaccmuEFT}}
\end{figure}

The dependence on the matching scales $\mu_W$
and $\mu_c$ is illustrated on Fig.~\ref{figure:etaccmuw}. This 
illustrates the  strong dependence of the LO result on the matching scales 
and the much milder dependence at NLO. 
This behaviour is similar to 
what is observed in the SM~\cite{Herrlich:1993yv,Buchalla:1995vs,Herrlich:1996vf} and it constitutes another significant check of our computation. In the case of the dependence on $\mu_c$, the relevant quantity is $N \bar \eta_{cc}$ with the normalisation factor given by
\begin{equation}
N= S^{LR}(x_c(\mu_c),\beta,\omega)/S^{LR}(x_c(m_c),\beta,\omega) \,,
\label{eq:noretamuc}
\end{equation}
considering that $S^{LR}(x_c(m_c))$ is the quantity multiplied by $\bar\eta_{cc}^{(LR)}$.
We also show the
dependence on the choice of the hadronic scale $\mu_h$ on the right panel of Fig.~\ref{figure:etaccmuEFT} for typical values between $1 < \mu_h < 2$~GeV. As can be
seen on the left panel of the same figure, there is a very mild dependence on the ratio of the masses of the $W'$ and $H$ bosons at NLO. 

\section{Discussion of the results}\label{sec:FinalResult}

We are now in a position to give our final results for the short-distance QCD corrections to $ K \bar{K} $ mixing
at NLO in LRM. Adding up our results from the previous sections yields the effective Hamiltonian:
\begin{eqnarray}
H &=& H^{SM} +\frac{G_F^2 M^2_W}{4\pi^2} 8 \beta h^2 Q_2^{LR} \sum_{U,V = c,t} \lambda_U^{LR} \lambda_V^{RL} \bar\eta_{UV}^{(LR)} \sqrt{x_U x_V} S^{LR}(x_U,x_V,\beta,\omega) \nonumber\\
&& - \frac{4 G_F}{\sqrt{2}} u \beta \omega Q_2^{LR} \sum_{U,V = c,t} \lambda_U^{LR} \lambda_V^{RL} \bar\eta_{UV}^{(H)} \sqrt{x_U x_V} \nonumber\\
&& + \frac{G_F^2 M_W^2}{4\pi^2} Q_2^{LR} \sum_{U,V = c,t} \lambda_U^{LR} \lambda_V^{RL} \bar\eta_{UV}^{(H^\pm \rm box)} S^{H}_{LR}(x_U,x_V,\beta \omega) + h.c. ,
\label{eq:LRMHam}
\end{eqnarray}
where $H^{SM}$ is given in Eq.~\eqref{eq:SMEffHamiltonian}, and
\begin{eqnarray}\label{eq:ttSLR}
S^{LR}(x_c,x_t,\beta,\omega) & = & \frac{1}{4} \left[\frac{x_t - 4}{x_t - 1} \log (x_t) + \log (\beta) + F(\omega) \right] \, , \nonumber\\
S^{LR} (x_t, \beta, \omega) & = & \frac{1}{4} \left( \frac{x_t^2 - 2 x_t + 4}{(x_t - 1)^2} \log (x_t) + \frac{x_t - 4}{x_t - 1} + \log (\beta) + F(\omega) \right) .
\end{eqnarray}
$S^{LR} (x_c, \beta, \omega)$ and $S^{H}_{LR}(x_U,x_V,\beta \omega)$  are given in Eqs.~\eqref{eq:chargedHiggsLoopFunction} and \eqref{eq:ccSLR}, respectively.

In the MR model we add the contributions given in Table~\ref{tab:etaLR} for the three diagrams~\ref{fig:diagall}(a), (b), (c) with the relevant weights and we normalise the result to 
$S^{LR}(x_U,x_V,\beta,\omega)$ in order to get the result in the appropriate form (the same applies to the charged Higgs in the
box which corresponds to the third line in Eq.~\eqref{eq:LRMHam}).

\subsection{Short-range contributions for the $cc$ box}

Since we computed $\bar\eta_{cc}^{(LR)} $ in both approaches, we can  compare the EFT result with the MR calculation.
We get from Eq.~\eqref{eq:etaEFT} and Table~\ref{tab:etaLR} for  $\omega=0.1$ ($\omega=0.8$)
\begin{eqnarray}
\left.\bar\eta_{cc}^{(LR)}\right|_{EFT}&=&1.41 +0.67-0.43=1.65\qquad (3.41-0.17-1.03=2.21)\, , \\
\left.\bar\eta_{cc}^{(LR)}\right|_{MR}&=&1.16+0.13+0.03=1.32 \qquad (2.46+0.27-1.32=1.41)\, .
\label{eq:etaccLRMRnum}
\end{eqnarray}
For consistency, the MR result is obtained by applying  the same counting for LO, NLO and NNLO contributions 
as in the EFT approach, which means that the non-logarithmic NLO contributions shown in Tab.~\ref{tab:etaLR} are counted as NNLO and are not included in Eq.~\eqref{eq:etaccLRMRnum}.
As in the SM case, we see that the central values from the MR are only in broad agreement (around 30\%) with the EFT approach in the presence of large logarithms, and in this sense we could quote a 30\% uncertainty in Eq.~\eqref{eq:etaccLRMRnum}. Including this uncertainty in our result and considering the values obtained with resummation of $\log \beta$, we have
\begin{equation}
\left. \bar\eta^{(LR)}_{cc} \right|_{MR} = 1.35 \pm 0.41 \pm 0.08 \qquad (1.48 \pm 0.44 \pm 0.10) ,
\end{equation}
where the first error comes from the comparison of MR and EFT, and the
second error is obtained by considering the values obtained with and without the resummation of $\log \beta$.

The EFT NLO central value will be taken as our final result. At the scale $\mu=1$ GeV and for $\omega=0.1$ ($\omega=0.8$), we have:
\begin{equation}\label{eq:LRetacc}
\bar\eta_{cc}^{(LR)}= 1.65 \pm  0.50 \qquad  (2.21\pm 0.66)\,,
\end{equation}
where the conservative $30 \%$ error bar includes our estimate of higher-order terms, namely:
the contribution from $C_a^{op}$ (which turns out to be very small), contributions from the expansion of 
Eq.~\eqref{eq:etaeft} up to NNLO, an estimate of the NNLO term assuming a geometrical growth from  LO to NLO, the arbitrariness in the choice of $\mu_W$ when integrating out the $W$ and $W'$ bosons to match onto the four-flavour theory (we vary $\mu_W$ between the two high scales $M_W$ and $M_{W'}$), the dependence on the choice of the matching scales for the matching onto the three-flavour theory. Each of these uncertainties are of the order of a few percent. Furthermore we have not resummed the contributions $\log\beta$. This last error is clearly difficult to determine without an explicit calculation, however this logarithm $\log \beta$ is multiplied by a suppressing factor $\alpha_s(\mu_W)$, suggesting that the error should be smaller than our conservative estimate of $30 \%$.

\subsection{Short-range contributions  for the $ct$ and $tt$ boxes }

The short-distance contributions from the $ct$ and $tt$ boxes in the MR are: 
\begin{eqnarray}
\bar\eta_{ct}^{(LR)}&=&  2.74 \pm 0.82 \pm 0.05   \qquad (2.67 \pm 0.80 \pm 0.03)\,,\\
\bar\eta_{tt}^{(LR)}&=& 5.88 \pm 1.76 \pm 0.23 \qquad (5.55 \pm 1.67 \pm 0.11)\,,
\end{eqnarray}
where the central value and the second uncertainty are  obtained by considering the values obtained with or without a resummation of $\log\beta$. The first uncertainty is a conservative 30\% estimate of the uncertainty of the MR coming from our previous experience in the SM, in relation with the fact that the top quark is not treated on the same footing as other heavy degrees of freedom in this approach.
As indicated earlier, resumming or not $\log\beta$ yields a small uncertainty from a few percent in both cases (as expected, since the
potentially large logarithm $\log\beta$ is multiplied by a suppressing factor  $\alpha_s(\mu_W)$). Moreover, we can see that our result is very stable with respect to  $\omega$, which will allow us to neglect the dependence of QCD short-distance corrections on $\omega$ when discussing constraints on LRM coming from $K\bar{K}$ mixing~\cite{BDV:2016}.

\subsection{Short range contribution from neutral and charged Higgs exchange}
The values of the QCD short-distance corrections for the box containing a charged heavy Higgs (see Fig.~\ref{fig:diagall}) are 
\begin{eqnarray}
\bar\eta_{ct}^{(H^\pm \rm box)} & = & 2.76 \pm 0.83 \pm 0.07 \qquad (2.79 \pm 0.84 \pm 0.10) , \\
\bar\eta_{tt}^{(H^\pm \rm box)} & = & 5.85 \pm 1.76 \pm 0.20  \qquad (5.90 \pm 1.77 \pm 0.25) , \\
\bar\eta_{cc}^{(H^\pm \rm box)} & = & 1.29 \pm 0.39 \pm 0.01 ,
\end{eqnarray}
where the first uncertainty corresponds to a conservative 30\% error related to the MR method,\footnote{Note that we provide only one $ \bar\eta_{cc}^{(H^\pm \rm box)} $ since the dependence on $ \omega $ is negligible.} and the second uncertainty  corresponds to an average of the results with and without a resummation of $ \log\beta $. For the tree-level neutral Higgs exchange we have
\begin{eqnarray}
\bar\eta_{ct}^{(H)} & = & 2.70 \pm 0.09 , \\
\bar\eta_{tt}^{(H)} & = & 5.66 \pm 0.30 , \\
\bar\eta_{cc}^{(H)} & = & 1.28 \pm 0.04 ,
\end{eqnarray}
where the quoted uncertainty assesses conservatively the neglected NLO corrections coming from the matching at $ \mu_H $ and the NNLO corrections based on a geometrical progression of the perturbative series.

\section{Conclusion}\label{sec:concl}

Among the extensions of the Standard Model, Left-Right models provide an interesting solution to  the violation of parity coming from the weak interaction. These models exhibit both additional $W'$ and $Z'$ gauge bosons and an extended Higgs sector needed to trigger the breakdown of the left-right symmetry. They are significantly constrained by several kinds of observables, and in particular kaon mixing which is accurately measured and which gets contributions from tree-level neutral Higgs inducing flavour-changing neutral currents. 

Kaon mixing can be analysed in the framework of the effective Hamiltonian, separating short- and long-distance contributions. The latter yield matrix elements that can be evaluated at a hadronic scale of a few GeV using lattice QCD simulations. The short-distance contributions can be determined thanks to a matching onto the fundamental theory (SM or Left-Right model) at a high scale corresponding to the mass of the heavy degrees of freedom. The bridge between the two scales is provided by RGE, which allows one to perform a resummation of large logarithms stemming from QCD corrections.

These short-distance QCD corrections are relevant to compute kaon mixing accurately in the Standard Model. They have been computed in the SM using a rigorous EFT approach where heavy degrees of freedom are progressively integrated out as the scale is lowered, showing the importance of NLO corrections. Another, approximate, method has been devised in earlier times to compute these QCD corrections
at LO, consisting in determining the range of loop momenta responsible for the
large logarithms and introducing the relevant anomalous dimensions to resum these logarithms. This method of regions is admittedly approximate but is far less demanding in terms of computation, compared to the EFT approach (once the relevant anomalous dimensions have been computed).

We first recalled basic features of these two methods, before proposing an extension of the method of regions to include NLO corrections. We compared the results of the two methods in the case of the Standard Model,  finding a good agreement for SM diagrams dominated by a single mass, but a 30\% discrepancy between our extension of the method of regions and the EFT computation in the case of large logarithm. We then considered the corrections for the Left-Right models using the method of regions. For some of the contributions, the computation has a different structure, depending on whether $\log\beta$ is treated as a large logarithm or not.

Since the $cc$ box exhibits a large logarithm $\log x_c$ at LO and thus might suffer from a large uncertainty in the method of regions, we decided to compute the short-distance QCD correction within 
the EFT approach, following closely Refs.~\cite{Herrlich:1993yv,Herrlich:1994kh,Herrlich:1996vf,Buras:2000if}. We matched the LRM onto a four-flavour theory, which was run down to $m_c$ and matched onto a three-flavour theory, before reaching a low hadronic scale $\mu_{h}$. A large number of cross-checks have been performed on our results (independence of the QCD gauge, independence of the definition of the evanescent operators, independence of the infrared regulators). Our result for $\bar\eta^{(LR)}_{cc}$ at NLO in the EFT approach showed again a 30\% discrepancy with the method of regions. We finally provided an estimate of the uncertainty to attach to our EFT computation at NLO.

We considered also the case of $ct$ and $tt$ boxes, where another  logarithm, namely $\log\beta$, may or may not be considered as large. Within the method of regions, both
cases led to very similar results. We then provided estimates for
$\bar\eta^{(LR)}_{ct}$ and $\bar\eta^{(LR)}_{tt}$ at NLO, using conservative error estimates based on our previous comparisons between the two approaches.

These results can be extended to the mixing for $B_d$ and $B_s$ meson, and they
can be used in order to constrain Left-Right models. Other constraints, such as electroweak 
precision observables, flavour-changing charged currents and direct searches, 
have also proven important and call for a global analysis of these models within an appropriate statistical framework. This will be the object of future work to determine the viability of Left-Right models in the doublet case, their ability to solve the violation of parity occurring in the Standard Model and the possibility to find part of their spectrum in the next run of the LHC~\cite{BDV:2016}.

\acknowledgments

We would like to thank A.~Buras, M.~Knecht, H.~Sadzjian and G.~Senjanovi\'{c} for interesting and useful discussions.
LVS acknowledges funding by the P2IO LabEx (ANR-10-LABX-0038) in the framework ``Investissements d'Avenir" (ANR-11-IDEX-0003-01) managed by the French National Research Agency  (ANR). 

\appendix

\section{$|\Delta S|=2$ effective Hamiltonian in the SM \label{app:SMcomput}}

We outline the main steps of the derivation of the $|\Delta S|=2$ Hamiltonian in the Standard Model, borrowing heavily from Ref.~\cite{Herrlich:1996vf} (which should be consulted for any further detail) and neglecting penguin contributions for simplicity.

\subsection{Minimal operator basis}

One has the following Hamiltonian for $|\Delta S|=1$ transitions
\begin{eqnarray}
H^{|\Delta S|=1}_{\rm eff} &=& - \frac{G_F}{\sqrt{2}}  \sum_{i=1}^2
\sum_{U,V =u,c} V_{ks}^\ast  V_{ld}  C_i Q_i^{UV} 
\label{lags1}
\end{eqnarray}
with the two operators
\begin{equation}
Q_1^{UV} = \left(\bar{s} \gamma_{\mu} L U\right) \cdot \left(\bar{V} \gamma^{\mu} L d\right) \cdot \widetilde{\bf 1
}\qquad Q_2^{UV} = \left(\bar{s} \gamma_{\mu} L U\right) \cdot \left(\bar{V} \gamma^{\mu} L d\right) \cdot {\bf 1}\label{defQ}
\end{equation}
where $\bf 1$ and $\widetilde{\bf 1}$ denote colour singlet and antisinglet and $L=(1-\gamma_5)$.
The 2$\times$2 renormalization matrix
$Z^{-1}_{ij}$ is diagonal in the basis
\begin{equation}
Q_{\pm}^{UV} = \frac{1}{2} \left(Q_2^{UV} \pm Q_1^{UV}\right),
\end{equation}
provided one preserves Fierz symmetry in the renormalization process

The Hamiltonian for $|\Delta S|=2$ transitions reads
\begin{eqnarray}
H^{|\Delta S|=2}_{\rm eff} &=&
- \frac{G_F}{\sqrt{2}} \sum_{i=\pm} C_i
 [  \sum_{j=\pm}  Z^{-1}_{ij}
\sum_{U,V=u,c} V_{Us}^\ast  V_{Vd}   Q_j^{UV, \, {\rm bare}}   ] \nonumber \\
&& - \frac{G_F^2}{16\pi^2} \lambda_t^2 \tilde{C}_{S2}^{(t)} \tilde{Z}_{S2}^{-1}  \tilde{Q}_{S2}^{\rm bare}
- \frac{G_F^2}{2} \lambda_c \lambda_t
     [ \sum_{k,l=\pm}
   C_k C_l \tilde{Z}_{kl,7}^{-1} + \tilde{C}_7 \tilde{Z}_{77}^{-1} ]
  \tilde{Q}_7^{{\rm bare}}
\label{lags2}
\end{eqnarray}
where counterterms proportional to evanescent operators are not displayed and 
 local operators absorb the divergences arising from the charm-top and top-top boxes:
\begin{eqnarray}
\tilde{Q}_7 &=& \frac{m_c^2}{g^2 \mu^{2 \epsilon}}  \tilde{Q}_{S2}
= \frac{m_c^2}{g^2 \mu^{2 \epsilon}} \cdot
     \bar{s} \gamma_\mu L d \cdot \bar{s} \gamma^\mu L d \, .
\label{defqseven}
\end{eqnarray}
Since the charm is still dynamical, the $\tilde{Q}_7$ operator gets two types of divergences, corresponding to graphs with two insertions of $|\Delta S|=1$ operators with charm quarks, or to the single insertion of the local operator $\tilde{Q}_7$. Due to the GIM mechanism, there are no divergences in the SM for boxes with identical internal flavours, so that for top-top boxes, only the second type of contribution arises for $\tilde{Q}_{S2}$ whereas there are no such local operators for charm-charm boxes.

Evanescent operators must be introduced as counterterms above in order to make the one-loop diagrams with the insertion of $Q_j$ finite:
\begin{eqnarray}
\hspace{-2em}
E_1[Q_j] &\! = \! & \left[
\gamma_\mu \gamma_\nu \gamma_\eta L \otimes \gamma^\eta \gamma^\nu \gamma^\mu L
- \left(4+a_1\epsilon\right) \gamma_\mu L \otimes \gamma^\mu L
\right] K_{1j} , \hspace{1em} j=1,\ldots 2
\label{StdEvas1-12}\\
E_1[\tilde{Q}_7] &\! = \! & \frac{m_c^2}{g^2} \left[
\gamma_\mu \gamma_\nu \gamma_\eta L \otimes \gamma^\eta \gamma^\nu \gamma^\mu L
- \left(4+\hat{a}_1\epsilon\right) \gamma_\mu L \otimes \gamma^\mu L
\right] K_{12} ,
\label{StdEvas1-loc}\\
E_2[\tilde{Q}_7] &\! = \! & \frac{m_c^2}{g^2} \left[
\gamma_\mu \gamma_\nu \gamma_\eta \gamma_\sigma \gamma_\tau L \otimes
\gamma^\tau \gamma^\sigma \gamma^\eta \gamma^\nu \gamma^\mu L 
- \left[\left(4+\hat{a}_1\epsilon\right)^2+\hat{b}_1 \epsilon \right]
\gamma_\mu L \otimes \gamma^\mu L \right] K_{22} \, , \nonumber\\
\label{StdEvas2-loc}
\end{eqnarray}
with colour factors $K_{ij}$ being linear combinations of $ \widetilde{\bf 1}$ and $ {\bf 1}$
and arbitrary constants $a_{1,2},\hat{a}_1,\hat{b}_1$ defining these evanescent operators.

\subsection{Matching at the high scale}

The determination of the $|\Delta S|=1$ Wilson coefficients can be done at the high scale as
\begin{eqnarray}
C_{\pm}\left(\mu_{tW}\right) &=& 1 +
\frac{\alpha_s\left(\mu_{tW}\right)}{4\pi} \left[ \ln \frac{\mu_{tW}}{M_W}
\gamma^{(0)}_{\pm} + B_{\pm} \right] + \mathcal{O} (\alpha_s^2)
\end{eqnarray}
with the anomalous dimensions $\gamma_\pm^{(0)}$ of the $|\Delta S =1|$ operators defined in App.~\ref{app:anomdimgen}.
For $|\Delta S|=2$ Wilson coefficients, we must perform the matching of a $|\Delta S|=2$ Green function at the high scale in the full and the effective theory
\begin{eqnarray}
\!\!
\left\langle T
\exp\left[i \int d^D x H^{|\Delta S|=2}_{\rm eff}\left(x\right)\right]
\right\rangle_{|\Delta S|=2} &=&
-i \left\langle H^{c}+H^{t}+H^{ct}\right\rangle
+ \mathcal{O} \left( G_F^3 \right) ,
\label{GreenMix}
\end{eqnarray}
where
\begin{subequations}
\label{GreenMixH}
\begin{eqnarray}
H^{c}\left(x\right) &=& \lambda_c^2 \frac{G_F^2}{2}
\sum_{i,i',j,j'=\pm} C_i C_j
\underbrace{Z^{-1}_{ii'} Z^{-1}_{jj'}
	\mathcal{O}^{{\rm bare}}_{i'j'}\left(x\right)}
	_{\displaystyle \equiv \mathcal{O}_{ij}\left(x\right)},
\label{GreenMixHc}
\\
H^{t}\left(x\right) &=& \lambda_t^2 \frac{G_F^2}{16\pi^2} \tilde{C}_{S2}^{(t)}
\tilde{Z}_{S2}^{-1} \tilde{Q}_{S2}^{{\rm bare}}\left(x\right),
\label{GreenMixHt}
\\
H^{ct}\left(x\right) &=& \lambda_c \lambda_t\frac{G_F^2}{2}\Biggl[\!\!
\;\sum_{i,j=\pm}\!\!\! C_i C_j\!
\underbrace{\left(
	\sum_{i',j'=\pm} 
	Z^{-1}_{ii'} Z^{-1}_{jj'}
	\mathcal{R}^{{\rm bare}}_{i'j'}\left(x\right)
	+
	\tilde{Z}_{ij,7}^{-1} \tilde{Q}_7^{{\rm bare}}\left(x\right)
	\right)}
	_{\displaystyle \equiv \mathcal{R}_{ij}\left(x\right)}
+  \tilde{C}_7 \tilde{Z}_{77}^{-1} \tilde{Q}_7^{{\rm bare}}\left(x\right) \Biggr].\nonumber\\
\label{GreenMixHct}
\end{eqnarray}
\end{subequations}
Here, the bare $\mathcal{O}_{ij}$ and $\mathcal{R}_{ij}$ combinations denote the bilocal structures composed of two $|\Delta S|=1$ operators.
In each case (charm-charm, charm-top, or top-top box), the computation of the above 
Green function allows one to determine the values of the Wilson coefficients for the $|\Delta S|=2$ operators.

\subsection{RG evolution of the Wilson coefficients from the high scale down to $\mu=m_c$}

The renormalisation is again discussed in a different manner for single and double insertions. In the first case, the derivation can be obtained from the RG equation
\begin{eqnarray}
\sum_{j=\pm} \left[ \delta_{jk}\, \frac{d}{d\mu} - \gamma_{jk} \right] C_j = 0
\qquad
\gamma_{ij}\left(g\left(\mu\right)\right) =
\sum_{k=\pm}  Z^{-1}_{ik} \mu \frac{d}{d\mu} Z_{kj}
\label{RGsingle}
\end{eqnarray}
for the Wilson coefficient functions $C_j$, where $\gamma$
is the anomalous dimension matrix of the $|\Delta S|=1$ operators $Q_k$ (we recall that we neglect penguin operators). In the case of $Q_\pm$, $\tilde{Q}_7$ or $\tilde{Q}_{S2}$ which do not mix with other operators, this matrix reduces to simple numbers. Attention should be paid for the crossing of thresholds (such as $\mu=m_b$).

We expand the renormalization matrix $Z^{-1}$ as
\begin{eqnarray}
Z^{-1} &=& 1
+\frac{\alpha_s}{4\pi} Z^{-1,(1)}
+\left(\frac{\alpha_s}{4\pi}\right)^2 Z^{-1,(2)}
+\ldots,
\qquad
Z^{-1,(n)} =
\sum_{r=0}^{n} \frac{1}{\epsilon^r} Z_r^{-1,(n)}\,.
\label{ExpZSingle}
\end{eqnarray}
To deal with the evanescent operators, $Z^{-1}$ contains a finite
renormalization piece.  The coefficients of the
perturbative expansion of
\begin{eqnarray}
\gamma &=&
\frac{\alpha_s}{4\pi} \gamma^{(0)}
+\left(\frac{\alpha_s}{4\pi}\right)^2 \gamma^{(1)}
+\ldots,
\label{ExpAnomSingle}
\end{eqnarray}
are obtained as
\begin{eqnarray}
\gamma^{(0)} &=& 2 Z^{-1,(1)}_1 + 2\epsilon Z^{-1,(1)}_0 \\
\gamma^{(1)} &=& 4 Z^{-1,(2)}_1 + 2\left\{Z^{-1,(1)}_0,Z^{-1,(1)}_1\right\} +
2 \beta_0 Z^{-1,(1)}_0\,.
\end{eqnarray}

The local operator counterterms proportional to
$\widetilde{Z}_{kl,7}^{-1}\left(\mu\right)$  do not influence the RG
evolution of the coefficients $C_l$, but they modify the running of
$\tilde{Q}_7$.  The independence of the $|\Delta S|=2$ effective Hamiltonian on $\mu$ yields the following RG equation
\begin{eqnarray}
\frac{d}{d\mu}\tilde{C}_7\left(\mu\right) &=&
\tilde{C}_7\left(\mu\right) \tilde{\gamma}_{77}
+ \sum_{k,k'=\pm} 
   C_{k}\left(\mu\right) C_{k'}\left(\mu\right) \tilde{\gamma}_{kk',7}
\label{RGdouble1}
\end{eqnarray}
with the anomalous dimension tensor
\begin{eqnarray}
\tilde{\gamma}_{kn,7} &=&
\frac{\alpha_s}{4\pi} \tilde{\gamma}_{kn,7}^{\left(0\right)}
+ \left(\frac{\alpha_s}{4\pi}\right)^2 \tilde{\gamma}_{kn,7}^{\left(1\right)}
+ \ldots
\nonumber \\
&=&
- \sum_{k^\prime,n^\prime=\pm}
\left[\gamma_{kk'} \delta_{nn'} + \delta_{kk'} \gamma_{nn'}\right]
\tilde{Z}_{k'n',7}^{-1} \tilde{Z}_{77}
-\left[\mu \frac{d}{d\mu} \tilde{Z}_{kn,7}^{-1}\right] \tilde{Z}_{77}.
\label{AnomDimTens}
\end{eqnarray}
Its first perturbative coefficients are
\begin{eqnarray}
\tilde{\gamma}_{kn,7}^{(0)} &=&
2 [\tilde{Z}^{-1,(1)}_{1}]_{kn,7} + 2\epsilon [\tilde{Z}^{-1,(1)}_{0}]_{kn,7} \\
\tilde{\gamma}_{kn,7}^{(1)} &=&
4 [\tilde{Z}^{-1,(2)}_{1}]_{kn,7} + 2\beta_0 [\tilde{Z}^{-1,(1)}_{0}]_{kn,7}
\nonumber \\
& & -2 [\tilde{Z}^{-1,(1)}_{0}]_{kn,7} [\tilde{Z}^{-1,(1)}_{1}]_{77}
	-2 [\tilde{Z}^{-1,(1)}_{1}]_{kn,7} [\tilde{Z}^{-1,(1)}_{0}]_{77}
\nonumber \\
& & -2 \sum_{k^\prime,n^\prime=1}^2  \left\{
\left( [Z^{-1,(1)}_{0}]_{kk'} \delta_{nn'}
	+ \delta_{kk'} [Z^{-1,(1)}_{0}]_{nn'} \right)
	[\tilde{Z}^{-1,(1)}_{1}]_{k'n',7}    \right.
\nonumber \\
& &\qquad\left.
+ \left( \left[Z^{-1,(1)}_{1}\right]_{kk'} \delta_{nn'}
	+ \delta_{kk'} \left[Z^{-1,(1)}_{1}\right]_{nn'} \right)
	[\tilde{Z}^{-1,(1)}_{0}]_{k'n',7} \right\}\,.
\end{eqnarray}
The above equations include finite
renormalisation constants (subscript 0), which
appear  when counterterms proportional to
evanescent operators must be included. The extra terms involving the finite renormalisation
constants can be simply included into the calculation by multiplying
all one-loop diagrams containing a finite counterterm by a factor of 1/2.

The value of the anomalous
dimension tensor $\tilde{\gamma}_{\pm j,7}$ governs the mixing from double
insertions to $\tilde{C}_7$.  This tensor is determined from the
renormalization factor $\tilde{Z}_{ij,7}^{-1}$, which can be determined from the finiteness of the Green function $-i\langle H^{ct}\rangle$
\begin{equation}
[\tilde{Z}^{-1,(1)}]_{ij,7} \left\langle \tilde{Q}_7 \right\rangle^{(0)} = -\left\langle \mathcal{R}_{ij} \right\rangle^{(0),{\rm bare}},
\end{equation}
and similarly for higher orders, requiring the evaluation of $\langle \mathcal{R}_{ij} \rangle^{{\rm bare}}$ and $\langle \tilde{Q}_7\rangle ^{{\rm bare}}$ up to the relevant order. Standard methods can then be used to solve the differential equation Eq.~\eqref{RGdouble1} (especially when $i,j=\pm$ leading to diagonal expressions).

\subsection{Matching at $\mu=m_c$}

At the scale $\mu=m_c$, one can then match this theory to the effective three-quark theory
\begin{equation}
H =
-\frac{G_F^2}{16\pi^2} \left[
\lambda_c^2 \tilde{C}_{S2}^{(cc)}\left(\mu\right)
+\lambda_t^2 \tilde{C}_{S2}^{(tt)}\left(\mu\right)
+\lambda_c \lambda_t \tilde{C}_{S2}^{(ct)}\left(\mu\right)
\right] \tilde{Z}_{S2}^{-1}\left(\mu\right) \tilde{Q}_{S2}^{\rm bare} \, .
\label{lags2c}
\end{equation}
Equating the Green function Eq.~\eqref{GreenMix} in both four-quark and three-quark theories yields
the values of the Wilson coefficients in this theory at the scale $\mu=\mu_c$.
In the charm-top case, one gets
\begin{equation}
\sum_{i,j=\pm} C_i ( \mu_c ) C_j ( \mu_c ) 
              \langle {\mathcal R}_{ij} \rangle ( \mu_c ) 
+ \tilde{C}_7 ( \mu_c ) \langle \tilde{Q}_7 \rangle ( \mu_c )
 = 
\frac{1}{8 \pi^2} \tilde{C}_{S2}^{(ct)} ( \mu_c ) \langle \tilde{Q}_{S2} \rangle
   ( \mu_c ).
\label{matchct}
\end{equation}
$\tilde{C}_7(\mu_c)$ is already nonzero in the LO due to its mixing with
$C_2$,  whereas the two insertion contribution starts at NLO only
\begin{eqnarray}
\left\langle\mathcal{R}_{ij}\left(\mu\right)\right\rangle^{(0)}
&=&
\frac{m_c^2\left(\mu\right)}{16\pi^2} \, 2 \, r_{ij,S2}\left(\mu\right)
\left\langle\tilde{Q}_{S2}\right\rangle^{(0)} \, ,
\label{DefRmatch}
\end{eqnarray}
with $r_{ij,S2}$ given by the finite part of the diagrams $D_i$ and $L_i$
leading to
\begin{equation}
\tilde{C}_{S2}^{(ct)}\left(\mu_c\right) =
m_c^2\left(\mu_c\right) \left[\frac{1}{2}
\frac{4\pi}{\alpha_s\left(\mu_c\right)} \tilde{C}_7\left(\mu_c\right)
+ \sum_{i=\pm} \sum_{j=1}^{6} r_{ij,S2}\left(\mu_c\right)
C_i\left(\mu_c\right) C_j\left(\mu_c\right) \right].
\label{cllNLOct}
\end{equation}
In the top-top case, the three-quark and four-quark theories are completely identical up to the running of the strong coupling constant, making the determination of the Wilson coefficient $\tilde{C}_{S2}(\mu)$ very simple. In the charm-charm case, only two insertions of $|\Delta S|=1$ operators contribute in the four-quark theory, leading to a simple parametrisation of the matching
\begin{eqnarray}
\left\langle\mathcal{O}_{ij}\left(\mu\right)\right\rangle &=&
\frac{m_c^2\left(\mu\right)}{16\pi^2} \, 2 \, d_{ij,S2}\left(\mu\right)
\left\langle \tilde{Q}_{S2}\left(\mu\right)\right\rangle.
\label{DefDmatch}
\end{eqnarray}

\subsection{RG evolution of the Wilson coefficients from $\mu=m_c$ down to the low scale}

The running of the Wilson coefficients according to 
the RG equation in the three-flavour theory is then trivial, limited to the single operator $\tilde{Q}_{S2}$, with the expression
\begin{eqnarray}
\tilde{C}_{S2}^{(j)}\left(\mu\right) &=&
\tilde{C}_{S2}^{(j)}\left(\mu_c\right)
\left[\frac{\alpha_s\left(\mu_c\right)}{\alpha_s\left(\mu\right)}\right]^{d_{+}^{[3]}}
\left(1-J_{+}^{[3]}
	\frac{\alpha_s\left(\mu_c\right)-\alpha_s\left(\mu\right)}{4\pi}\right),
\label{EvolS2}
\end{eqnarray}
where $d_{+}^{[3]}$ and $J_{+}^{[3]}$ are the RG quantities for three
active flavours which can be determined from the results in App.~\ref{app:anomdimgenS2}.

The results at the low scale allow then to determine the expression of the short-distance QCD corrections for the three different boxes in the SM case.

\section{SM case at NLO with the method of regions \label{app:SMVy}}

We want to apply the method of regions as explained in Sec.~\ref{sec:SMLO} in order
to determine the short-distance corrections $\bar\eta$ at NLO. We start with
 the behaviour of the one-loop integrals.
In the SM these integrals are given by the following functions
\begin{eqnarray}
S^{LL}(x_t)&=&x_t\biggl(\frac{1}{4}+\frac{9}{4}\frac{1}{1-x_t}
-\frac{3}{2}\frac{1}{(1-x_t)^2}\biggr) -\frac{3}{2}\biggl[\frac{x_t}{(1-x_t)}
 \biggr]^3 \log x_t \,,
\nonumber \\
S^{LL}(x_c)&=& x_c+ {\cal O}(x_c^2)\,,
\label{eq:inami}\\
S^{LL}(x_c,x_t)&=& -x_c \log x_c+ x_c F(x_t)+ {\cal O}(x_c^2 \log x_c)  \, , \nonumber \\
&& \quad F(x_t)=\frac{x_t^2 - 8 x_t +4}{4(1-x_t)^2} \log x_t +\frac{3}{4} \frac{x_t}{(x_t-1)}\,.
\nonumber
\end{eqnarray}
Clearly the leading behaviour of the one-loop
integral for $\bar\eta_{tt}$ is ${\cal O}(1)$, for  $\bar\eta_{cc}$ ${\cal O}( x_c)$ and
for  $\bar\eta_{ct}$ ${\cal O}( x_c \log(x_c) )$. Following the method of regions,
the remaining integration over the momentum $k$ leads to  $m_t^2$ in the
first case, and $m_c^2$ in the second, as already discussed in Sec.~\ref{sec:SMLO}. For $ct$
 one has to introduce the function $R(\gamma,m_1,m_2)$ defined
in Eq.~\eqref{eq:Rlog} at
LO. At NLO the quantity $x_c F(x_t)$ contributes to $\bar\eta_{ct}$, so that
the result of the integration is $m_c^2$, similarly to  $\bar\eta_{cc}$.

One has then to determine  the anomalous dimensions of the operators which appear in the calculation of the box diagrams. These anomalous dimensions are well known up to NLO, for instance see Ref.~\cite{Buras:2000if}. We have to combine the contributions of the $|\Delta S|=1$ operators (hence the presence of $d_r$ and $d_l$) between $\mu^2_W$ and $k^2$ with the term from the $|\Delta S|=2$ operator between  $k^2$ and $\mu_{h}$ (leading to $d_V$). Setting $k^2=m_t^2$, we obtain 
the following formula for the scale-independent correction $\eta_{tt}$

\begin{equation}
\eta_{tt} = \frac{(\eta_{tt}^{(WW)} + \eta_{tt}^{(GG)} x_t^2 / 4) I_2 (x_t,x_t,1) - \eta_{tt}^{(WG)} 2 x_t^2 I_1 (x_t,x_t,1)}{(1 + x_t^2 / 4) I_2 (x_t,x_t,1) - 2 x_t^2 I_1 (x_t,x_t,1)}
\label{eq:etattMR}
\end{equation}
where $ I_{1,2} $ are the Inami-Lim functions of Eq.~\eqref{eq:InamiLimfunctionsGeneral}. The superscripts $ (WW), (WG)$, and $ (GG) $ indicate respectively the contributions from a box containing two $ W $ bosons, one $ W $ boson and one Goldstone $ G $, and two Goldstones $ G $ in the 't Hooft-Feynman gauge (the last two come at higher order on $ m_c / M_W $ in the $ ct $ and $ cc $ cases). The corresponding short-distance corrections are given by

\begin{eqnarray}
\eta_{tt}^{(WW)}&=&\sum_{r,l=\pm}
 \left({\alpha_s(m_c)}\right)^{d_V^{(3)}}\left(\frac{\alpha_s(m_t)}{\alpha_s(\mu_5)}\right)^{d_V^{(5)}}\left(\frac{\alpha_s(\mu_5)}{\alpha_s(m_c)}\right)^{d_V^{(4)}}
    \left(\frac{\alpha_s(\mu_W)}{\alpha_s(m_t)}\right)^{d_r^{(5)}+d_l^{(5)}}a^{(WW)}_{rl}
\\\nonumber
&&    \left(1-\frac{\alpha_s(m_c)}{4\pi} (J_V^{(3)}-J_V^{(4)})-\frac{\alpha_s(\mu_5)}{4\pi} (J_V^{(4)}-J_V^{(5)})+\frac{\alpha_s
(m_t)}{4\pi}(J_r^{(5)}+J_l^{(5)}-J_V^{(5)})\right. 
\\\nonumber
&& \left. \qquad\qquad-\frac{\alpha_s(\mu_W)}{4\pi}(J_l^{(5)}+J_r^{(5)}-B_r-B_l)\right)\,,
\end{eqnarray}

\begin{eqnarray}
\eta_{tt}^{(WG)}&=&\sum^{2}_{i,j,k,p,q=1} \sum_{r=\pm}
 \left({\alpha_s(m_c)}\right)^{d_V^{(3)}}\left(\frac{\alpha_s(m_t)}{\alpha_s(\mu_5)}\right)^{d_V^{(5)}}\left(\frac{\alpha_s(\mu_5)}{\alpha_s(m_c)}\right)^{d_V^{(4)}}
\\\nonumber
&& \left(\frac{\alpha_s(\mu_W)}{\alpha_s(m_t)}\right)^{d_r^{(5)}+2 d_m^{(5)} + d^{(5)}_j} (\hat{a}^{(WG)}_{r})_i \delta_{ip} \hat{V}_{pj} \hat{V}^{-1}_{jq} \delta_{qk} (C_0)_k
\\\nonumber
&& \qquad \left(1-\frac{\alpha_s(m_c)}{4\pi} (J_V^{(3)}-J_V^{(4)})-\frac{\alpha_s(\mu_5)}{4\pi} (J_V^{(4)}-J_V^{(5)}) \right.
\\\nonumber
&& \qquad\qquad \left. +\frac{\alpha_s
(m_t)}{4\pi}(J_r^{(5)}+2 J_m^{(5)}-J_V^{(5)} + ( \hat{J}^{(5)} )_{ip})\right. 
\\\nonumber
&& \left. \qquad\qquad\qquad -\frac{\alpha_s(\mu_W)}{4\pi}(J_r^{(5)}-B_r + 2 J^{(5)}_m + (\hat{J}^{(5)})_{qk})\right)\,,
\end{eqnarray}

\begin{eqnarray}
\eta_{tt}^{(GG)}&=&\sum^{2}_{i,j,k,p,q=1} \;\; \sum^{2}_{i',j',k',p',q'=1} 
 \left({\alpha_s(m_c)}\right)^{d_V^{(3)}}\left(\frac{\alpha_s(m_t)}{\alpha_s(\mu_5)}\right)^{d_V^{(5)}}\left(\frac{\alpha_s(\mu_5)}{\alpha_s(m_c)}\right)^{d_V^{(4)}}
\\\nonumber
&& \left(\frac{\alpha_s(\mu_W)}{\alpha_s(m_t)}\right)^{4 d_m^{(5)} + d^{(5)}_j + d^{(5)}_{j'}} (\hat{a}^{(GG)})_{i i'} \delta_{ip} \hat{V}_{pj} \hat{V}^{-1}_{jq} \delta_{qk} (C_0)_k \delta_{i' p'} \hat{V}_{p' j'} \hat{V}^{-1}_{j' q'} \delta_{q' k'} (C_0)_{k'}
\\\nonumber
&& \qquad \left(1-\frac{\alpha_s(m_c)}{4\pi} (J_V^{(3)}-J_V^{(4)})-\frac{\alpha_s(\mu_5)}{4\pi} (J_V^{(4)}-J_V^{(5)}) \right.
\\\nonumber
&& \qquad\qquad \left. +\frac{\alpha_s
(m_t)}{4\pi}(4 J_m^{(5)}-J_V^{(5)} + ( \hat{J}^{(5)} )_{ip} + ( \hat{J}^{(5)} )_{i' p'})\right. 
\\\nonumber
&& \left. \qquad\qquad\qquad -\frac{\alpha_s(\mu_W)}{4\pi}(4 J^{(5)}_m + (\hat{J}^{(5)})_{qk} + (\hat{J}^{(5)})_{q' k'})\right)\,,
\end{eqnarray}
where

\begin{equation}
 a^{(WW)}_{rl} = t_{rl} \, , \quad \hat{a}^{(WG)}_{r} = \begin{pmatrix} -(1 + N r) \\ -(1 + r) \\ \end{pmatrix} \, , 
\quad  \hat{a}^{(GG)} = 4 \begin{pmatrix} N & 1 \\ 1 & 1 \\ \end{pmatrix} \, , 
\quad C_0 = \begin{pmatrix} 0 \\ -1/2 \\ \end{pmatrix} \, , 
\end{equation}
\noindent
with $t_{rl}$ defined in Eq.~\eqref{eq:resumv}.
Above, the $J$'s arise from the RGE evolution as described in App.~\ref{app:anomdimgen}, where the definition of all the quantities appearing here are given and the exponents denote the number of active flavours. The thresholds are explicitly shown: 
 $\mu_5$ is the threshold for the integration of the $b$ quark, and $\mu_4=m_c$ for the $c$ quark. Since the formulae become rather large once including
the thresholds explicitely we will not give their
expressions  in the following, but it is rather straightforward to implement them and their effect is included in our final results.

It is interesting to compare the MR result, Eq.~\eqref{eq:etattMR}, with the one obtained at NLO in EFT~\cite{Buras:1990fn}. There, contrary to what is done in the Method of Regions where one keeps explicitely the top quark degree of freedom, one ignores the difference between the two scales $\mu_t \equiv m_t$ and $\mu_W$, and integrates at the same time both the top and the $ W $. The EFT approach leads to a much simpler expression since in 
this case only the $ \vert \Delta S \vert = 2 $ operator survives
\begin{eqnarray}
\eta_{tt}^{EFT,NLO}&=&
 \left({\alpha_s(m_c)}\right)^{d_V^{(3)}}\left(\frac{\alpha_s(\mu_W)}{\alpha_s(\mu_5)}\right)^{d_V^{(5)}}\left(\frac{\alpha_s(\mu_5)}{\alpha_s(m_c)}\right)^{d_V^{(4)}}
\\\nonumber
&&  \!\!\!\!\! \!\!\!\!\!  \left(1-\frac{\alpha_s(m_c)}{4\pi} (J_V^{(3)}-J_V^{(4)})-\frac{\alpha_s(\mu_5)}{4\pi} (J_V^{(4)}-J_V^{(5)})-\frac{\alpha_s(\mu_W)}{4\pi}(J_V^{(5)}-Y(x_t)-R)\right)\,.
\end{eqnarray}
The last two terms in this equation stem from  the NLO matching on the full theory at the high scale $\mu_W = \mathcal{O} (m_t, M_W)$  ($5.8<Y(x_t)+R<13.4$ for  $1 \leq x_t\leq 4.6$). We refer the reader to ~\cite{Buras:1990fn}
for more details. At LO, taking $m_t= \mu_W$ in the MR expressions above, it is
easy to show that
$\eta_{tt}^{(WW)}=\eta_{tt}^{(WG)}=\eta_{tt}^{(GG)}=\eta_{tt}^{EFT,NLO}$.  
At NLO, one would have to replace the contributions from $B_{l,r}$, which come
from the matching onto two 
$\vert \Delta S \vert =1$ local operators, by $Y(x_t)+R$. Clearly the difference between the two approaches involves  the ratio   $\alpha_s(m_t)/\alpha_s(\mu_W)$
and terms of  ${\cal O}(\alpha_s(\mu_W)/(4\pi))$, which are effects of a few 
percent, as detailed in Sec.~\ref{sec:MRNLO}.

For $\eta_{ct}$, we have two different types of contributions: a large logarithm $\log x_c$ and a constant term. Since we want to resum contributions of the form $\alpha_s\log x_c$, the first can be formally counted as coming one order earlier than the latter in the power counting. We can take this into account by treating differently the resummation of the large logarithm and the constant term
\begin{eqnarray}
\eta_{ct}&=&\frac{1}{-\log x_c + F(x_t)}\alpha_s(m_c)^{d_V} \sum_{r,l=\pm } a_{rl}
\left(\frac{\alpha_s(\mu_W)}{\alpha_s(m_c)}\right)^{d_l+d_r} 
 \\\nonumber
&&\times 
\biggl(-\log x_c R^{NLO}_{\log} \Bigg[-d_l-d_r+d_V+2 d_m,
 u_{rl}, j_{rl} ; m_c,\mu_W \Bigg] 
+ F(x_t)  \biggr)\,,\nonumber
\end{eqnarray}
with 
\begin{eqnarray}
a_{rl}&=&[1+r+l+3rl]/4
\nonumber\\
u_{rl}&=&1+ 2\frac{\alpha_s(m_c)}{4\pi} J_m-\frac{\alpha_s(\mu_W)}{4\pi}( J_l +J_r-B_l-B_r), \nonumber\\
j_{rl} &=& J_l+J_r-J_V-2J_m\,,
\end{eqnarray}
and
\begin{equation}
R^{NLO}_{\log}(\gamma,U,J;m_1,m_2)=\log^{-1}\frac{m_2^2}{m_1^2}  \left(\frac{\alpha_s(m_1)}{\alpha_s(\mu)}\right)^{-\gamma} \int_{m_1^2}^{m_2^2} \frac{dk^2}{k^2}  \left(\frac{\alpha_s(k)}{\alpha_s(\mu)}\right)^{\gamma} \left[U+\frac{\alpha_s(k)}{4\pi}J\right]\,,
\label{eq:rnlo}
\end{equation}
where $U$ does not depend on $k$, yielding for $ \gamma \neq 0, 1 $
\begin{eqnarray}
&& R^{NLO}_{\log}(\gamma,U,J;m_1,m_2)=\frac{1}{\log ( m_2^2/m_1^2)}\frac{4\pi}{\beta_0 \alpha_s(m_1)}\\ \nonumber
&&\qquad\qquad\times\biggl[ \frac{1}{1-\gamma}
\biggl\{ \biggl( 
\frac{\alpha_s(m_2)}{\alpha_s(m_1)} 
\biggr)^{\gamma-1} -1
 \biggr\}U +\frac{\alpha_s(m_1)}{4\pi}\frac{1}{\gamma}\left[\frac{\beta_1}{\beta_0}U-J\right]\biggl\{\biggl (\frac{\alpha_s(m_2)}{\alpha_s(m_1)}\biggr)^{\gamma} -1\biggr\} 
\biggr]\,.
\end{eqnarray}
An analytic comparison with the EFT result in this case would be much more difficult due to the complexity of the expressions. We refer to Sec.~\ref{sec:MRNLO} for a numerical comparison.
The previous cases, where a single mass scale $m_1$ dominates the integral, can be described using the averaging function
\begin{equation}\label{eq:rnlo1}
R^{NLO}_1(\gamma,U,J;m_1,m_2)=\left[U+\frac{\alpha_s(m_1)}{4\pi}J\right] \, .
\end{equation}

\section{Operators and anomalous dimensions}
\label{app:anomdimgen}

\subsection{$|\Delta S|=1$ operators}\label{app:anomdimgenS1}

We have the $|\Delta S|=1$ vector operators for the SM case~\cite{Herrlich:1996vf,Buras:2000if}
\begin{eqnarray}
O_1^{VLL}&=&(\bar{d}^\alpha \gamma_\mu P_L s^\beta)(\bar{V}^\beta\gamma^\mu P_L U^\alpha)\,,
\qquad O_2^{VLL}=(\bar{d}\gamma_\mu P_L s)(\bar{V}\gamma^\mu P_L U)\,,\\
O_1^{VLR}&=&(\bar{d}\gamma_\mu P_L s)(\bar{V}\gamma^\mu P_R U)\,,
\qquad\qquad O_2^{VLR}=(\bar{d}^\alpha\gamma_\mu P_L s^\beta)(\bar{V}^\beta\gamma^\mu P_R U^\alpha)\,,
\end{eqnarray}
where $U$ and $V$ can be any up-type fermions.
The anomalous dimensions for the vector-vector operators is simpler for~\cite{Buras:2000if}
\begin{equation}
O_\pm=\frac{O_1\pm O_2}{2}\,,
\end{equation}
which are the following
\begin{eqnarray}
&&\gamma_\pm^{(0)} = \pm 6\frac{N\mp 1}{N}\,,\qquad  \gamma_\pm^{(1)} = \frac{N \mp 1}{2 N} \biggl(- 21 \pm 
\frac{57}{N} \mp 19\frac {N}{3} \pm \frac{4}{3} f \biggr)\,,
\nonumber
\\
&&\gamma_m^{(0)}=6 C_F\,, \qquad\qquad \qquad \gamma_m^{(1)}= C_F \biggl( 3 \,  C_F +\frac{97}{3}N - \frac{10}{3} f \biggr)\,,
\end{eqnarray}
where the second line corresponds to the anomalous dimensions for masses with \linebreak $C_F=(N^2-1)/2N$, and for $N=3$, $\gamma_+^{(0)}=4, \gamma_-^{(0)}=-8,\gamma_m^{(0)}=8$.

We introduce the correction of the anomalous dimensions
\begin{eqnarray}
J_\pm &=&\frac{d_\pm \beta_1}{\beta_0} -\frac{\gamma^{(1)}_\pm}{2 \beta_0}\,, \qquad d_\pm=\frac{\gamma^{(0)}_\pm}{2 \beta_0}\,,\\
J_m &=& \frac{d_m \beta_1}{\beta_0} -\frac{\gamma^{(1)}_m}{2 \beta_0}\,, \qquad d_m=\frac{\gamma^{(0)}_m}{2 \beta_0}\,,
\end{eqnarray}
and the value of the Wilson coefficients at the high scale
$C_{\pm}(\mu_W)$
defined in 
Ref.~\cite{Buchalla:1995vs}
\begin{equation}\label{eq:cpmuw}
C_{\pm}(\mu_W)=1 + \frac{\alpha_s (\mu_W)}{4 \pi} \biggl ( \log \frac{\mu_W}{M_W} \gamma_\pm^{(0)} + B_{\pm} \biggr) +{\cal O}(\alpha_s^2)\,,
\end{equation}
with 
\begin{equation}
B_{\pm}= -\frac{11}{2 N} \pm \frac{11}{2}\,,
\end{equation}
leading to the evolution
\begin{eqnarray}
C^{NLO}_\pm(\mu;\mu_0)=\left(1 + \frac{\alpha_s(\mu)}{4 \pi} J_\pm \right) \biggl( \frac{\alpha_s(\mu_0)}{\alpha_s(\mu)}\biggr)^{d_\pm}\left(1 - \frac{\alpha_s(\mu_0)}{4 \pi}[J_\pm-B_\pm] \right)\,, \\
C^{NLO}_m(\mu;\mu_0)=\left(1 + \frac{\alpha_s(\mu)}{4 \pi} J_m \right) \biggl( \frac{\alpha_s(\mu_0)}{\alpha_s(\mu)}\biggr)^{d_m} \left(1 -\frac{\alpha_s(\mu_0)}{4 \pi} J_m \right) \,.
\end{eqnarray}
We have 
\begin{equation}
d_m=4/\beta_0 \qquad d_+=2/\beta_0 \qquad d_-=-4/\beta_0\,.
\end{equation}
The same equations can be written for $O_i^{VRR}$ which will be useful for the discussion of the LRM, with identical results for the anomalous dimensions.

One may also consider the running of the $|\Delta S|=1$ local operators VLR. In the basis $O_1^{VLR},O_2^{VLR}$, the anomalous dimensions are
{\small\begin{eqnarray}
\hat\gamma^{(0)}_{VLR}&=&\left[\begin{array}{cc} 6/N & -6\\ 0 & -6N+6/N\end{array}\right]\,,\\
\hat\gamma^{(1)}_{VLR}&=&\left[\begin{array}{cc} \frac{137}{6}+\frac{15}{2N^2}-\frac{22}{3N} f & - \frac{100N}{3}+\frac{3}{N}+ \frac{22}{3} f\\\nonumber
-\frac{71}{2} N -\frac{18}{N}+4f & - \frac{203}{6} N^2+\frac{479}{6}+\frac{15}{2N^2}+\frac{10}{3} Nf -\frac{22}{3N} f  \end{array}\right]\,.
\end{eqnarray}}
Introducing
\begin{eqnarray}
\hat{V}&=&\left(\begin{array}{cc} 3/2 & 0 \\ -1/2 & -1/2 \end{array}\right)\,,\\
\hat\gamma^{(0)}_D&=& \hat{V}^{-1} \hat\gamma^{(0)T}_{VLR} \hat{V}
   =\left(\begin{array}{cc} 6/N & 0\\ 0 & -6N+6/N\end{array}\right)\,,\qquad \gamma^{(0)}_{1}=2\,,\qquad \gamma^{(0)}_{2}=-16\,,\\
\hat{G}&=& \hat{V}^{-1} \hat\gamma^{(1)T}_{VLR} \hat{V}\,,\\
\hat{H}_{ij}&=& \delta_{ij} \gamma^{(0)}_i \frac{\beta_1}{2\beta_0^2} - \frac{\hat{G}_{ij}}{2\beta_0 + \gamma^{(0)}_i - \gamma^{(0)}_j} \qquad\qquad ({2\beta_0 + \gamma^{(0)}_i - \gamma^{(0)}_j} \neq 0)\,,\\
\hat{J} &=& \hat{V} \hat{H} \hat{V}^{-1}\,,
\end{eqnarray}
one can write down the evolution
\begin{eqnarray}
\vec{C}^{LR}(\mu;\mu_0)&=&
   \left(1+\frac{\alpha_s(\mu)}{4\pi} \hat{J}\right) \hat{V} D(\mu;\mu_0) \hat{V}^{-1} 
   \left(1-\frac{\alpha_s(\mu_0)}{4\pi} \hat{J}\right)
   \vec{C}^{LR}(\mu_0)\,,\\
 D(\mu;\mu_0)&=&\left(\begin{array}{cc} (\alpha_s(\mu_0)/\alpha_s(\mu))^{d_1} & 0\\
 0&  (\alpha_s(\mu_0)/\alpha_s(\mu))^{d_2} \end{array}\right)\,,
\end{eqnarray}
with $d_i=\gamma^{(0)}_i/(2\beta_0)$.

\subsection{$|\Delta S|=2$ operators}\label{app:anomdimgenS2}

For $|\Delta S|=2$ operators, we recall the anomalous dimensions associated with the operator $Q_{V}$
\begin{equation}
Q_{V}=(\bar{s}^\alpha \gamma_\mu P_L d^\alpha)(\bar{s}^\beta \gamma^\mu P_L d^\beta)\,,
\end{equation}
with
\begin{eqnarray}
\gamma^{(0)}_V&=&6-6/N\,, \\
\gamma^{(1)}_V&=&-19/6N-22/3+39/N-57/(2N^2)+2/3f-2/(3N)f\,,\\
J_V &=&\frac{d_V \beta_1}{\beta_0} -\frac{\gamma^{(1)}_V}{2 \beta_0}\,,
\qquad\qquad\qquad d_V=\frac{\gamma^{(0)}_V}{2\beta_0}\,,
\end{eqnarray}
and we can write down a similar evolution for the $|\Delta S|=2$ local operators $Q_1^{LR},Q_2^{LR}$ 
\begin{eqnarray}
Q_1^{LR}&=&(\bar{s}^\alpha \gamma_\mu P_L d^\alpha)(\bar{s}^\beta \gamma_\mu P_R d^\beta)\,,
\qquad
Q_2^{LR}=(\bar{s}^\alpha P_L d^\alpha)(\bar{s}^\beta P_R d^\beta)\,,
\end{eqnarray}
with the anomalous dimensions
{\small
\begin{eqnarray}\label{eq:anoQLR}
\hat\gamma^{(0)}_{LR}&=&\left[\begin{array}{cc} 6/N & 12\\ 0 & -6N+6/N\end{array}\right]\,,\\
\hat\gamma^{(1)}_{LR}&=&\left[\begin{array}{cc}  \frac{137}{6}+\frac{15}{2N^2}-\frac{22}{3N} f & \frac{200N}{3}-\frac{6}{N}-\frac{44}{3} f\\\nonumber
\frac{71}{4} N + \frac{9}{N}-2f & - \frac{203}{6} N^2+ \frac{479}{6}+\frac{15}{2N^2}+\frac{10}{3} Nf -\frac{22}{3N} f  \end{array}\right]\,.
\end{eqnarray}
}
Introducing
\begin{eqnarray}
\hat{W}&=&\left(\begin{array}{cc} 3/2 & 0 \\ 1 & 1 \end{array}\right)\,,\\
\hat\gamma^{(0)}_D&=& \hat{W}^{-1} \hat\gamma^{(0)T}_{LR}\hat{W}
   =\left(\begin{array}{cc} 6/N & 0\\ 0 & -6N+6/N\end{array}\right)\qquad \gamma^{(0)}_{1}=2\qquad \gamma^{(0)}_{2}=-16\,,\\
\hat{G}&=& \hat{W}^{-1} \hat\gamma^{(1)T}_{LR}\hat{W}\,,\\
\hat{H}_{ij}&=& \delta_{ij} \gamma^{(0)}_i \frac{\beta_1}{2\beta_0^2} - \frac{\hat{G}_{ij}}{2\beta_0 + \gamma^{(0)}_i - \gamma^{(0)}_j} \qquad\qquad ({2\beta_0 + \gamma^{(0)}_i - \gamma^{(0)}_j} \neq 0)\,,\\
\hat{K} &=& \hat{W} \hat{H} \hat{W}^{-1}\,,
\end{eqnarray}
one can write down the evolution 
\begin{eqnarray}
\vec{C}^{LR}(\mu;\mu_0)&=&
   \left(1+\frac{\alpha_s(\mu)}{4\pi} \hat{K}\right) \hat{W} D(\mu;\mu_0) \hat{W}^{-1} 
   \left(1-\frac{\alpha_s(\mu_0)}{4\pi} \hat{K}\right)
   \vec{C}^{LR}(\mu_0)\,,\\
 D(\mu;\mu_0)&=&\left(\begin{array}{cc} (\alpha_s(\mu_0)/\alpha_s(\mu))^{d_1} & 0\\
 0&  (\alpha_s(\mu_0)/\alpha_s(\mu))^{d_2} \end{array}\right)\,,
\end{eqnarray}
with $d_i=\gamma^{(0)}_i/(2\beta_0)$.
The associated LO anomalous dimensions are 
\begin{equation}
 \gamma^{(0)}_{1}=2\,,\qquad \gamma^{(0)}_{2}=-16\,,\\
\end{equation}
and we have 
\begin{equation}
d_1=1/\beta_0\,, \qquad d_2=-8/\beta_0\,,\qquad d_V=2/\beta_0\,.
\end{equation}

\section{LR case at NLO with the method of regions}\label{lrregion}

\subsection{Contributions with $\log\beta$}

{Following Ref.~\cite{Ecker:1985vv},} if we consider the box with the Goldstone boson associated to $W$ together with $W'$,
the masses stem from the Goldstone boson coupling (evaluated at the scale $\mu_W$), whereas the largest contribution to $I_2$ comes  from the range between $\mu_W$ and $\mu_R$.
We obtain
\begin{eqnarray}\nonumber
&&\xi^{(W'2)}_{a,UV}[R]=\sum_{r=\pm,i,j=1,2}
 \left(\frac{\alpha_s(\mu_W)}{\alpha_s(\mu_{h})}\right)^{-d_r+d_i+2d_m} 
  \left(\frac{\alpha_s(m_U)}{\alpha_s(\mu_{h})}\right)^{-d_m} 
    \left(\frac{\alpha_s(m_V)}{\alpha_s(\mu_{h})}\right)^{-d_m} 
    \left(\frac{\alpha_s(\mu_R)}{\alpha_s(\mu_{h})}\right)^{d_r}\\\nonumber
&& \quad\times 
\left[\left(1+\frac{\alpha_s(\mu_{h})}{4\pi} \hat{K}\right) \hat{W}\right]_{ai}\\\nonumber
&& \quad\times 
 R^{NLO}\Bigg(-d_r+d_i-d_j,\\\nonumber
&&  \qquad\qquad\qquad
  \left[\hat{W}^{-1}\hat{a}^{(W'2)}_r \hat{V}\right]_{ij} [\hat{V}^{-1}\vec{C}_0]_j  \\\nonumber
&&  \qquad\qquad\qquad\qquad\qquad
\times  \left(1 -\frac{\alpha_s(\mu_R)}{4\pi}[J_r-B_r]-\frac{\alpha_s(\mu_W)}{4\pi}2J_m
   +\frac{\alpha_s(m_U)+\alpha_s(m_V)}{4\pi}J_m\right)\\\nonumber
&&   \qquad\qquad\qquad\qquad  -\frac{\alpha_s(\mu_W)}{4\pi}
   \left[\hat{W}^{-1}\hat{a}^{(W'2)}_r \hat{V}\right]_{ij} [\hat{V}^{-1}\hat{J}\vec{C}_0]_j 
   , \\\nonumber
&&   \qquad\qquad\qquad
   \left[\hat{W}^{-1}[\hat{a}^{(W'2)}_r \hat{J}-\hat{K}\hat{a}^{(W'2)}_r]\hat{V}\right]_{ij} [\hat{V}^{-1}\vec{C}_0]_j 
   + \left[\hat{W}^{-1}\hat{a}^{(W'2)}_r \hat{V}\right]_{ij} [\hat{V}^{-1}\vec{C}_0]_j  J_r,
      \\
&&  \qquad\qquad\qquad\qquad
 \mu_W, \mu_R\Bigg)\,,
\end{eqnarray}
with the initial conditions for the evolution of the operators $O_{1,2}^{VLR}$ and the
coefficients for the matching from the two-point function of $O_\pm^{VRR}$  and $O_{1,2}^{VLR}$  to the local operators $Q_{1,2}^{LR}$ at $\mu=k^2$.
\begin{equation}
\vec{C}_0=\left(\begin{array}{c} 0\\ -1/2\end{array}\right)\,,
 \qquad
C_a^{LR}  \leftrightarrow \sum_{r,i} (\hat{a}^{(W'2)}_{r})_{ai} C_i^{VLR} C_r^{VRR}\,,
\qquad
\hat{a}^{(W'2)}_r=\left(\begin{array}{cc} (3r+1)/2 & r/2\\ 0 & -1\end{array}\right) \,.
\end{equation}

If we consider the box with $W$ and a charged Higgs boson $H$,
the masses stem from the Higgs couplings (to be evaluated at a high scale $\mu_H$), whereas the largest contribution to $I_2$ comes  from the range between $\mu_W$ and $M_H$. We obtain
\begin{eqnarray}\nonumber
\xi^{(H2)}_{a,UV}[R]\!\!\!&=&\!\!\!\!\!\!\sum_{l=\pm,i,j=1,2}
 \left(\frac{\alpha_s(\mu_W)}{\alpha_s(\mu_{h})}\right)^{d_i-d_j} 
  \left(\frac{\alpha_s(m_U)}{\alpha_s(\mu_{h})}\right)^{-d_m} 
    \left(\frac{\alpha_s(m_V)}{\alpha_s(\mu_{h})}\right)^{-d_m} 
    \left(\frac{\alpha_s(\mu_H)}{\alpha_s(\mu_{h})}\right)^{d_j+2d_m}\\\nonumber
&& \quad\times 
\left[\left(1+\frac{\alpha_s(\mu_{h})}{4\pi} \hat{K}\right) \hat{W}\right]_{ai}\\\nonumber
&& \quad\times 
 R^{NLO}\Bigg(-d_l+d_i-d_j,\\\nonumber
&&  \qquad\qquad\qquad
  \left[\hat{W}^{-1}\hat{a}^{(H2)}_l \hat{V}\right]_{ij} [\hat{V}^{-1}\vec{C}_0]_j  \\\nonumber
&&  \qquad\qquad\qquad
\times  \left(1 -\frac{\alpha_s(\mu_W)}{4\pi}[J_l-B_l]-\frac{\alpha_s(\mu_H)}{4\pi}2J_m
   +\frac{\alpha_s(m_U)+\alpha_s(m_V)}{4\pi}J_m\right), \\\nonumber
&&   \qquad\qquad\qquad
   \left[\hat{W}^{-1}[\hat{a}^{(H2)}_l \hat{J}-\hat{K}\hat{a}^{(H2)}_l ]\hat{V}\right]_{ij} [\hat{V}^{-1}\vec{C}_0]_j 
   - \left[\hat{W}^{-1}\hat{a}^{(H2)}_l \hat{V}\right]_{ij} [\hat{V}^{-1}\hat{J}\vec{C}_0]_j  \\
&&  \qquad\qquad\qquad\qquad
   + \left[\hat{W}^{-1}\hat{a}^{(H2)}_l \hat{V}\right]_{ij} [\hat{V}^{-1}\vec{C}_0]_j  J_l, \mu_W, \mu_H\Bigg)\,,
\end{eqnarray}
with the same initial conditions for the evolution of the operators $Q_{1,2}^{VLR}$  and the
coefficients for the matching from the two-point function of $O_\pm^{VLL}$ and $O_{1,2}^{VRL}$  to the local operators $Q_{1,2}^{LR}$ at $\mu=k^2$.
\begin{equation}
C_a^{LR}  \leftrightarrow \sum_{l,j} (\hat{a}^{(H2)}_{l})_{ai} C_j^{VRL} C_l^{VLL}\,,
\qquad
\hat{a}^{(H2)}_l=\hat{a}^{(W'2)}_{r=l}\,.
\end{equation}
One can check that the expressions from Ref.~\cite{Ecker:1985vv} are recovered at leading order.

If we consider $\log\beta$ as small (``small $\log\beta$ approach''), we see that the diagrams are dominated by the region $k^2=\mathcal{O}(m_t^2,\mu_W^2)$ in all cases: this is obvious for $tt$ and $ct$ boxes, whereas the $cc$ box receives only suppressed contributions from the region $k^2=\mathcal{O}(m_c^2)$. We obtain thus expressions involving the averaging weight for constant terms $R^{NLO}_1$
\begin{equation}
\bar\eta^{(W'2)}_{a,UV}=\xi^{(W'2)}_{a,UV}[R^{NLO}_1]\,,
\qquad\qquad \bar\eta^{(H2)}_{a,UV}=\xi^{(H2)}_{a,UV}[R^{NLO}_1]\,,
\end{equation}
where we have identified the two scales for the integration $\mu_W=\mu_R$ to a common average value (this is similar to the treatment of the region between $m_t$ and $M_W$ in the SM case).

In the case of a large $\log\beta$ (``large $\log\beta$ approach''), we want to perform the resummation of the large $\log\beta$ with $R^{NLO}_{\log}$ and consider the rest of the contribution as dominated by the region $k^2=\mathcal{O}(m_t^2,\mu_W^2)$. In the case of $(W'2)$ we obtain
\begin{eqnarray}
\bar\eta^{(W'2)}_{a,UV} &=& \Bigg[F^{(W'2)}_{UV} \\
& \times & \sum_{r=\pm, \, i,j=1,2}
\left(\frac{\alpha_s(\mu_W)}{\alpha_s(\mu_{h})}\right)^{-d_r+d_i+2d_m}
  \left(\frac{\alpha_s(m_U)}{\alpha_s(\mu_{h})}\right)^{-d_m} 
    \left(\frac{\alpha_s(m_V)}{\alpha_s(\mu_{h})}\right)^{-d_m} 
    \left(\frac{\alpha_s(\mu_R)}{\alpha_s(\mu_{h})}\right)^{d_r} \nonumber\\
\qquad & \times & \hat{W}_{ai} \left[\hat{W}^{-1}\hat{a}^{(W'2)}_r \hat{V}\right]_{ij} [\hat{V}^{-1}\vec{C}_0]_j
+\log(\beta) \times \xi^{(W'2)}_{a,UV}[R^{NLO}_{\log}]
     \Bigg] \frac{1}{\log(\beta) + F^{(W'2)}_{UV}}\qquad\qquad \nonumber
\end{eqnarray}
with the contributions from the constant term
\begin{equation}
F^{(W'2)}_{tt} =  \frac{x_t^2 - 2 x_t}{(x_t - 1)^2} \log (x_t) + \frac{x_t}{x_t - 1} \,,\quad
F^{(W'2)}_{ct}=  \frac{x_t}{x_t - 1} \log (x_t) \,,\quad
F^{(W'2)}_{cc} = 0 \,,
\end{equation}
and similarly for $(H2)$
\begin{eqnarray}
\bar\eta^{(H2)}_{a,UV} &=& \Bigg[  F^{(H2)}_{UV} \\
& \times & \sum_{l=\pm, \, i,j=1,2}
\left(\frac{\alpha_s(\mu_W)}{\alpha_s(\mu_{h})}\right)^{d_i-d_j}   \left(\frac{\alpha_s(m_U)}{\alpha_s(\mu_{h})}\right)^{-d_m} 
    \left(\frac{\alpha_s(m_V)}{\alpha_s(\mu_{h})}\right)^{-d_m} 
    \left(\frac{\alpha_s(\mu_H)}{\alpha_s(\mu_{h})}\right)^{d_j+2d_m} \nonumber\\
\qquad &\times &    \hat{W}_{ai} \left[\hat{W}^{-1}\hat{a}^{(H2)}_l \hat{V}\right]_{ij} [\hat{V}^{-1}\vec{C}_0]_j
   +\log(\beta \omega) \times  \xi^{(H2)}_{a,UV}[R^{NLO}_{\log}]   \Bigg] \frac{1}{\log(\beta \omega) + F^{(H2)}_{UV}}\qquad\qquad \nonumber
\end{eqnarray}
with the contributions from the constant term
\begin{equation}
F^{(H2)}_{tt} =  x_t \frac{x_t + (x_t - 2) \log (x_t) - 1}{(x_t - 1)^2} \,,\qquad
F^{(H2)}_{ct}  =  \frac{x_t}{x_t - 1} \log (x_t)\,,\qquad
F^{(H2)}_{cc}  =  0 \,.
\end{equation}

\subsection{Contributions without $\log\beta$}

If we consider the box with the Goldstone associated with $W$ and a charged Higgs boson $H$, 
the masses stem from the Higgs couplings, the Goldstone boson couplings and the propagator, whereas the largest contribution to $I_1$ comes  from the range between $m_V$ and $\mu_W$. We obtain
\begin{eqnarray}\nonumber
\!\!\!\!\!\!&&\bar\eta^{(H1)}_{a,UV}\!\!\!=\!\!\!\!\!\!\!\!\!\sum_{b,i,j,j',k,k'=1,2}
  \left(\frac{\alpha_s(m_U)}{\alpha_s(\mu_{h})}\right)^{-3d_m} \!\!\!
    \left(\frac{\alpha_s(m_V)}{\alpha_s(\mu_{h})}\right)^{d_i-d_k-d_{k'}-d_m} \!\!\!
     \left(\frac{\alpha_s(\mu_W)}{\alpha_s(\mu_{h})}\right)^{d_k+2dm} \!\!\!
    \left(\frac{\alpha_s(\mu_H)}{\alpha_s(\mu_{h})}\right)^{d_{k'}+2d_m}\\\nonumber
&& \quad\times 
\bar{a}^{(H1)}_{b,jj'}
\left[\left(1+\frac{\alpha_s(\mu_{h})}{4\pi} \hat{K}\right) \hat{W}\right]_{ai}
\left[\hat{V}^{-1}\left(1-\frac{\alpha_s(\mu_W)}{4\pi} \hat{J}\right) \vec{C}_0\right]_k
\left[\hat{V}^{-1}\left(1-\frac{\alpha_s(\mu_H)}{4\pi} \hat{J}\right) \vec{C}_0\right]_{k'}
\\\nonumber
&& \quad\times 
 R^{NLO}\Bigg(d_i-d_k-d_{k'}+2d_m,\\\nonumber
&&  \qquad\qquad\qquad
 \hat{W}^{-1}_{ib} \hat{V}_{jk} \hat{V}_{j'k'} 
 \times  \left(1 -\frac{\alpha_s(\mu_W)}{4\pi}2J_m-\frac{\alpha_s(\mu_H)}{4\pi}2J_m
   +\frac{\alpha_s(m_U)+\alpha_s(m_V)}{4\pi}3J_m\right), \\\nonumber
&&   \qquad\qquad\qquad
 -2J_m \hat{W}^{-1}_{ib} \hat{V}_{jk} \hat{V}_{j'k'}
 - (\hat{W}^{-1}\hat{K})_{ib} \hat{V}_{jk} \hat{V}_{j'k'}
 +\hat{W}^{-1}_{ib} (\hat{J}\hat{V})_{jk} \hat{V}_{j'k'}
 + \hat{W}^{-1}_{ib} \hat{V}_{jk} (\hat{J}\hat{V})_{j'k'},\\
&&   \qquad\qquad\qquad
   m_V,\mu_W\Bigg)\,,
\end{eqnarray}
where $\bar{a}^{(H1)}_{a,ij}$ provides the
coefficients for the matching from the two-point function of $O_{1,2}^{VLR}$  to the local operators $Q_{1,2}^{LR}$ at $\mu=k^2$:
\begin{equation}
C^{LR}_a  \leftrightarrow \sum_{ij} \bar{a}^{(H1)}_{a,ij} C_i^{VLR} C_j^{VRL}\,,
\end{equation}
with the non-vanishing entries
\begin{equation}
 \bar{a}^{(H1)}_{1,12}=-2\,, \qquad  \bar{a}^{(H1)}_{1,21}=-2\,, \qquad  \bar{a}^{(H1)}_{1,11}=-6\,,
 \qquad  \bar{a}^{(H1)}_{2,22}=4\, .
\end{equation}
The only relevant case is $tt$, where $R^{NLO}$ can be replaced by $R^{NLO}_1$.

If we consider tree-level $H^0$ exchanges, we have
\begin{eqnarray}\nonumber
\bar\eta^{(H)}_{a,UV}&=& \left(\frac{\alpha_s(m_U)}{\alpha_s(\mu_{h})}\right)^{-d_m} 
    \left(\frac{\alpha_s(m_V)}{\alpha_s(\mu_{h})}\right)^{-d_m} 
    \left(\frac{\alpha_s(\mu_H)}{\alpha_s(\mu_{h})}\right)^{2d_m} \\
&&\qquad\times   \left(1-\frac{\alpha_s(\mu_H)}{4\pi}2J_m
   +\frac{\alpha_s(m_U)+\alpha_s(m_V)}{4\pi}J_m\right) \\
&&\qquad\times 
 \left[\left(1+\frac{\alpha_s(\mu_{h})}{4\pi} \hat{K}\right) \hat{W}
 \left(\frac{\alpha_s(\mu_H)}{\alpha_s(\mu_{h})}\right)^{\vec{d}}
 \hat{W}^{-1}\left(1-\frac{\alpha_s(\mu_H)}{4\pi} \hat{K}\right)\vec{C}_0
 \right]_a\,, \nonumber
\end{eqnarray}
where the matching yields the value of the Wilson coefficients for the $|\Delta S|=2$ operators at the high scale. One can check that the expressions from Ref.~\cite{Ecker:1985vv} are recovered at leading order.

\begin{table}[t!]
\begin{center}
{\renewcommand{\arraystretch}{2}
\begin{tabular}{|c| c   |}
\hline
& $1/\epsilon$\\
\hline
$d_1$&$-2  ( 6 R -7 +\xi (2 R -1))$\\
\hline
$d_2$ &$ \left(\left(24-\frac{\bar{b}}{4}\right) \lambda +\frac{\bar{b}}{4}+\xi  (4-4 R)-30\right)$ \\
\hline
$d_3$ &$ \left( - 3 + \frac{\bar{b}}{8} + 2  \xi \left(2 \log\left(\frac{m_s^2}{\mu^2}\right)+1\right)\right)$\\
\hline
$\tilde{d}_3$ & $ - (3 +\xi)$ \\
\hline
$d_4$ &$ \left(\left(48-\frac{\bar{b}}{2}\right) \lambda +\frac{\bar{b}-72}{4}+4 \xi \right)$\\
\hline
$d_5$ &$ 2 (7 -\xi)$\\
\hline
$d_6$ &$(-3 + 2\xi)$\\
\hline
$d_7$ &$ \left(\left(24-\frac{\bar{b}}{4}\right) \lambda +\frac{\bar{b}}{8}-2 \xi -32\right)$\\
\hline
\end{tabular}}
\end{center}
\caption{\small\it Divergences $d_i, \tilde d_i$ of the two-loop diagrams $D_i$ leading to $Q_i$. $\lambda$ multiplies the contribution from the evanescent operators which vanish for the standard value  $\bar{b}=b=96$, see Eq.~\eqref{eq:stvalai}. The exact definition of the divergences can be found in Eq.~\eqref{eq:defDi}.}
\label{tab:div}
\end{table}

\section{Result for the individual diagrams}\label{app:indgraph}

In order to evaluate the diagrams necessary to determine the short-distance QCD 
corrections for meson mixing in Left-Right models we used the packages Feyncalc and 
TARCER~\cite{Mertig:1998vk}. 

\subsection{Diagrams $D_i$}

The two-loop diagrams in Fig.~\ref{fig:diagsDi} have the following structure
\begin{eqnarray}
D_i^{rl}&=& -i\frac{m_c^2}{64 \pi^2} \frac{\alpha_s}{4 \pi} \biggl(\biggl[
-\frac{1}{\epsilon} \biggl(C_i^{rl} d_i - 2 \widetilde{C_i}^{rl} \tilde d_i \biggr)+
 (C_i^{rl} A_i - 2 \widetilde{C_i}^{rl} B_i ) \biggr] P_R \otimes P_L 
\nonumber\\
&+& \left[-\frac{1}{\epsilon}  \left( \tilde{d}_i C_i^{rl} -  \widetilde{C_i}^{rl}  d_i /2 \right) + 
     ( B_i C_i^{rl} -  \widetilde{C_i}^{rl}  A_i /2 )\right] \gamma_\mu P_R \otimes  \gamma_\mu P_L \cdots\biggr)
\label{eq:defDi}
\end{eqnarray}
where the ellipsis stands for possible other operators (and $1/\epsilon^2$ poles) uninteresting for our
purpose. The coefficients $d_i$ and $\tilde d_i$ of the $1/\epsilon$ term are given in Tab.~\ref{tab:div} while the 
$C_i^{rl}$ and  $\widetilde{C_i}^{rl}$ are colour factors given in Tab.~\ref{tab:Dcolour}. 
The diagram $D_8=0$ for zero external momenta. Other classes can be obtained
through either a rotation of 90 or 180 degrees, or a left-right reflection
(resulting in the exchange $r \leftrightarrow l$ in the colour factors in some cases, see Tab.~\ref{tab:Dcolour}).

\begin{table}[t!]
\begin{center}
{\renewcommand{\arraystretch}{1.4}
\begin{tabular}{|c| c c c c |}
\hline
 & $D_0$& $D_1$ & $D_2$ &$D_3$   \\
\hline
$C^{r l}$ & 1 & $\frac{N^2-1}{2 N}-  \tau_{rl}$ &  $-\frac{1}{2 N}$ &
 $-\frac{
1}{2 N}- \tau_{rl}$ \\
\hline
$\widetilde{C}^{rl}$ &$-2 \tau_{rl}$ & $\frac{ \tau_{rl}}{ N}$ &  $\frac{
1}{2 }-\frac{N^2-1}{N}  \tau_{rl}$ &  $\frac{
1}{ 2}+\frac{1}{N}  \tau_{rl}$   \\
\hline
\hline
 & $D_4$ & $D_5$ & $D_6$& $D_7$ \\
\hline
$C^{r l}$  &  $-\frac{1}{2 N}$  & $\frac{N^2-1}{2N}$ &  $\frac{N^2+ r N -1}{2N}$ &$\frac{ r N -1}{2N}$\\
\hline
$\widetilde{C}^{r l}$  &  $\frac{
1}{2}-\frac{N^2-1}{N}  \tau_{rl}$ &  $-\frac{
(N^2-1)}{ N}  \tau_{rl}$ &  $\frac{(N^2 -1)l-r}{2N} $ &$\frac{
l (N^2-1)+N- r }{2 N}$  \\
\hline
\end{tabular}}
\end{center}
\caption{\small\it Colour factors for the diagrams  $D_i$. $r,l$ can have the values  $\pm 1$
and $ \tau_{rl}$ is defined in Eq.~\eqref{eq:taurl}.}
\label{tab:Dcolour}
\end{table}

The finite gauge-independent part is given by
\begin{eqnarray}
A_1&=&6\bigl(-2 (R-2) \log \left(m_c^2/\mu^2\right)+ \log ^2\left(m_c^2
/\mu^2\bigr)+2
   R/3- R_2-\pi ^2/6+8/3\right)\,,\nonumber\\
A_2&=&-6 \big( \log \left(m_c^2/\mu^2\right)+ R+1/2\bigr)\,,\nonumber\\
A_3&=& \frac{3}{2} \left(12 \log(m_s^2/\mu^2)-13 \right)\,, \nonumber\\
B_3&=&-3  \big( \log(m_s^2/\mu^2)+ \log \left(m_c^2/\mu^2\right)+7/6\bigr)\,,\nonumber\\
A_4&=&12 \log \left(m_c^2/\mu^2\right)+59\,, \nonumber\\
A_5&=&-2  \bigl(3 \log ^2\left(m_c^2/\mu^2\right)+4 \log \left(m_c^2/\mu^2\right)+3\bigr)\,,\nonumber\\
A_6&=&-6  \bigr( \log \left(m_c^2/\mu^2\right)+7/12\bigr)\,,\nonumber\\
A_7&=& 6 \log ^2\left(m_c^2/\mu^2\right)-16 \log \left(m_c^2/\mu^2\right)+5\,,
\end{eqnarray}
while the gauge-dependent part
\begin{eqnarray}
A_1^{\xi}&=&2 \bigl(-2 (R-2) \log \left(m_c^2/\mu^2\right)+ \log
   ^2\left(m_c^2/\mu^2\right)- 2 R - R_2+7-\pi ^2 /6\bigr)\,,
\nonumber\\
A_2^{\xi}&=&-4\bigl( (R-1) \log \left(m_c^2/\mu^2\right)+ R+ R_2/2-1+\pi
   ^2/12\bigr)\,,
\nonumber\\
A_3^{\xi}&=&2 \bigl( \log(m_s^2/\mu^2)^2+2 (\log(m_s^2/\mu^2)-1) \log
   \left(m_c^2/\mu^2\right)+4 \log(m_s^2/\mu^2)-\bigr. 
\nonumber\\
&&\bigl. \log
   ^2\left(m_c^2/\mu^2\right)+\pi ^2/6-5\bigr),\nonumber\\
  B_3^{\xi}&=&-\log(m_s^2/\mu^2)-\log \left(m_c^2/\mu^2\right)-1/2\,,
\nonumber\\
A_4^{\xi}&=&-4 \bigl(\log ^2\left(m_c^2/\mu^2\right)+2 \log
   \left(m_c^2/\mu^2\right)+5\bigr)\,,
\nonumber\\
A_5^{\xi}&=&2 \bigl(\log ^2\left(m_c^2/\mu^2\right)+2 \log
   \left(m_c^2/\mu^2\right)+5\bigr)\,,
\nonumber\\
A_6^{\xi}&=&-2 \bigl(\log ^2\left(m_c^2/\mu^2\right)+2 \log
   \left(m_c^2/\mu^2\right)+5\bigr)\,,
\nonumber\\
A_7^{\xi}&=&2 \bigl(\log ^2\left(m_c^2/\mu^2\right)+2 \log
   \left(m_c^2/\mu^2\right)+5\bigr)\,.
\end{eqnarray}
$R$ and $R_2$ are defined as
\begin{eqnarray}
R&=& \frac{1} {m_s^2- m_d^2} \bigl( m_s^2  \log(m_s^2/\mu^2) - m_d^2 \log(m_d^2/\mu^2)\bigr)\,,
\nonumber\\
R_2&=&  \frac{1} {m_s^2- m_d^2} \bigl(m_s^2  \log^2(m_s^2/\mu^2) - m_d^2 \log^2(m_d^2/\mu^2)\bigr)\,.
\end{eqnarray}

They add up to
\begin{equation}
\langle O_{rl} (\mu) \rangle^{(1)} = \langle O_{rl} (\mu) \rangle^{(0)} - \frac{m_c^2 (\mu)}{64 \pi^2} \frac{\alpha_s (\mu)}{4 \pi} \sum_{i=1}^2 (\langle Q^{LR}_i (\mu) \rangle^{(0)} d_i^{rl}(\mu) + \cdots),
\end{equation}
with
\begin{eqnarray}
&& N d_1^{r l}(\mu)= e_1^{r l}(\mu) + \xi \biggl[ \log \left(\frac{m_c^2}{\mu ^2}\right)
\nonumber \\
&&\quad \left(\left(N+2 \tau_{rl}\right) \log \left(\frac{m_d^2 m_s^2}{\mu ^4}\right)+R \left(4
   \left(N^2-2\right) \tau_{rl}-2 N\right)-2 \left(2 N^2+N-4\right)
   \tau_{rl}+2 N-1\right) \biggr.
\nonumber \\
&&\quad+\left((4-N) \tau_{rl}+2
   N-\frac{1}{2}\right) \log \left(\frac{m_d^2 m_s^2}{\mu ^4}\right)+R_2
   \left(2 \left(N^2-2\right) \tau_{rl}-N\right)
\nonumber \\
&&\quad+R \left(4
   \left(N^2-2\right) \tau_{rl}-2 N\right)
\nonumber \\
&&\quad\biggl. +\left(\frac{1}{3} \left(\pi
   ^2-12\right) N^2-N-\frac{\pi ^2}{3}+8\right) \tau_{rl}+T
   \left(\frac{N}{2}+\tau_{rl}\right)+2 N-\frac{1}{2} \biggr]\,,\\
&& N d_2^{r l}(\mu)=  e_2^{r l}(\mu)+\xi  \biggl[\log \left(\frac{m_c^2}{\mu ^2}\right)
\nonumber\\
&&\quad \left(\left(4 N \tau_{rl}+2\right) \log \left(\frac{m_d^2 m_s^2}{\mu ^4}\right)+R
   \left(4 \left(N^2-2\right)-8 N \tau_{rl}\right)-2 \left(2
   N^2+N-4\right)+(8 N-4) \tau_{rl}\right) \biggr.
\nonumber\\
&&\quad+\left((8 N-2) \tau_{rl}-N+4\right) \log \left(\frac{m_d^2 m_s^2}{\mu ^4}\right)+R
   \left(4 \left(N^2-2\right)-8 N \tau_{rl}\right)
\\
&&\quad\biggl. +R_2 \left(2
   \left(N^2-2\right)-4 N \tau_{rl}\right) +\frac{1}{3} \left(\pi
   ^2-12\right) N^2+(2 N (T+4)-2) \tau_{rl}-N+T-\frac{\pi ^2}{3}+8 \biggr]\,.\nonumber
\end{eqnarray}
$\xi =0$ corresponds to the gauge-independent results,  $T=\log^2(m_d^2/\mu^2)+\log^2( m_s^2/\mu^2)$
and $\beta_{rl}=l+r$.
The gauge-independent parts $ e_i^{r l}(\mu)$ are given by:
\begin{eqnarray}
 e_1^{r l}(\mu)&=&
\log \left(\frac{m_c^2}{\mu ^2}\right) \left(-11 \left(N^2-2\right) \beta_{rl}+\left(8 N^2-6 N+16\right) \tau_{rl}-16 N-12 R \tau_{rl}-3\right)
\nonumber\\
&&+\log ^2\left(\frac{m_c^2}{\mu ^2}\right) \left(3
   \left(N^2-2\right) \beta_{rl}+6 N^2 \tau_{rl}+6
   N\right)
\nonumber\\
&&+\left((9-3 N) \tau_{rl}+\frac{3}{2} (3 N-1)\right) \log
   \left(\frac{m_d^2 m_s^2}{\mu ^4}\right) +R \left(\left(6 N^2-2\right) \tau_{rl}-3 N\right)
-6 R_2 \tau_{rl}
\nonumber\\
&&+\frac{3}{4} \left(N^2-2\right) \beta_{rl}+\left(-\frac{41 N^2}{2}-7 N-\pi ^2+17\right) \tau_{rl}+\frac{1}{2} (17 N-7)\,,
\label{eq:c1}
\end{eqnarray}
\begin{eqnarray}
 e_2^{r l}(\mu)&=&\log \left(\frac{m_c^2}{\mu ^2}\right) (R \left(12 \left(N^2-1\right)-24
   N \tau_{rl}\right)+22 N \beta_{rl}+(48 N-12) \tau_{rl}
\nonumber\\
&&-2 (N (2 N+3)+14)) 
\nonumber\\
&&+\log ^2\left(\frac{m_c^2}{\mu
   ^2}\right) \left(-6 N \beta_{rl}+12 N \tau_{rl}+12\right)+\left((18 N-6) \tau_{rl}-3 N+9\right) \log
   \left(\frac{m_d^2 m_s^2}{\mu ^4}\right)
\nonumber\\
&&+R \left(-4 N^2+8 N \tau_{rl}-2\right)+R_2 \left(6 \left(N^2-1\right)-12 N \tau_{rl}\right)
\nonumber\\
&&-\frac{ 3 N \beta_{rl}}{2}-\left(\left(7+2 \pi
   ^2\right) N+14\right) \tau_{rl}+N \left(\left(\pi ^2-3\right)
   N-7\right)-\pi ^2+20\,.
\label{eq:c2}
\end{eqnarray}

\subsubsection{Contributions of the diagrams $L_i$}

The diagrams $L_i$ have different types of contributions depending on the operators
involved.

\phantom{xxx}
\hspace{0.2cm}$\bullet$ {\bf Contribution from the operators $Q_i$}

\phantom{xxx}

The three diagrams of Fig.~\ref{fig:diagsLi} have to be evaluated with insertions of
the operators  $Q_i/\epsilon$ ($i=1,2$) defined in Eq.~\eqref{eq:opQi}.
Considering also the other members of each class of diagrams obtained by left-right and up-down reflections, we get
\begin{equation} 	
\langle Q_i (\mu) \rangle^{(1)} = \langle Q_i (\mu) \rangle^{(0)} +\frac {\alpha_s(\mu)}{4 \pi} \sum_j \biggl( \frac{h_{Q_i}}{\epsilon} \delta_{ij} +b_{ji}(\mu) \biggr) \langle Q_j (\mu) \rangle^{(0)}\, ,
\label{eq:Litot}
\end{equation}
where the divergent parts are 
\begin{eqnarray}\label{eq:qiinf}
h_{Q1} &=&\frac{3 R \tau_{rl}}{N}+\frac{1}{2}\left(\frac{4}{N}-3 N+3\right) \tau_{rl}+\frac{3}{4 N}+\frac{3}{2}
-\frac{\xi}{2} 
\biggl(\left(\frac{\tau_{rl}}{N}+\frac{1}{2}\right) \log \left(\frac{m_d^2 m_s^2}{\mu
   ^4}\right)
\biggr.
\nonumber\\
&&
\biggl.+R \left(\left(2 N-\frac{4}{N}\right) \tau_{rl}-1\right)+\left(\frac{4}{N}-2 N-1\right) \tau_{rl}-\frac{1}{2 N}+1
\biggr)\,,\nonumber\\
h_{Q2}&=& 3 R \left(- N+\frac{1}{N}+2 \tau_{rl}\right)+ N+\frac{2}{N}
+\frac{3}{2} +\left( \frac{3}{N}+1 \right) \tau_{rl}
-\frac{\xi}{2} \biggl(
\left(\frac{1}{N}+2 \tau_{rl}\right) \log \left(\frac{m_d^2 m_s^2}{\mu ^4}\right)
\biggr.
\nonumber\\
&&
\biggl.
+ 2 R
   \left( N-\frac{2}{N}-2 \tau_{rl}\right)-2 N+\frac{4}{N}-1+2 \left(2 -\frac{1}{N} \right)\tau_{rl}   
\biggr)\,.
\end{eqnarray}

\begin{table}[t!]
\begin{center}
{\renewcommand{\arraystretch}{1.4}
\begin{tabular}{|c| c c c  |}
\hline
$L_k$ & $L_1$ & $L_2$ &$L_3$   \\
\hline
$C^{k}$ &  $\frac{N^2-1}{2 N}$ &  $-\frac{1}{2N}$&  $-\frac{1}{2 N}$  \\
\hline
$\widetilde{C}^{k}$ &$0$ & $\frac{1}{2 }$ &  $\frac{1}{2}$   \\
\hline
\end{tabular}}
\end{center}
\caption{\small\it Colour factors for the diagrams  $L_k$.}
\label{tab:Lcolour}
\end{table}

The finite parts of the diagrams in Fig.~\ref{fig:diagsLi} with insertions from the operators 
$Q_i$ divided by $\epsilon$ can be written in the following way
\begin{eqnarray}
&& Q_{ii}^{(1)}= \sum_{k=1}^3(\bar{A}_{ii}^k C_k + f_i \bar{A}_{ji}^k \widetilde{C_k}  )  Q_i \,,\quad Q_{ij}^{(1)}= \sum_{k=1}^3( \bar{A}_{ii}^k  \widetilde{ C_k}/f_i  + 
\bar{A}_{ji}^k C_k ) Q_j, \nonumber\\
&& Q_{i}^{(1)} = Q_{ii}^{(1)} + Q_{ij}^{(1)} = \sum_m b_{mi} Q_{m}\,,
\label{eq:Qij}
\end{eqnarray}
(no sum on repeated indices) where $k$ denotes the diagram $k$ and $j=2,1$. $C_k$ and $ \widetilde{C_k}$ are the colour factors
given in Tab.~\ref{tab:Lcolour} and $f_i$ are coefficients coming from
the Fierz transformation, $f_1=-1/2$ and $f_2=-2$.
The $2 \times 2$ matrices $\bar{A}^{1,2}$ turn out to be diagonal. 
One has:
\begin{equation}\bar{A}^1= \left( \begin{array}{cc}
\frac{3}{2}R-\frac{5}{4} +G_a \xi & 0 \\
0 &  G^{\rm{ind}} + G_a  \xi
\end{array} \right) , \end{equation}
\begin{equation}\bar{A}^2= \left( \begin{array}{cc}
 G^{\rm{ind}}  +G_a \xi & 0 \\
0 &  \frac{3}{2}R-\frac{9}{4} + G_a  \xi
\end{array} \right) \, , \label{eq:ll}  \end{equation}
and 
\begin{equation}\bar{A}^3= \left( \begin{array}{cc}
  \frac{3}{2} \left( \log \left(\frac{m_s^2}{\mu^2}\right)-\frac{1}{2}\right) +G_b\xi & 
\frac{1}{4} \left(3 \log \left( \frac{m_s^2}{\mu^2} \right)-
\frac{5}{2}\right)+ \frac{1}{4} \xi\left[  \log \left( \frac{m_s^2}{\mu^2} \right)-\frac{3}{2}\right] \\
3 \log \left( \frac{m_s^2}{\mu^2} \right)-\frac{11}{2}+\xi
\left[ \log \left( \frac{m_s^2}{\mu^2} \right)-\frac{5}{2}\right] &  \frac{3}{2} \left( \log \left(\frac{m_s^2}{\mu^2}\right)-\frac{1}{2}\right)+G_b\xi
\end{array} \right) \end{equation}
with
\begin{eqnarray}
G^{\rm{ind}}&=& \frac{1}{2} \left( -2 R+3 R_2+ \frac{\pi^2}{2} +2 \right) , \quad G_a = \frac{1}{4} \left( -4 R+2 R_2 + \frac{\pi ^2}{3}+4 \right) ,
\nonumber\\ 
G_b&=&-\frac{1}{2} \left(\log ^2\left(\frac{m_s^2}{\mu ^2}\right)+\frac{\pi ^2}{6}\right) .
\end{eqnarray}
Note that the 
graph $L_2$ can be obtained 
from $L_1$ by a Fierz transformation. It is easy to check that this implies 
that $\bar{A}_1$ and $\bar{A}_2$ are obtained from one another by interchanging their 
diagonal elements. One can see that this is indeed the case for the gauge-dependent terms but not for the terms independent of the regularisation
in the gauge-independent ones. This comes from the fact that the relations for
the Fierz transformation are generally valid only in 4 dimensions. The corrections in $D$
dimensions define the evanescent operators $E_5$ and $E_6$, Eq.~\eqref{eq:eva1}.

One can perform a similar computation inserting $Q_i$ (without additional $1/\epsilon$ contribution). We get the following finite parts for the first diagram $L_1$
\begin{equation}A^1= \left( \begin{array}{cc}
 -\frac{3}{2}+\xi(1-R)& 0 \\
0 & 1-3 R +\xi(1-R)
\end{array} \right).  \label{eq:a1threeflav}
\end{equation} 
$A^2$ can be obtained from $A^1$ by interchanging the diagonal elements. 
This can be understood easily, since
$L_2$ can be obtained from $L_1$ by a Fierz 
transformation and the evanescent operators have been defined so 
as to  conserve the Fierz relations.
Evaluating $L_3$
one gets
\begin{equation}A^3= \left( \begin{array}{cc}
 -\frac{3}{2}+\xi \log \left(\frac{m_s^2}{\mu ^2}\right)& -(3+\xi)/4 \\
-3-\xi & -\frac{3}{2}+\xi \log \left(\frac{m_s^2}{\mu ^2}\right)
\end{array} \right).  \label{eq:a3threeflav}
\end{equation}
The infinite parts of these diagrams are related to the LO anomalous 
dimensions of the operators $\tilde{Q}^{LR}_{1,2}$ and $Q^{LR}_{1,2}$. We have checked that they agree with the 
ones obtained in Ref.~\cite{Buras:2000if}.

Adding up these contributions, the elements $(ij)$ of the gauge-independent finite part of the matrix $4 N  b(\mu)$ 
in Eq.~\eqref{eq:Litot} are given by
\begin{eqnarray}
(11)&=&-3 (N+1) \log \left(\frac{m_d^2 m_s^2}{\mu^4}\right)+2\left(3 N^2-1\right) R-6 R_2 -5 N^2+11 N -\pi ^2 +8\,,
\nonumber\\
(12)&=&-\frac{3}{2} (N+1) \log \left(\frac{m_d^2 m_s^2}{\mu^4}\right)-3 N R+6 N+\frac{5}{2} \,,
\\
(21)&=&-6 (N+1) \log \left(\frac{m_d^2 m_s^2}{\mu^4}\right)+8 N R-12 N R_2+2 \left(\left(3-\pi ^2\right)N+11\right)  \,,
\nonumber\\
(22)&=& -3 (N+1) \log \left(\frac{m_d^2 m_s^2}{\mu^4}\right)-2\left(2 N^2+1\right) R+6 \left(N^2-1\right) 
R_2 \nonumber\\
&& + N(4N+5)+8+ \pi^2\left(N^2 -1\right)\,, \nonumber
\label{eq:Licoefftot}
\end{eqnarray} 
and the gauge-dependent ones by
\begin{eqnarray}
(11)&=&
\xi \biggl(T-N \log \left(\frac{m_d^2 m_s^2}{\mu^4}\right)+4\left(2- N^2\right) R+2 \left(N^2-2\right) R_2 \nonumber\\
&& + N(4 N+5)-8+\frac{\pi^2}{3} (N^2-1) \biggr) , \,
\nonumber\\
(12)&=&\xi \biggl(\frac{N}{2} T -\frac{1}{2} \log \left(\frac{m_d^2 m_s^2}{\mu^4}\right)+2 N R-N R_2 -2N +\frac{3}{2}  \biggr)\,,
\\
(21)&=& \xi \biggl(2 N T-2 \log \left(\frac{m_d^2 m_s^2}{\mu^4}\right)+8 N R-4 N R_2 -8N +10 \biggr)\,,
\nonumber\\
(22)&=&\xi \biggl(T-N \log \left(\frac{m_d^2 m_s^2}{\mu^4}\right)+4\left(2- N^2\right) R+2 \left(N^2-2\right) R_2 \nonumber\\
&&+ N(4 N+3)-8+\frac{\pi^2}{3}
 (N^2-1) \biggr)  \, .\nonumber
\label{eq:Licoefftotxi}
\end{eqnarray}

\subsubsection{Insertion of $E_i$}

The contribution of the evanescent operators $E_{1,3,5}$ 
have also to be evaluated. In principle, a finite
contribution could be added to these evanescent operators in the same way as
for the $C_i$. However, as indicated earlier, it has been shown in Refs.~\cite{Herrlich:1994kh,Dugan:1990df} that the result should not depend on the value of the constant
coefficients and that one can choose a regularisation scheme where these
contributions cancel. Summing all the diagrams $L_i$ together
with all the members of the same class (not shown, obtained by left-right and
up-down reflections)
one gets both finite and infinite parts, with a similar structure to Eq.~\eqref{eq:Litot}. The divergent pieces are
\begin{eqnarray} \label{eq:evaninf}
h_{E5,1} &=& -12 +\frac{\bar{b}}{4} \, , \qquad \qquad\qquad\quad h_{E5,2}= - 24 N + \frac{\bar{b}}{2 N} \,,
\nonumber \\
h_{E1,1} &=& 0 \, , \qquad \qquad \qquad \qquad \qquad h_{E1,2} = 0\,,
\nonumber \\
h_{E6,1} &=& -12 N+\frac{\bar{b} \left(N^2-2\right)}{8 N}\, , \quad \quad h_{E6,2} = -\frac{\bar{b}+96}{4}\,,
\end{eqnarray}
while the finite parts read
\begin{eqnarray}
b_{E1,1}&=& \biggl( 3 + 2 N +  \xi \biggr)  \frac{1}{4}  \,, \nonumber\\
b_{E6,1}&=&  \left( \frac{6}{N} \log \left( \frac{m_d^2 m_s^2 }{\mu^4} \right) + 
\frac{12}{N} R +14 N -\frac{24}{N} + 3+ \xi \right) \,,\nonumber\\
b_{E5,1}&=&   -6\biggl ( \log \left(\frac{m_d^2 m_s^2 }{\mu^4}\right) -3 \biggr) \,, \nonumber\\
b_{E1,2}&=& \biggl( 3 + 2 N +  \xi \biggr) \left( \frac{1}{2 N} \right) \,,\\
b_{E6,2}&=& 12  \biggl(\log \left(\frac{m_d^2 m_s^2 }{\mu^4}\right)+ 2 R +\frac{1}{2 N} -\frac{5}{3}+ \frac{1}{6 N} \xi \biggr)\,, \nonumber\\
b_{E5,2}&=&  \frac{12}{N} \biggl(- \log \left(\frac{m_d^2 m_s^2 }{\mu^4}\right) +2(N^2-1) R - N^2+4\biggr) \,. \nonumber
\end{eqnarray}

\end{document}